\documentclass[aps,prb,amsmath,amssymb,reprint,superscriptaddress,preprintnumbers,showpacs,intlimits,shortbibliography]{revtex4-2}
\pdfoutput=1
\bibpunct{[}{]}{,}{n}{}{}

\usepackage{bm,latexsym,mathrsfs,enumerate}
\usepackage{xcolor}
\usepackage{mathtools}
\usepackage[breaklinks=true,unicode=true,urlcolor = blue,colorlinks = true,citecolor = blue,linkcolor = blue]{hyperref}
\usepackage{pgf,tikz}
\usepackage{graphicx}
\usepackage[most]{tcolorbox}
\usepackage{multirow}
\usepackage{wasysym}
\usepackage[mathcal]{euscript}
\usepackage{siunitx}
\renewcommand{\vec}[1]{\bm{#1}}

\newcommand{\dd}{\mathrm{d}}

\graphicspath{{./figs/}}
\usepackage[normalem]{ulem}
\begin{document}
%
%
\title{Nonlinear dynamics of skyrmion strings}
%
%

\author{Volodymyr P. Kravchuk}
\email{v.kravchuk@ifw-dresden.de}
\affiliation{Leibniz-Institut f\"ur Festk\"orper- und Werkstoffforschung, IFW Dresden, 01171 Dresden, Germany}
\affiliation{Institut f\"ur Theoretische Festk\"orperphysik, Karlsruher Institut f\"ur Technologie, 76131 Karlsruhe, Germany}
\affiliation{Bogolyubov Institute for Theoretical Physics of the National Academy of Sciences of Ukraine, 03143 Kyiv, Ukraine}
%


%
%
%
%
\begin{abstract}
	The skyrmion core, percolating the volume of the magnet, forms a skyrmion string -- the topological Dirac-string-like object. Here we analyze the nonlinear dynamics of skyrmion string in a low-energy regime by means of the collective variables approach which we generalized for the case of strings. Using the perturbative method of multiple scales (both in space and time), we show that the weakly nonlinear dynamics of the translational mode propagating along the string is captured by the nonlinear Schr{\"o}dinger equation of the focusing type. As a result, the basic  ``planar-wave'' solution, which has a form of a helix-shaped wave, experiences modulational instability. The latter leads to the formation of cnoidal waves. Both types of cnoidal waves, dn- and cn-waves, as well as the separatrix soliton solution, are confirmed by the micromagnetic simulations. Beyond the class of the traveling-wave solutions, we found Ma-breather propagating along the string. Finally, we proposed a generalized approach, which enables one to describe nonlinear dynamics of the modes of different symmetries, e.g. radially symmetrical or elliptical.
\end{abstract}
\maketitle
%
%
%
%
\section{Introduction}
\begin{figure}
	\includegraphics[width=\columnwidth]{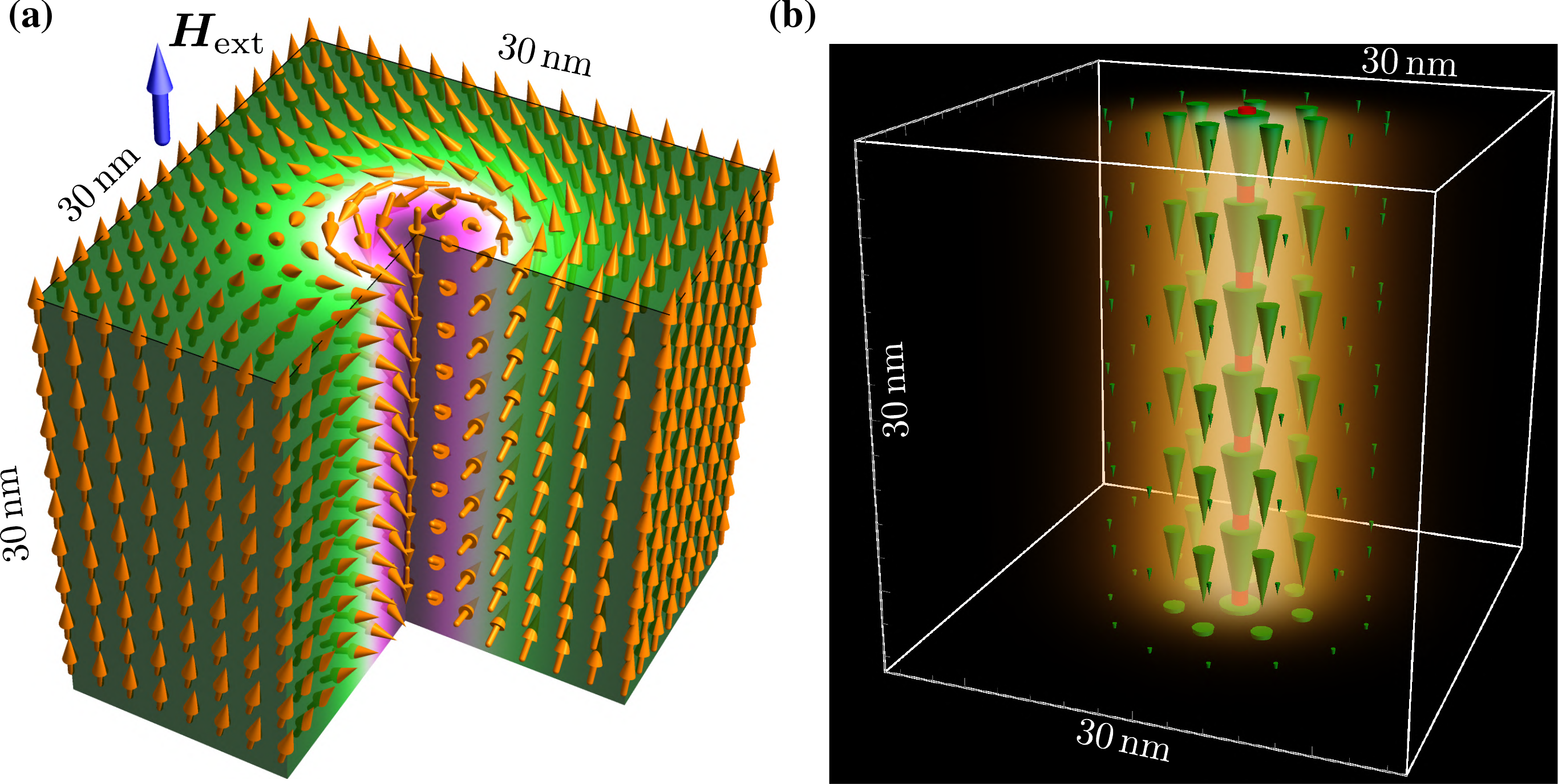}
	\caption{Representation of skyrmion string in terms of magnetization and the emergent magnetic field $\vec{B}^{\text{e}}$~\cite{Schulz12}  are shown in panels (a) and (b), respectively. On panel (a), arrows show the distribution of the magnetization $\vec{m}$ and color corresponds to the component $m_z$. On panel (b), arrows show the distribution of $\vec{B}^{\text{e}}\propto\vec{g}$ with $|\vec{B}^{\text{e}}|$ reflected by the color intensity, see Eq.~\eqref{eq:g}. The central red line shows position of the string center defined in \eqref{eq:Xi}. The data is obtained by means of micromagnetic simulations for FeGe in external field $\mu_0H_{\text{ext}}=0.8$~T.
	}\label{fig:string}
\end{figure}

Magnetic skyrmions~\cite{Seki16,Liu20c,Back20,Nagaosa13,Fert17,Wiesendanger16,Bogdanov20a} are traditionally considered as two-dimensional topological solitons existing in magnetic films with Dzyaloshinskii-Moriya interaction (DMI). During the last decade, skyrmionics demonstrated an explosive development, which, in part, is due to the number of skyrmion properties potentially useful for the application in spintronic devices, namely topological protection and controllability by electrical currents \cite{Sampaio13,Zhang15c,Fert13}. Although, the first experimental observation of skyrmion lattices~\cite{Muehlbauer09} implied the existence of the skyrmion strings, aligned along the applied magnetic field, the three-dimensional structure of skyrmions was not considered in the most studies. However, recent advances in the experimental techniques enabled the real-space imaging of skyrmion strings in noncentrosymmetric bulk magnets~\cite{Seki21,Birch20,Wolf21}. Skyrmion string (tube) is a skyrmion core percolating the volume of the magnet, see Fig.~\ref{fig:string}(a). It is quite analogous to vortex filaments in superfluids~\cite{Pismen99,Sonin87}, superconductors~\cite{Blatter94}, Bose-Einstein condensates~\cite{Madison00}. Skyrmion strings carry an emergent magnetic field~\cite{Schulz12} which is the source of the topological Hall effect~\cite{Bruno04} experienced by the conducting electron, see Fig.~\ref{fig:string}(b). The skyrmion string-based realization of the Dirac strings in condensed matter was discussed~\cite{Lin16}. Note that termination of the skyrmion string results in creation of the pair of Dirac monopole and antimonopole known as Bloch points~\cite{Lin16,Braun12,Milde13,Schuette14b}. 

A number of physical effects are already established for skyrmion strings. It was shown that the spin excitations can propagate along the string on distance of tens of micrometers~\cite{Seki20}. Thus, the strings can be considered as magnonic waveguides for the information transfer. The nucleation and annihilation of skyrmion strings can be effectively controlled by the external magnetic field~\cite{Mathur21}, as well as by electrical current~\cite{Birch22}.  Skyrmion strings can be moved by the spin-polarized current applied perpendicularly to the string~\cite{Yokouchi18,Jonietz10,Yu12,Tang21}. This current-induced dynamics has a threshold character caused by the effects of pinning on the impurities. The longitudinal current, however, leads to the string instability~\cite{Okumura23}. Skyrmion strings can merge or unwind by means of Bloch points~\cite{Milde13,Kagawa17,Yu20,Xia22}. Position of the Bloch point along the string can be controlled by the current pulses, opening up a range of design concepts for future 3D spintronic
devices~\cite{Birch22}. A bunch of skyrmion strings immersed into the conical phase can twist and create a braid superstructure~\cite{Zheng21}. It worth mentioning that in addition to the skyrmion strings there is a number other string-like topological objects in magnets that are of interest from both fundamental and applied points of view, e.g. vortex strings~\cite{Yan07,Ding14,Guslienko22,Finizio22,Volkov23} and screw dislocations~\cite{Azhar22}.

The linear spin excitations propagating along skyrmion strings are well studied both theoretically~\cite{Lin16,Kravchuk20,Seki20} and  experimentally~\cite{Seki20}. Previously we also reported on finding a nonlinear solution in form of the solitrary wave propagating along skyrmion string~\cite{Kravchuk20}. Here we report on a systematic study of the possible nonlinear low-energy string dynamics. To this end we generalized the collective variables approach and obtain a Thiele-like equation for the string. Next, using the perturbative method of multiple scales~\cite{Ryskin00,Nayfeh08} we demonstrate that for the case of the low-amplitude dynamics, the string equation of motion is reduced to the nonlinear Schr{\"o}dinger equation (NLSE) of focusing type. Next, by means of the full-scale micromagnetic simulations we found a number of well-known solutions of NLSE, namely nonlinear cnoidal waves, solitons, and breathers. These numerically found solutions agrees very well with predictions of our model. Finally, we suggest the generalization of the used collective variables approach for the string-like collective variables of an arbitrary meaning.

\section{Definition of skyrmion string and its equation of motion}

We consider a cubic chiral ferromagnet saturated by an external magnetic field along $z$-axis. Such a magnet can host magnetic skyrmion as an excitation of the uniform ground state $\vec{m}(\vec{r})=\hat{\vec{z}}$~\cite{Bogdanov89r,Bogdanov94}. Here and below, the unit vector $\vec{m}$ denotes the dimensionless magnetization. In the case of a bulk sample, skyrmion core penetrates the magnet volume forming a string-like object. In equilibrium, the string is oriented along $\hat{\vec{z}}$. Deviation of the string shape from the equilibrium straight line results in the string dynamics. Our aim here is to describe this dynamics in the low-amplitude limit. 

We define the skyrmion string as a time dependent 3D curve $\vec{\gamma}(z,t)=\hat{\vec{x}}X_1(z,t)+\hat{\vec{y}}X_2(z,t)+z\hat{\vec{z}}$, where
\begin{equation}\label{eq:Xi}
	X_{i}(z,t)=\frac{1}{N_{\text{top}}}\iint x_ig_{z}(\vec{r},t)\dd x\dd y
\end{equation}
are the first moments of the topological charge density $g_{z}=\frac{1}{4\pi}\vec{m}\cdot\left[\partial_x\vec{m}\times\partial_y\vec{m}\right]$. Here $x_1=x$ and $x_2=y$. The total topological charge $N_{\text{top}}=\iint g_{z}\dd x\dd y$ is a constant integer number. Since the magnetization evolution is assumed  to be continuous in space and time, and the boundary conditions $\vec{m}=\hat{\vec{z}}$ are fixed for $\varrho=\sqrt{x^2+y^2}\to\infty$, the topological charge $N_{\text{top}}$ does not depend neither on time nor on $z$ coordinate. Note that $g_z$ is vertical (along the ground state) component of the gyrovector density
\begin{equation}\label{eq:g}
	\vec{g}=\frac{\vec{e}_i\epsilon_{ijk}}{8\pi}\vec{m}\cdot[\partial_j\vec{m}\times\partial_k\vec{m}]
\end{equation}
which determines the emergent magnetic field $\vec{B}^{\text{e}}\propto\vec{g}$~\cite{Schulz12}.

We base our study on Landau-Lifshitz equation $\partial_t{\vec{m}}=\frac{\gamma_0}{M_s}[\vec{m}\times\frac{\delta H}{\delta\vec{m}}]$, which equivalently can be written in the Hamiltonian form $\partial_t{\vec{m}}=\{\vec{m},H\}$ supplemented with the Poisson brackets for the magnetization components
\begin{equation}\label{eq:m-poiss}
	\{m_i(\vec{r}),m_j(\vec{r}')\}=-\frac{\gamma_0}{M_s}\epsilon_{ijk}m_k(\vec{r})\delta(\vec{r}-\vec{r}').
\end{equation}
Here $H$ is the Hamiltonian, $\gamma_0$ is gyromagnetic ratio and $M_s$ is the saturation magnetization. Using \eqref{eq:m-poiss} and definition \eqref{eq:Xi} we obtain
\begin{equation}\label{eq:X-poiss}
	\{X_i(z),X_j(z')\}=-\epsilon_{ij}\frac{\gamma_0}{M_s}\frac{\delta(z-z')}{4\pi N_{\text{top}}},
\end{equation}
for details see Appendix~\ref{app:PB}. This is the generalization of the previously obtained Poisson brackets for coordinates of 2D topological solitons \cite{Papanicolaou91}. In the following, we restrict ourselves by the low-energy limit and therefore we take into account only the collective variables $X_1$ and $X_2$, which are associated with the low-energy translation mode. In this case, the string equations of motion are $\partial_t{X}_i=\{X_i,H\}$. By means of \eqref{eq:X-poiss} we obtain the following explicit form for the equations of motion $\partial_tX_i=-\epsilon_{ij}G^{-1}\delta H/\delta X_j$, where $G=4\pi N_{\text{top}}M_s/\gamma_0$ 
and $H=\int\mathcal{H}(\partial_zX_i,\partial_{z}^2X_i,\dots)\dd z$. Note that due to the transnational invariance, the longitudinal Hamiltonian density $\mathcal{H}$ does not depend on $X_i$ but only on its derivatives. For a number of problems, it can be convenient to introduce vector $\vec{X}=X_1\hat{\vec{x}}+X_2\hat{\vec{y}}$. The corresponding equations of motion has the Thiele-like form~\cite{Kravchuk20,Ding14,Guslienko22} $\left[\partial_t{\vec{X}}\times\vec{G}\right]=\delta H/\delta\vec{X}$, where $\vec{G}=G\hat{\vec{z}}$ is the string gyrovector. In contrast to the conventional Thiele equation~\cite{Thiele73}: (i) the skyrmion position $\vec{X}(z,t)$ depends not only on time but also on the coordinate $z$, (ii) right hand side of the equation of motion contains functional derivative instead of the partial one.

In what follows, we, however, use the alternative representation by means the complex-valued function $\Psi=X_1+iX_2$. The corresponding equation of motion is of the Schr{\"o}dinger-like form
\begin{equation}\label{eq:Thiele-psi}
 i\partial_t{\Psi}=-\frac{2}{G}\frac{\delta H}{\delta\Psi^*}.
\end{equation}
Although the general form of equation of motion \eqref{eq:Thiele-psi} enables one to make a conclusion about some integrals of motion, e.g. total energy $E=H$ or linear momentum $P=G\frac{i}{4}\int(\Psi\partial_z\Psi^*-\Psi^*\partial_z\Psi)\dd z$, in order to obtain a concrete solution, one needs to know the structure of Hamiltonian $H=\int\mathcal{H}\dd z$. The concrete dependence of the effective Hamiltonian density $\mathcal{H}(\partial_z\Psi,\partial_z\Psi^*,\partial_z^2\Psi,\dots)$ on the collective string-variable $\Psi$ is determined by the magnetic interactions present in the system. In the following, we consider the case of cubic chiral ferromagnet (e.g. FeGe, MnSi, Cu$_2$OSeO$_3$) immersed in external magnetic field $\vec{H}_{\text{ext}}=H_{\text{ext}}\hat{\vec{z}}$. The corresponding Hamiltonian
\begin{equation}\label{eq:H}
H=\int\left(A\mathscr{H}_{\text{ex}}+D\mathscr{H}_{\textsc{dmi}}+\mu_0H_{\text{ext}}M_s\mathscr{H}_{\text{z}}\right)\dd\vec{r}	
\end{equation} 
collects three contributions, namely the exchange energy with $\mathscr{H}_{\text{ex}}=\partial_i\vec{m}\cdot\partial_i\vec{m}$, where $i=x,y,z$, Dzyaloshinskii-Moriya energy with $\mathscr{H}_{\textsc{dmi}}=\vec{m}\cdot[\vec{\nabla}\times\vec{m}]$, and Zeeman energy with $\mathscr{H}_{\text{z}}=1-m_z$. 

Typical length- and time-scales of the system \eqref{eq:H} are determined by the wave-vector $Q=D/(2A)$ and frequency $\omega_{c2}=\gamma_0D^2/(2AM_s)$, respectively. In terms of the dimensionless units $\tilde{\vec{r}}=\vec{r}Q$ and $\tilde{t}=t\omega_{c2}$, system  \eqref{eq:H} is controlled by a single parameter which is the dimensionless magnetic field $h=H_{\text{ext}}/H_{c2}$, where $\mu_0H_{c2}=\omega_{c2}/\gamma_0$. In the following, we consider the regime $H_{\text{ext}}>H_{c2}$ in which the ground state is uniformly polarized along the field. Such a polarized magnet can host isolated skyrmion strings as topologically protected excitations~\cite{Kravchuk20}. In the following we proceed to the dimensionless order parameter $\psi=\Psi Q$. 

Here we discuss two ways of derivation of the structure of the string Hamiltonian. The simplest way is based on the gradient expansion of the Hamiltonian density with respect to $\psi$, $\psi^*$ and their derivatives. In the expansion, we keep only real-valued terms, which are not total derivatives with respect to $z$ and which do not violate the translational and $U(1)$ symmetries. Due to the translational symmetry the expansion terms can depend only on derivatives $\psi^{(n)}$ and ${\psi^*}^{(m)}$, with $n,m=1,2,\dots$. The $U(1)$ symmetry is the consequence of the isotropy of the model \eqref{eq:H} within $xy$-plane. As a result, the string Hamiltonian must be invariant with respect to the arbitrary rotations within the $xy$-plane, i.e. with respect to the replacement $\psi\to e^{i\alpha}\psi$. The latter implies that the only quadratic blocks $\psi^{(n)}{\psi^*}^{(m)}$ are allowed in the string Hamiltonian expansion. This means that only even terms are allowed in the expansion, i.e. the leading nonlinear terms are of the 4th order. Finally, we present the string Hamiltonian as follows
\begin{equation}\label{eq:H-ser}
	\tilde{H}=\int\dd\tilde{z}\left[\mathscr{H}^{(0)}+\mathscr{H}^{(2)}+\mathscr{H}^{(4)}+\dots\right],
\end{equation}
where $\tilde{H}=H/(8\pi AQ^{-1})$ and $\tilde{z}=zQ$. Here $\mathscr{H}^{(0)}$ is the longitudinal energy density of the vertical unperturbed string, it does not depend on the collective string variable $\psi$. The harmonic and the leading nonlinear terms are as follows
\begin{equation}\label{eq:Bloch-H}
	\begin{split}
		&\mathscr{H}^{(2)}=\frac{a_1}{2}|\psi'|^2+i\sigma\frac{a_2}{4}(\psi'{\psi^*}''-{\psi^*}'\psi'')+\frac{a_3}{2}|\psi''|^2+\dots,\\
		&\mathscr{H}^{(4)}=-\frac{b_1}{4}|\psi'|^4-i\sigma\frac{b_2}{8}|\psi'|^2(\psi'{\psi^*}''-{\psi^*}'\psi'')+\dots,
	\end{split}
\end{equation}
where prime denotes the derivative with respect to $\tilde{z}$. The terms proportional to $(\psi'{\psi^*}''-{\psi^*}'\psi'')$ are responsible for the nonreciprocal effects since they are not invariant with respect to the transformation $\tilde{z}\to-\tilde{z}$. Since the presence of the derivatives $\partial_{\tilde{z}}$ in DMI is the only source of the non-reciprocity in the initial model \eqref{eq:H}, the nonreciprocal terms in \eqref{eq:Bloch-H} are proportional to  $\sigma=\text{sign}(D)=\pm1$. Note however, that DMI is not the only contributor to the non-reciprocity coefficients $a_2$ and $b_2$. As it will be shown latter (see Appendix~\ref{app:energies}), the exchange and Zeeman interactions also contribute due to the helicity $\varphi_0=\pm\pi/2$ of the Bloch skyrmion, such that $\sigma=\sin\varphi_0$. Indeed, due to the certain circulation of magnetization of the Bloch skyrmion string (clockwise or counter-clockwise), the directions $\pm\hat{\vec{z}}$ not equivalent. For N{\'e}el skyrmion string, the nonreciprocal terms are absent in the string Hamiltonian.

\begin{figure}
	\includegraphics[width=\columnwidth]{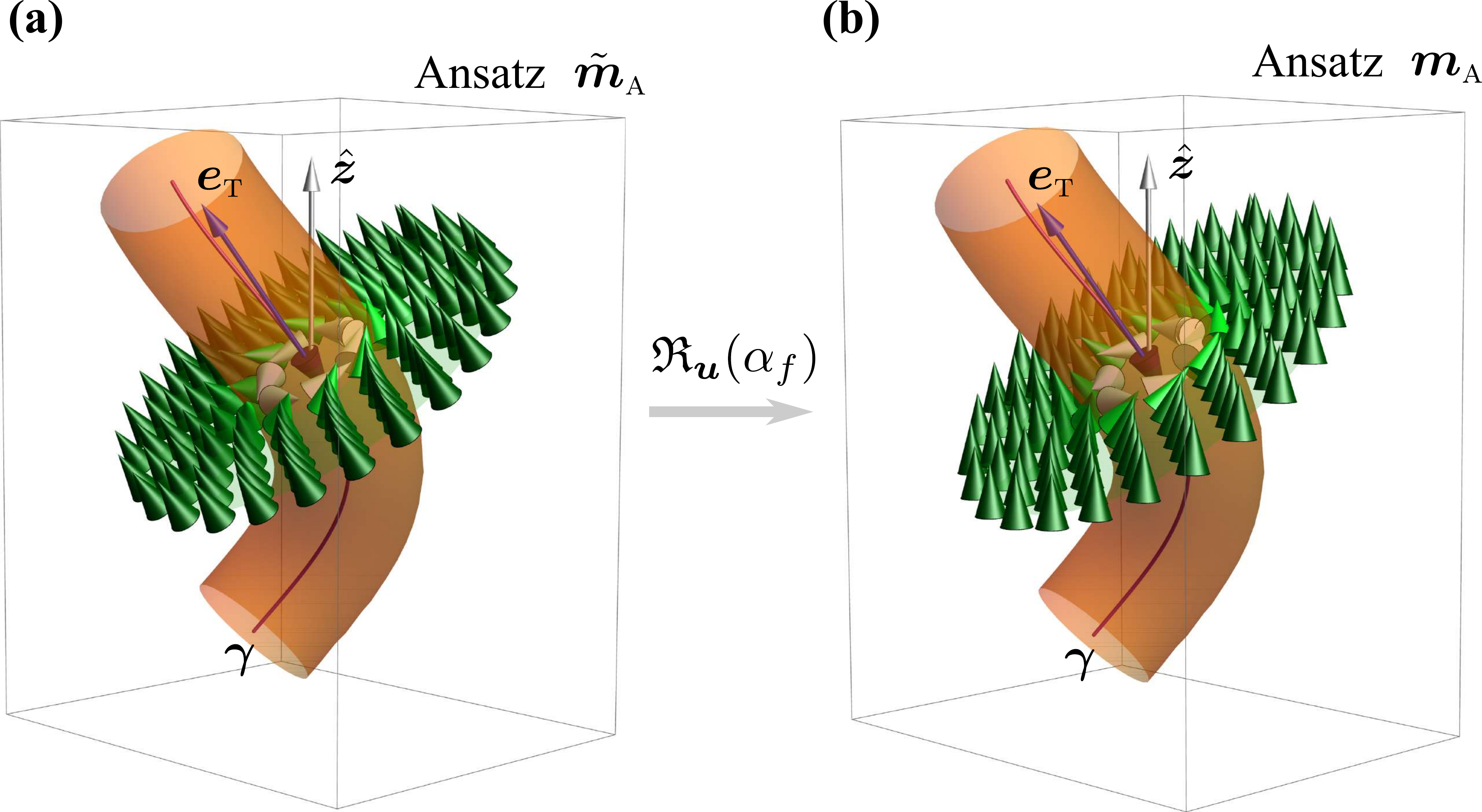}
	\caption{Two-step formation of the skyrmion string Ansatz. Models \eqref{eq:m-tilde} and \eqref{eq:Ansatz} are shown in panels (a) and (b), respectively. For Ansatz $\tilde{\vec{m}}_{\textsc{a}}$, magnetization within a perpendicular cross-section (shown by cones) coincides with the equilibrium 2D skyrmion solution. In Ansatz $\vec{m}_{\textsc{a}}$, the magnetization within each cross-section is inhomogeneously rotated such that $\vec{m}_{\textsc{a}}$ meets the ground state with a distance from the string, see Appendix~\ref{app:Ansatz} for details.}\label{fig:Ansatz}
\end{figure}

The second way of derivation of the structure of the string  Hamiltonian is based on the skyrmion string Ansatz which is explained in details in Appendix~\ref{app:Ansatz}. The string Ansatz is build as a two-step deformation of the magnetization field $\vec{m}_0(x,y)$ of the unperturbed vertical string in equilibrium. On the first step, we consider a model $\tilde{\vec{m}}_{\textsc{a}}(\vec{r})$, such that the magnetization within each cross-section perpendicular to the string coincides with $\vec{m}_0$ (in the reference frame, defined on the section plane), see Fig.~\ref{fig:Ansatz}(a). Although the model $\tilde{\vec{m}}_{\textsc{a}}$ is intuitively clear, it can not be used because $\tilde{\vec{m}}_{\textsc{a}}$ is not uniquely defined, if the distance to the string is larger than the curvature radius, and also $\tilde{\vec{m}}_{\textsc{a}}$ does not meat the ground state at large distances. For these reasons, on the second step, we apply a spatially-dependent rotation transformation $\vec{m}_{\textsc{a}}=\vec{\mathfrak{R}}_{\vec{u}}(\alpha_f)\tilde{\vec{m}}_{\textsc{a}}$ within each perpendicular cross-section. The rotation is performed around unit vector  $\vec{u}=\vec{e}_{\textsc{t}}\times\hat{\vec{z}}/|\vec{e}_{\textsc{t}}\times\hat{\vec{z}}|$, where $\vec{e}_{\textsc{t}}$ is the unit vector, tangential to the string. The rotation magnitude $\alpha_f$ depends on the distance $\rho$ to the string center (within each cross-section) and it is such that $\alpha_f\to\angle(\vec{e}_{\textsc{t}},\hat{\vec{z}})$ everywhere except small region $\rho<\rho^*$ around the string. Here $\rho^*\kappa\ll1$ with $\kappa$ being the string curvature. It is assumed that the angle $\alpha_f$ depends only on $\rho$, this dependence is captured by some unknown function $f(\rho)$, for details see Appendix~\ref{app:Ansatz}. The resulting magnetization is shown in Fig.~\ref{fig:Ansatz}(b).

The substitution of the string Ansatz $\vec{m}_{\textsc{a}}$ into the Hamiltonian \eqref{eq:H} and the integration over the coordinates perpendicular to the string, enables us to write the string Hamiltonian in the form \eqref{eq:H-ser}-\eqref{eq:Bloch-H}, for details see Appendix~\ref{app:energies}. The coefficients $a_n$ and $b_n$ are functionals of the skyrmion profile $\theta_0(\rho)$ and function $f(\rho)$. A rough estimation for the function $f(\rho)$ is discussed in Appendix~\ref{app:energies} and is shown in Fig.~\ref{fig:f}. For details and the explicit form of the coefficients $a_n$ and $b_n$ see Appendix~\ref{app:energies}. Remarkably, the Hamiltonian obtained by means of the Ansatz completely satisfies the symmetry requirements discussed above.

In the limit of large magnetic fields (infinitely thin strings) one has $\mathscr{H}^{(2)}\approx\frac{1}{2}|\psi'|^2$ and $\mathscr{H}^{(4)}\approx-\frac{1}{8}|\psi'|^4$, see Appendix~\ref{app:energies}. It reflects the increase of the string energy due to increase of the total string length $\int\sqrt{1+|\psi'|^2}\dd\tilde{z}$, since $\sqrt{1+|\psi'|^2}\approx1+\frac{1}{2}|\psi'|^2-\frac{1}{8}|\psi'|^4$. In this limit, the exchange contribution to the string energy dominates.

In the dimensionless units, equation of motion \eqref{eq:Thiele-psi} reads $\frac{i}{2}\dot{\psi}=\frac{\delta\tilde{H}}{\delta\psi^*}$, where the overdot indicates the derivative with respect to dimensionless time $\tilde{t}$ and we took into account that $N_{\text{top}}=-1$. With \eqref{eq:H-ser} and \eqref{eq:Bloch-H}, we write this equation of motion in the following explicit form
\begin{equation}\label{eq:eq-motion}
	\begin{split}
	-i\dot{\psi}=&a_1\psi''-ia_2\sigma\psi'''-a_3\psi^{\textsc{(iv)}}+\dots\\
	&-b_1\left(\psi'^2{\psi^*}'\right)'+ib_2\sigma\left(\psi'\psi''{\psi^*}'\right)'+\dots
	\end{split}
\end{equation}

The limit case of Eq.~\eqref{eq:eq-motion} with $a_{n>1}=0$ and $b_{n>1}=0$ was previously obtained and discussed in Ref.~\onlinecite{Kravchuk20}.

\section{Helical wave}\label{sec:hw}

\begin{figure}
	\includegraphics[width=\columnwidth]{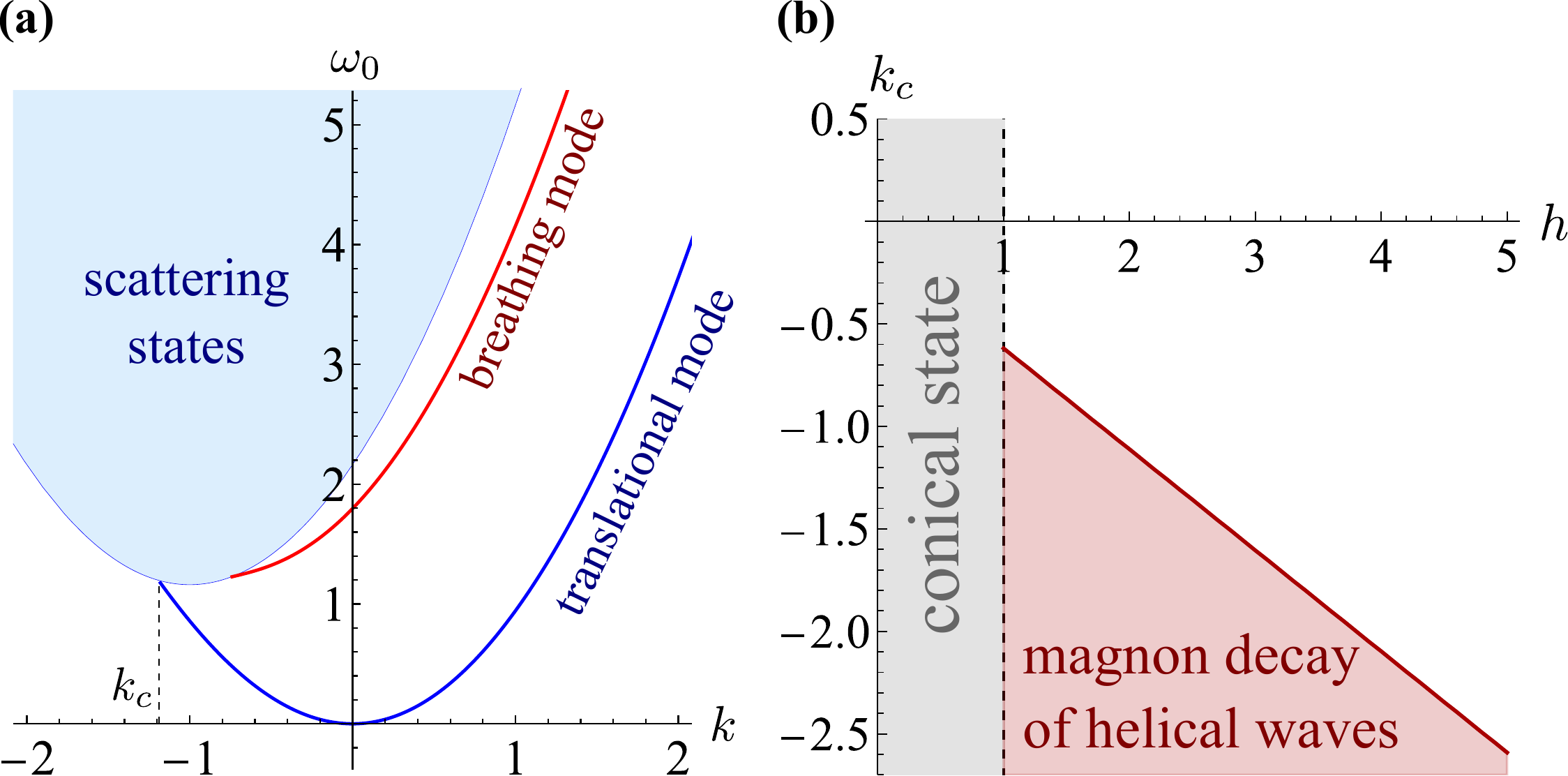}
	\caption{(a) Dispersion relation for the translational and breathing linear modes propagating along the string in field $h=2.16$ ($\mu_0H_{\text{ext}}=0.8$~T) and for the case $\sigma>0$. The data are obtained by means of diagonalization of the spin wave Hamiltonian as explained in Ref.~\onlinecite{Kravchuk20}. For $k<k_c$, the localized translational mode does not exist. The dependence of the critical wave-vector $k_c$ on the applied magnetic field is shown in panel (b). Helical waves \eqref{eq:h-wave} exist for $k>k_c$ only. }\label{fig:kc}
\end{figure}

\begin{figure}
	\includegraphics[width=\columnwidth]{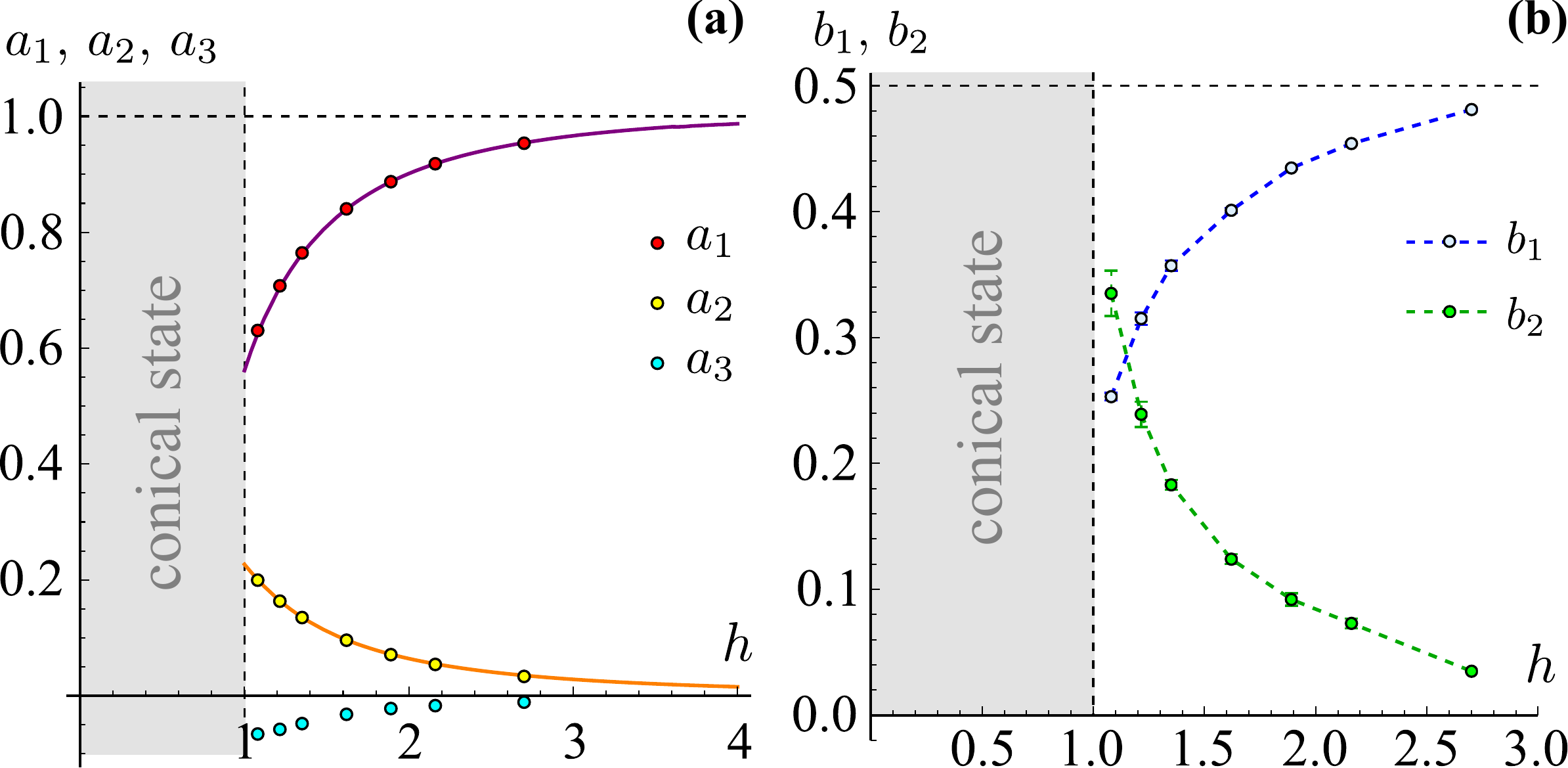}
	\caption{The field dependence of the linear $a_n$ and nonlinear $b_n$ coefficients in dispersion relation \eqref{eq:helix-disp}. Markers show the coefficient values obtained by means of numerical simulations, for details see Appendix~\ref{app:simuls}. Solid lines show the dependencies of $a_1(h)$ and $a_2(h)$ previously extracted from the dispersion of the translational magnon mode, see Fig.~\ref{fig:kc}(a) and Ref.~\onlinecite{Kravchuk20} for details.  }\label{fig:an-bn-vs-b}
\end{figure}

In spite of a complicated general form, Eq.~\eqref{eq:eq-motion} has a simple exact solution in form of nonlinear helical wave 
\begin{equation}\label{eq:h-wave}
	\psi=\mathcal{R}e^{i(k\tilde{z}-\omega\tilde{t})},
\end{equation}
where real constant $\mathcal{R}$ plays role of the helix radius. Helical wave \eqref{eq:h-wave} has the following dispersion relation
\begin{subequations}\label{eq:helix-disp}
	\begin{align}
\label{eq:disp-main}	\omega=\,&\omega_0(k)-\mathcal{R}^2k^2\omega_1(k)+\dots,\\
\label{eq:w0}	&\omega_0(k)=a_1k^2+a_2\sigma k^3+a_3k^4+\dots,\\
\label{eq:w1}	&\omega_1(k)=b_1k^2+b_2\sigma k^3+\dots.
	\end{align}
\end{subequations}
 In the limit $\mathcal{R}\to0$, the helical wave is transformed into the translational magnon mode with the dispersion $\omega_0(k)$ shown in Fig.~\ref{fig:kc}(a). For finite $\mathcal{R}$, the leading nonlinear term in the dispersion is represented by $\omega_1(k)$. The known forms of dispersions $\omega_0(k)$ and $\omega_1(k)$ enable one to determine coefficients $a_n$ and $b_n$ by means of numerical simulations of the helical wave dynamics for various $\mathcal{R}$ and $k$, for details see Appendix~\ref{app:simuls}. The dependencies of several first coefficients $a_n$ and $b_n$ on applied magnetic field are shown in Fig.~\ref{fig:an-bn-vs-b}.  Note that the normalized magnetic field $h$ is the only parameter which controls system \eqref{eq:H}.  We should emphasize that the numerical values of the coefficients $a_n$ and $b_n$ presented in Fig.~\ref{fig:an-bn-vs-b} are universal and valid for all cubic chiral magnets.

Wave vector $k$ of the helical wave \eqref{eq:h-wave} can not be arbitrary, it is limited by the domain of existence of the localized translational magnon mode propagating along the string, i.e. $k>k_c(h)$, see Fig.~\ref{fig:kc}. For the case $k<k_c$, the stationary helical wave \eqref{eq:h-wave} does not exist. In this regime the helix radius $\mathcal{R}$ rapidly decays, this process is accompanied by the magnon emission. The regime of the magnon decay of the helical wave is shown in Fig.~\ref{fig:kc}(b) by the red shadowing.  

Based on dispersions \eqref{eq:w0} and \eqref{eq:w1}, and on the numerically obtained dependencies $a_n(h)$ and $b_n(h)$ we found that the Lighthill criterion \cite{LIGHTHILL65,Zakharov09}
\begin{equation}\label{eq:Lighthill}
	\omega_0''(k)\omega_1(k)>0
\end{equation}
is satisfied for $k>k_c$~\footnote{Note that the nonlinear part of the dispersion relation \eqref{eq:disp-main} comes with the negative sign.}, meaning the modulational instability of the nonlinear helical wave \eqref{eq:h-wave}. This effect and its consequences are discussed in the next section.

\begin{figure}
	\includegraphics[width=\columnwidth]{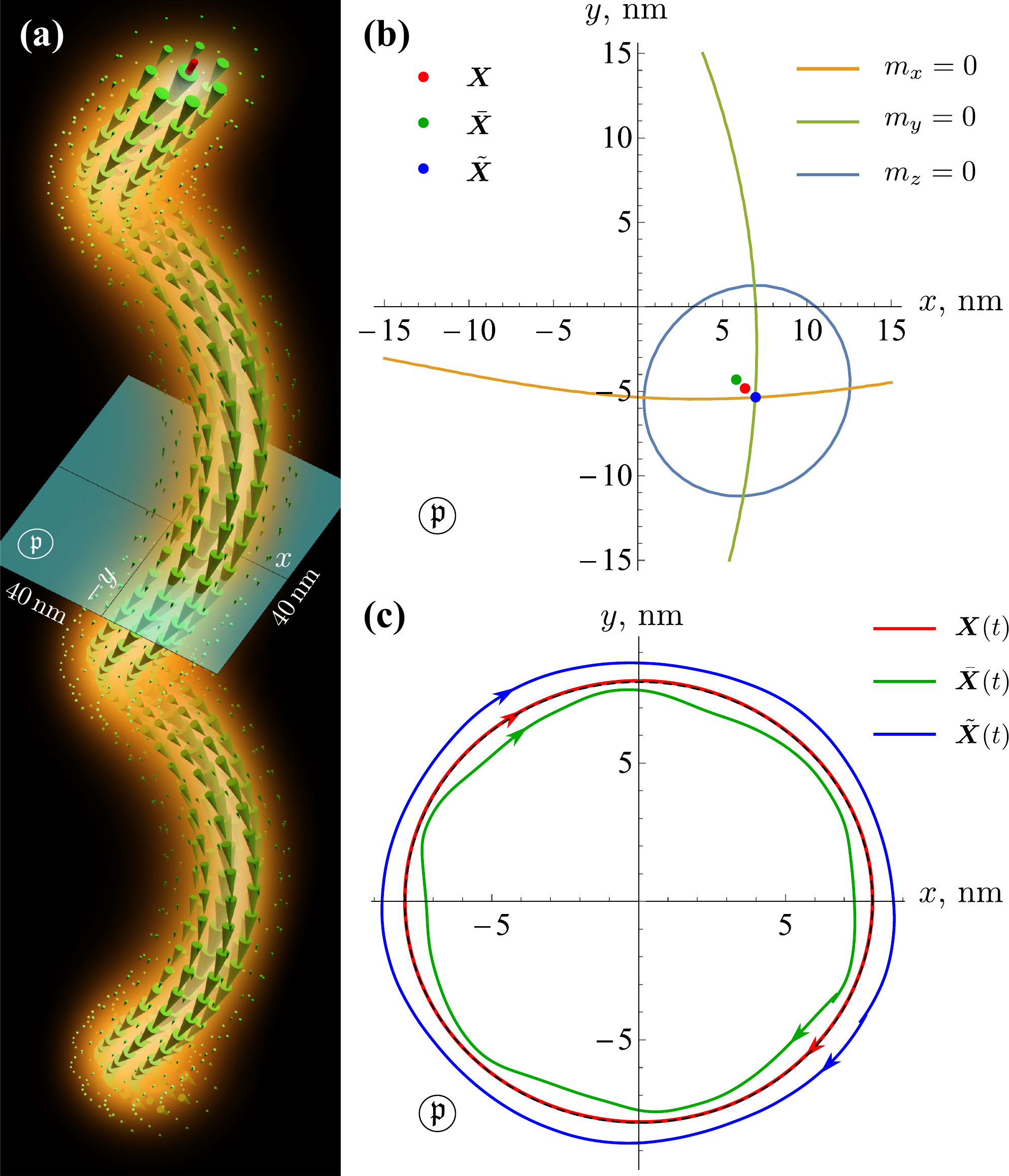}
	\caption{(a) A snapshot of a helical wave propagating along skyrmion string is shown in terms of the emergent field $\vec{B}^{\text{e}}$. Notations are the same as in Fig.~\ref{fig:string}(b). The helix radius and pitch are 8~nm and 120~nm ($k>0$), respectively. Data is obtained from the micromagnetic simulations for FeGe in field $\mu_0H_{\text{ext}}=0.7$~T. (b) Magentization isolines within the horizontal cross-section by the plane $\mathfrak{p}$. Dots represent the scross-sections with the string central lines defined in different manners (explained in text). Time evolution of these markers is shown on panel (c). The black dashed line shows the geometrical circle of radius 8~nm. The clockwise direction of the motion along the trajectories takes place also for $k<0$. The helical wave dynamics for both signs of $k$ is illustrated by supplemental movies 1 and 2~\cite{Note2}.}\label{fig:helix}
\end{figure}

An example of the skyrmion filament with the propagating helical wave is shown in Fig.~\ref{fig:helix}(a) in terms of the emergent magnetic field. A specificity of the preparation of the initial magnetization distribution used for the simulations (see Appendix~\ref{app:simuls}) is such that we can control the helix shape \eqref{eq:h-wave} of the central line defined  in \eqref{eq:Xi}, however the overall magnetization structure $\vec{m}(\vec{r})$ can slightly deviate from the real solution determined by the Landau-Lifshitz equation. As a result, a number of higher skyrmion modes, e.g. breathing or CCW mode, are excited together with the helical wave. These modes are very well recognizable on the supplemental movies~1 and 2~\footnote{Reference to the supplemental materials is provided by the Publisher.}. We use the presence of the additional magnon excitations in order to compare different definitions of the string central line alternative to \eqref{eq:Xi}. We consider two alternative definitions of the skyrmion guiding center which are the most common in the literature, namely the first moment of $m_z$ component
\begin{equation}\label{eq:Xim}
	\bar{X}_i(z,t)=\frac{\iint x_i[1-m_z(x,y,z,t)]\dd x\dd y}{\iint [1-m_z(x,y,z,t)]\dd x\dd y},
\end{equation}
and $\tilde{X}_i$: $m_z(\tilde{X}_1,\tilde{X}_2)=-1$. 
The comparison of different types of dynamics of the skyrmion guide-centers defined as \eqref{eq:Xi} and \eqref{eq:Xim} is discussed in a number of previous works~\cite{Komineas15c,Kravchuk18,Wu22a}, in which it was shown that the guiding center \eqref{eq:Xi} demonstrates the massless Thiele dynamics, while the guiding center defined in  \eqref{eq:Xim} show the additional oscillations typical for a massive particle. The definition $\tilde{X}_i$ was widely used for studying the dynamics of merons \cite{Mertens97,Hertel06,Hertel07}, and it was shown that the guiding center $\tilde{\vec{X}}$ also demonstrates a complicated dynamics with several additional high-frequency oscillations such that the massive, as well as the higher order terms are required in the corresponding equation of motion for $\tilde{\vec{X}}(t)$~\cite{Mertens97,Kovalev03a}. Here, using the helical wave as an example, we demonstrate that the string definitions $\vec{X}(z,t)$, $\bar{\vec{X}}(z,t)$ and $\tilde{\vec{X}}(z,t)$ are different, see Fig.~\ref{fig:helix}(b,c). It is important that within an arbitrary horizontal cross-section $z=z_0$, the trajectory $\vec{X}(z_0,t)$ is a circle, while trajectories $\bar{\vec{X}}(z_0,t)$ and $\tilde{\vec{X}}(z_0,t)$ exhibit additional cycloidal oscillations, see Fig.~\ref{fig:helix}(c). This is in agreement with the discussed above results for two-dimensional topological solitons. The circular trajectory for $\vec{X}(z_0,t)$ is consistent with the helix solution \eqref{eq:h-wave}, and this \emph{a posteriori} justifies the initial massless equation \eqref{eq:Thiele-psi}, in which we implicitly assumed vanishing of the Poisson brackets of $X_i$ with the amplitudes of higher magnon modes. However, as it follows from Fig.~\ref{fig:helix}(c), this assumption can not be applied for strings $\bar{\vec{X}}(z,t)$ and $\tilde{\vec{X}}(z,t)$.

\begin{figure*}
	\includegraphics[width=\linewidth]{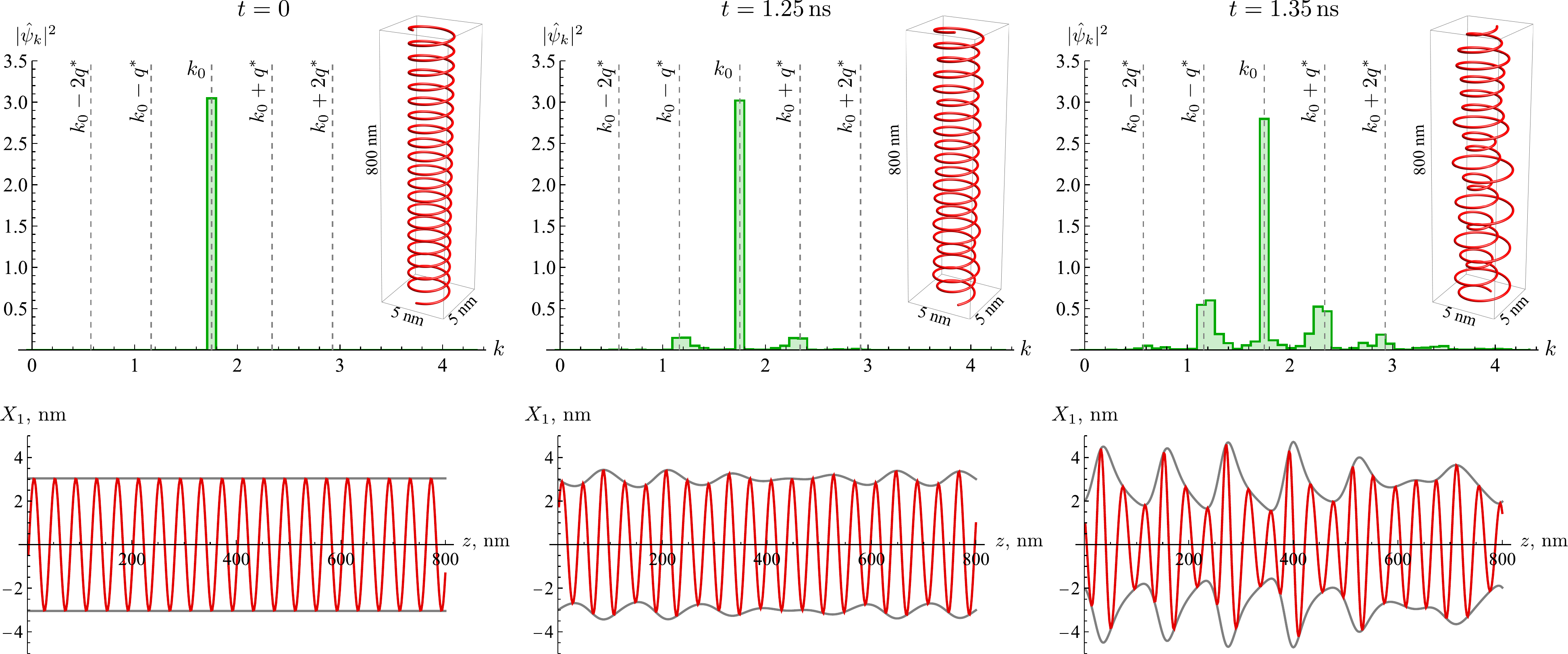}
	\caption{Instability development of helical wave \eqref{eq:h-wave} obtained by means of micromagnetic simulations for FeGe in field $h=2.16$ (0.8~T). The helix radius and wave-vector are $\mathcal{R}_0=0.27$ (3~nm) and $k=k_0=1.75$ ($2\pi/40\,\text{mn}^{-1}$), respectively. For the considered parameters $q^*\approx\mathcal{R}_0k_0^2\sqrt{b_1/a_1}\approx0.59$. The dependencies $X_1(z)$ shown in the bottom raw (red line) are extracted from the simulation data with the use of Eq.~\eqref{eq:Xi}. The envelope $R(z)=\sqrt{X_1^2+X_2^2}$ is shown by the gray line.}\label{fig:instab}
\end{figure*}

\section{Weakly nonlinear dynamics of the string}
Here we consider solutions of equation of motion \eqref{eq:eq-motion} in form $\psi=\mathcal{A}(\tilde{z},\tilde{t})e^{i(k_0\tilde{z}-\omega_0\tilde{t})}$ with $\mathcal{A}\in\mathbb{C}$ in so called adiabatic approximation which implies $|\dot{\mathcal{A}}|\ll|\omega_0\mathcal{A}|$, $|\mathcal{A}'|\ll|k_0\mathcal{A}|\ll1$. I.e., we consider a modulated wave with the slowly varied (in space and time) envelope profile. It is known that for the case of the nonlinear dispersion \eqref{eq:disp-main}, in the low-amplitude limit, the envelope wave is governed by the cubic nonlinear Schr{\"o}dinger equation. The latter result can be obtained from the general Whitham approach in the low-amplitude limit \cite{Whitham65,Zakharov09,Whitham99}, or within formalism of nonlinear geometrical optics \cite{Karpman69,Karpman75}, or by means of the multiple scales method \cite{Ryskin00,Nayfeh08}. Application of the method of multiple scales to \eqref{eq:eq-motion} results in the following nonlinear Schr{\"o}dinger equation (NLSE)
\begin{equation}\label{eq:NLS-set}
		i\left[\dot{\mathcal{A}}+v_g(k_0)\mathcal{A}'\right]+\frac{\mu(k_0)}{2}\mathcal{A}''+\nu(k_0)\mathcal{A}|\mathcal{A}|^2=0
\end{equation}
Here the group velocity $v_g(k)=\partial_k\omega_0$ as well as coefficients $\mu(k)=\partial_k^2\omega_0$ and $\nu(k)=k^2\omega_1(k)$ are completely determined by nonlinear dispersion \eqref{eq:h-wave}. For details see Appendix~\ref{app:mult-scales}. 
In the reference frame $\tilde{z}'=\tilde{z}-v_g(k_0)\tilde{t}$ which moves with the group velocity, the equation for amplitude $\mathcal{A}$ has a form of classical NLSE
\begin{equation}\label{eq:NLS}
	i\dot{\mathcal{A}}+\frac{\mu(k_0)}{2}\mathcal{A}''+\nu(k_0)\mathcal{A}|\mathcal{A}|^2=0.
\end{equation}
Based on \eqref{eq:w0} and \eqref{eq:w1}, and on the numerically obtained dependencies $a_n(h)$ and $b_n(h)$, we verified that $\mu(k_0)>0$ and $\nu(k_0)>0$ for $k_0>k_c$, meaning that NLSE \eqref{eq:NLS} is of focusing type. The latter agrees with the Lighthill criterion \eqref{eq:Lighthill}.


Solutions of NLSE~\eqref{eq:NLS}, are well studied and classified \cite{Zakharov72,Akhmediev87,Dysthe99}. In the rest part of this section, we use the micromagnetic simulations in order to confirm the existence and verify the main properties of the string excitations in form of the solutions predicted by NLSE~\eqref{eq:NLS}.

\subsection{Instability of the nonlinear helical wave}

In the following, it is convenient to present the wave envelope in form $\mathcal{A}=\mathcal{R}(\tilde{z}',\tilde{t})e^{i\phi(\tilde{z}',\tilde{t})}$ where $\mathcal{R},\,\phi\in\mathbb{R}$. Equation \eqref{eq:NLS} has spatially uniform solution which corresponds to the helical wave considered in Section~\ref{sec:hw}. In this case, $\mathcal{R}=\mathcal{R}_0=\text{const}$ and $\phi=\Omega\tilde{t}$, where $\Omega=\mathcal{R}_0^2\nu(k_0)=\mathcal{R}_0^2k_0^2\omega_1(k_0)$ is nonlinear shift of frequency of the helical wave. Introducing small deviations $\tilde{\mathcal{R}}$ and $\tilde{\phi}$ such that $\mathcal{R}=\mathcal{R}_0+\tilde{\mathcal{R}}$ and $\phi=\Omega\tilde{t}+\tilde{\phi}$, we obtain from \eqref{eq:NLS} the corresponding linearized equations for the deviations, whose solutions $\tilde{\mathcal{R}},\tilde{\phi}\propto e^{i(q\tilde{z}-\tilde{\omega}\tilde{t})}$ are characterized by the dispersion relation 
\begin{equation}\label{eq:disp-hel}
	\tilde{\omega}=|q|\sqrt{q^2\mu^2(k_0)/4-\mu(k_0)\nu(k_0)\mathcal{R}_0^2}.
\end{equation}
Thus, if the condition \eqref{eq:Lighthill} holds, the helical wave is unstable for $|q|<q_0=2\mathcal{R}_0\sqrt{\nu(k_0)/\mu(k_0)}=2\mathcal{R}_0|k_0|\sqrt{\omega_1(k_0)/\omega_0''(k_0)}$ and the maximum of the instability increment $\varkappa_{\text{max}}=\mathcal{R}_0^2\nu(k_0)$ corresponds to $q^*=q_0/\sqrt{2}$. 

We verified these predictions on the helical wave instability by means of micromagnetic simulations~\cite{mumax3}. The initial magnetization state in form of the helical wave was prepared as explained in Appendix~\ref{app:simuls}. Using \eqref{eq:Xi} we extract the central string line $\vec{X}(z,t)$ from the simulation data of the magnetization dynamics. The time-development of the modulational instability of the helical wave is shown in Fig.~\ref{fig:instab} in terms of $\vec{X}(z,t)$ and in terms of the corresponding $\psi(\tilde{z},\tilde{t})$. Applying the Fourier transform $\hat{\psi}_k(\tilde{t})=L^{-1}_z\int_{0}^{L_z}\psi(\tilde{z},\tilde{t})e^{-ik\tilde{z}}\dd\tilde{z}$ we observe the development of the cascade of satellites at $k=k_0\pm nq^*$, which is a signature of the modulation instability \cite{Zakharov09}. Note a good agreement of the satellites positions with the predictions (vertical dashed lines).

Note that the typical time of the instability development $\tau^*=1/\varkappa_{\text{max}}\propto k^{-4}_0$ rapidly increases with the decrease of the wave vector. This feature was utilized for numerical determination of the coefficients $a_n$ and $b_n$ by means of the micromagnetic simulations. The values of $k_0$ used in these simulations were 2 -- 4 times smaller as $k_0$ in Fig.~\ref{fig:instab}, see Appendix~\ref{app:simuls}. Thus, the instability effects were negligible during the simulation time. Another way to avoid the helical wave instability is to simulate a short sample with the length $L_z=2\pi/|k_0|<2\pi/q_0$ and the periodic boundary conditions applied along $z$. In this case, the wave vectors $|q|<q_0$ of the unstable excitations does not exist in the system. 

\subsection{Cnoidal waves and solitons}

\begin{figure*}
	\includegraphics[width=\textwidth]{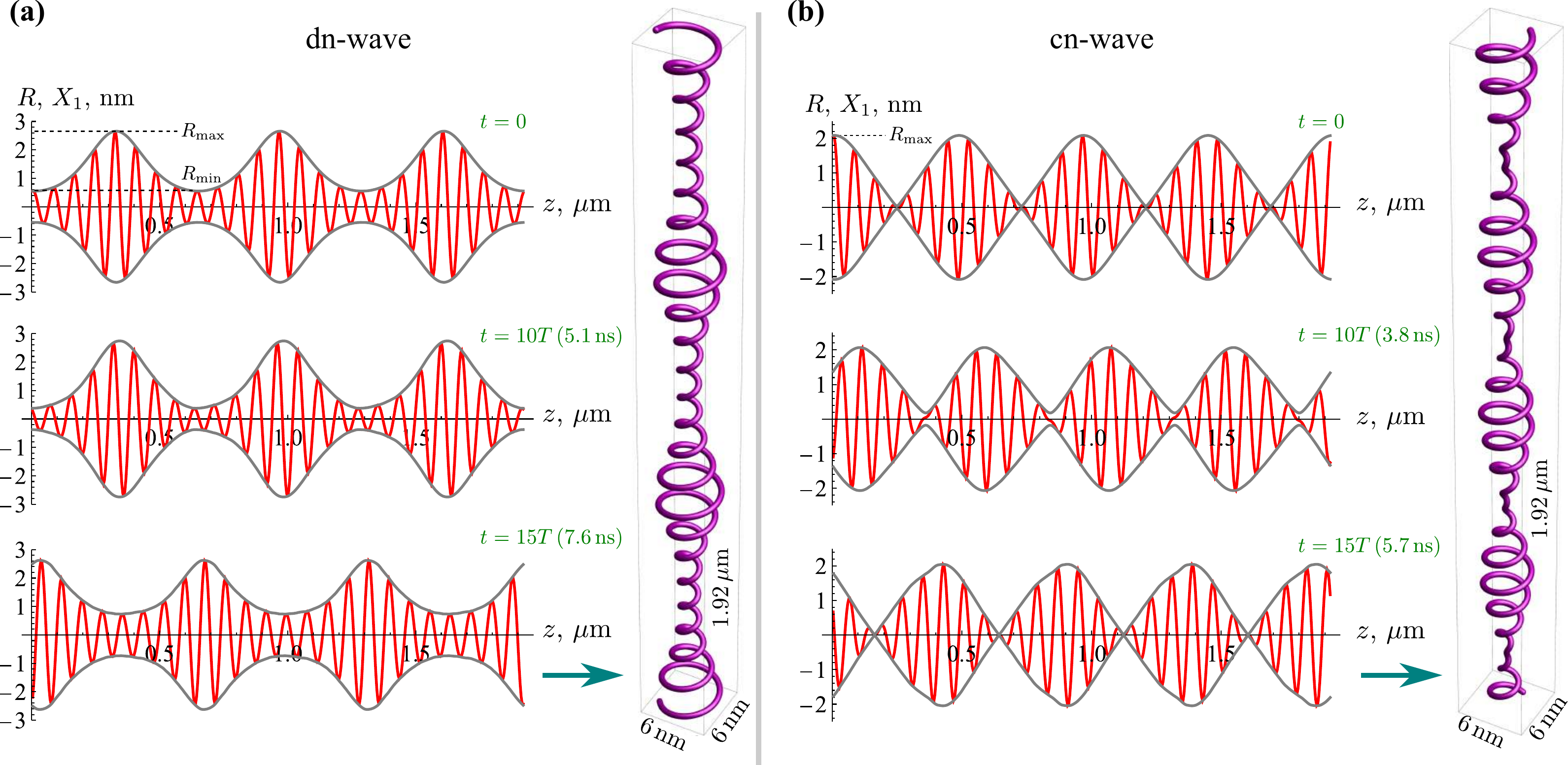}
	\caption{Nonlinear waves of skyrmion string in FeGe obtained by means of micromagnetic simulations for magnetic field $\mu_0H_{\text{ext}}=0.8$~T. Red and gray lines on the plots have the same meaning as in Fig.~\ref{fig:instab}. The upper plots ($t=0$) in insets a) and b) show the initial perturbation of the skyrmion string. The perturbations are close to solutions \eqref{eq:dn} and \eqref{eq:cn}, respectively. The lower plots demonstrate the snapshots taken at time moments $t=10T$ and $t=15T$ with $T=L/v_g$ being time it takes for a wave to cover a distance equal to its period $L$. Wave length of the carrying wave is 80~nm ($k_0=0.875$). For both panels, the vertical insets show the 3D string shape at  $t=15T$. Note that the horizontal sizes are magnified 30 times. Dynamics of the dn- and cn-waves is demonstrated in the supplemental movies~3 and 4, respectively~\cite{Note2}.}\label{fig:waves-simuls}
\end{figure*}

As a generalization of the helical wave solution $\mathcal{A}=\mathcal{R}_0e^{i\Omega\tilde{t}}$ with constant radius, NLSE \eqref{eq:NLS} has a class of traveling-wave solutions with coordinate dependent and static (in the moving reference frame) profiles $\mathcal{A}=\mathcal{R}(\tilde{z}')e^{i\Omega\tilde{t}}$. Here frequency $\Omega=\nu\mathcal{R}_0^2$ is the same as for the helical wave. The first integral of the corresponding equation for $\mathcal{R}(\tilde{z}')$ is $\mathcal{R}'^2+W(\mathcal{R})=\mathcal{E}_0$ where $W(\mathcal{R})=\frac{\nu}{\mu}\mathcal{R}^2(\mathcal{R}^2-2\mathcal{R}_0^2)$.
For NLSE  \eqref{eq:NLS} of focusing type, constant  $\mathcal{E}_0>-\frac{\nu}{\mu}\mathcal{R}_0^4$ can be interpreted as energy of an oscillator with potential well $W(\mathcal{R})$. Solutions for $\mathcal{R}(\tilde{z}')$ depend on two parameters $\mathcal{R}_0$, and $\mathcal{E}_0$ and they are divided on two classes depending on the sign of $\mathcal{E}_0$. The ``negative energy'' solutions have profile
\begin{equation}\label{eq:dn}
	\mathcal{R}_{\textsc{dn}}=\mathcal{R}_{\text{max}}\text{dn}\left(\tilde{z}'/\Delta_{\textsc{dn}},s_{\textsc{dn}}\right),
\end{equation}
where $\mathrm{dn}(x,s)$ is Jacobian elliptic function \cite{NIST10} with modulus $s_{\textsc{dn}}=\sqrt{1-\mathcal{R}_{\text{min}}^2/\mathcal{R}_{\text{max}}^2}$, and $\Delta_{\textsc{dn}}=\mathcal{R}^{-1}_{\text{max}}\sqrt{\mu/\nu}$. Here, as an alternative to parameters $\mathcal{R}_0$ and $\mathcal{E}_0$, we use parameters  $\mathcal{R}_{\text{max}}$ and $\mathcal{R}_{\text{min}}$, such that $\mathcal{R}_0^2=(\mathcal{R}_{\text{max}}^2+\mathcal{R}_{\text{min}}^2)/2$ and $\mathcal{E}_0=-\frac{\nu}{\mu}\mathcal{R}_{\text{max}}^2\mathcal{R}_{\text{min}}^2$. Parameters  $\mathcal{R}_{\text{max}}$ and $\mathcal{R}_{\text{min}}$ have meaning of maximal and minimal amplitudes of the dn-wave, see Fig.~\ref{fig:waves-simuls}(a). The wave period is $L_{\textsc{dn}}=2\Delta_{\textsc{dn}}\mathrm{K}(s_{\textsc{dn}})$ with $\mathrm{K}(x)$ being the complete elliptic integral of the first kind. In the limit case $\mathcal{R}_{\text{min}}\to\mathcal{R}_{\text{max}}$, the cnoidal wave \eqref{eq:dn} transforms to the helical wave with an excitation of vanishing amplitude $\tilde{\mathcal{R}}=(\mathcal{R}_{\text{max}}-\mathcal{R}_{\text{min}})/2$ and wave-vector $q=q_0$ corresponding to the edge of the helical wave instability. For the case $\mathcal{R}_{\text{min}}\ll\mathcal{R}_{\text{max}}$, the solution \eqref{eq:dn} has a form of a train of solitary waves separated by the distance $L_{\textsc{dn}}\approx2\Delta_{\textsc{dn}}\ln(2\mathcal{R}_{\text{max}}/\mathcal{R}_{\text{min}})$. In the limit case $\mathcal{R}_{\text{min}}=0$, the cnoidal wave \eqref{eq:dn} transforms to soliton 
\begin{equation}\label{eq:sol}
	\mathcal{R}_{\text{sol}}=\frac{\mathcal{R}_{\text{max}}}{\cosh(\tilde{z}'/\Delta_{\textsc{dn}})}
\end{equation}
with energy $\mathcal{E}_0=0$. So, soliton \eqref{eq:sol} is the separatrix solution between two classes of nonlinear waves with $\mathcal{E}_0<0$ and $\mathcal{E}_0>0$. In the latter case, the envelope profile is 
\begin{equation}\label{eq:cn}
	\mathcal{R}_{\textsc{cn}}=\mathcal{R}_{\text{max}}\text{cn}\left(\tilde{z}'/\Delta_{\textsc{cn}},s_{\textsc{cn}}\right),
\end{equation}
where $\Delta_{\textsc{cn}}=(\mathcal{R}_{\text{max}}^2+\mathcal{R}_{\text{min}}^2)^{-1/2}\sqrt{\mu/\nu}$ and $s_{\textsc{cn}}=1/\sqrt{1+\mathcal{R}_{\text{min}}^2/\mathcal{R}_{\text{max}}^2}$. Here the parameters  $\mathcal{R}_{\text{max}}$ and $\mathcal{R}_{\text{min}}$, are chosen such that $\mathcal{R}_0^2=(\mathcal{R}_{\text{max}}^2-\mathcal{R}_{\text{min}}^2)/2$ and $\mathcal{E}_0=\frac{\nu}{\mu}\mathcal{R}_{\text{max}}^2\mathcal{R}_{\text{min}}^2$. In contrast to dn-waves \eqref{eq:dn}, the profile of cn-wave \eqref{eq:cn} has nodes where $\mathcal{R}_{\textsc{cn}}=0$. The distance between nodes is $L_{\textsc{cn}}/2=2\Delta_{\textsc{cn}}\mathrm{K}(s_{\textsc{cn}})$. In the limit case $\mathcal{R}_{\text{min}}=0$, cn-wave \eqref{eq:cn} is transformed into soliton \eqref{eq:sol}. In literature, the nonlinear waves \eqref{eq:dn} and \eqref{eq:cn} described by the elliptical functions are known as ``cnoidal waves''. They are common for different physical media, the widely known examples are shallow water~\cite{Ryskin00,Whitham99} and atmosphere~\cite{Skyllingstad91}.

Solutions \eqref{eq:dn}, \eqref{eq:cn}, and \eqref{eq:sol} represent a partial case of the traveling-waves moving exactly with the group velocity $v_g(k_0)$. The generalization for the case of arbitrary velocity is realized by means of the replacement $\tilde{z}'\to\tilde{z}'-V\tilde{t}$. In this case $\phi=\Omega\tilde{t}+\frac{V}{\mu}\left(\tilde{z}'-\frac{V}{2}\tilde{t}\right)$. Note that $V$ is the envelope velocity in the moving reference frame $\tilde{z}'=\tilde{z}-v_g(k_0)\tilde{t}$. Three parameters $\mathcal{R}_{\text{max}}$, $\mathcal{R}_{\text{min}}$ and $V$ are stabilized by three integrals of motion, the number of excitations $N=\int_0^{L_z}\mathcal{R}^2\dd\tilde{z}'$, momentum $P=-\frac{V}{\mu}N$, and energy $E=\frac{V^2}{2\mu}N+\frac{1}{2}\int_0^{L_z}\left(\mu\mathcal{R}'^2-\nu\mathcal{R}^4\right)\dd\tilde{z}'$. 

\begin{figure}
	\includegraphics[width=\columnwidth]{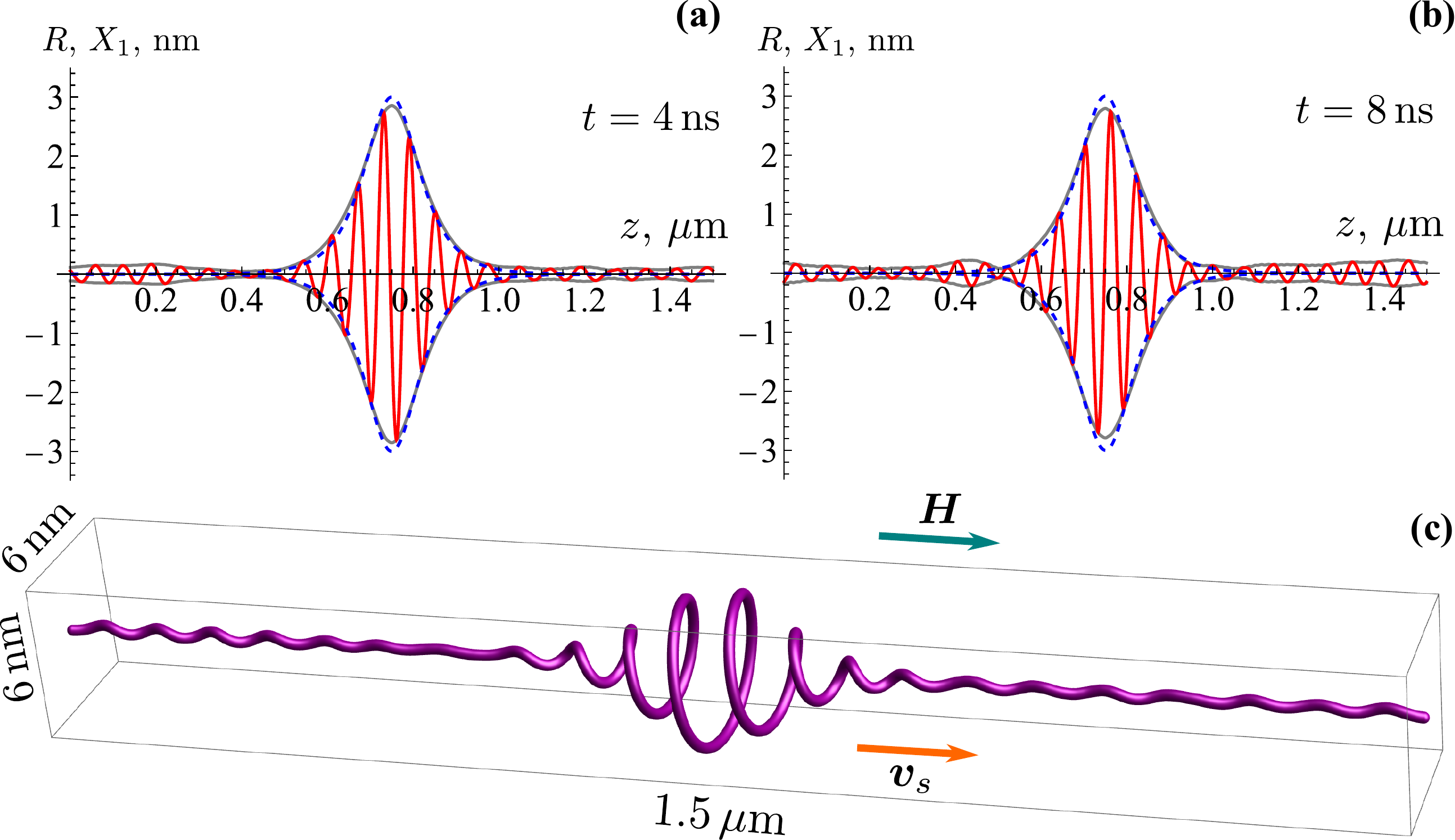}
	\caption{Soliton solution \eqref{eq:sol} obtained by means of micromagnetic simulations of skyrmion string in FeGe in magnetic field $\mu_0H=0.8$~T. The dashed blue line shows the initial soliton profile, the other notations are the same as in Fig.~\ref{fig:waves-simuls}. Insets (a) and (b) shows the snapshots taken in different moments of time starting from the beginning of the simulations. Panel (c) is the 3D version of the inset (a). Wave vector of the carrying wave $k_0=1.17$ (the wave length 60 nm) corresponds to the group velocity $v_g=2.2$ ($1.64\times10^3$~m/s). The latter is close to the soliton velocity $v_s=2.1$ ($1.52\times10^3$~m/s) found in the simulations. The soliton dynamics is show in Movie~5~\cite{Note2}.}\label{fig:soliton}
\end{figure}

\begin{figure*}
	\includegraphics[width=\textwidth]{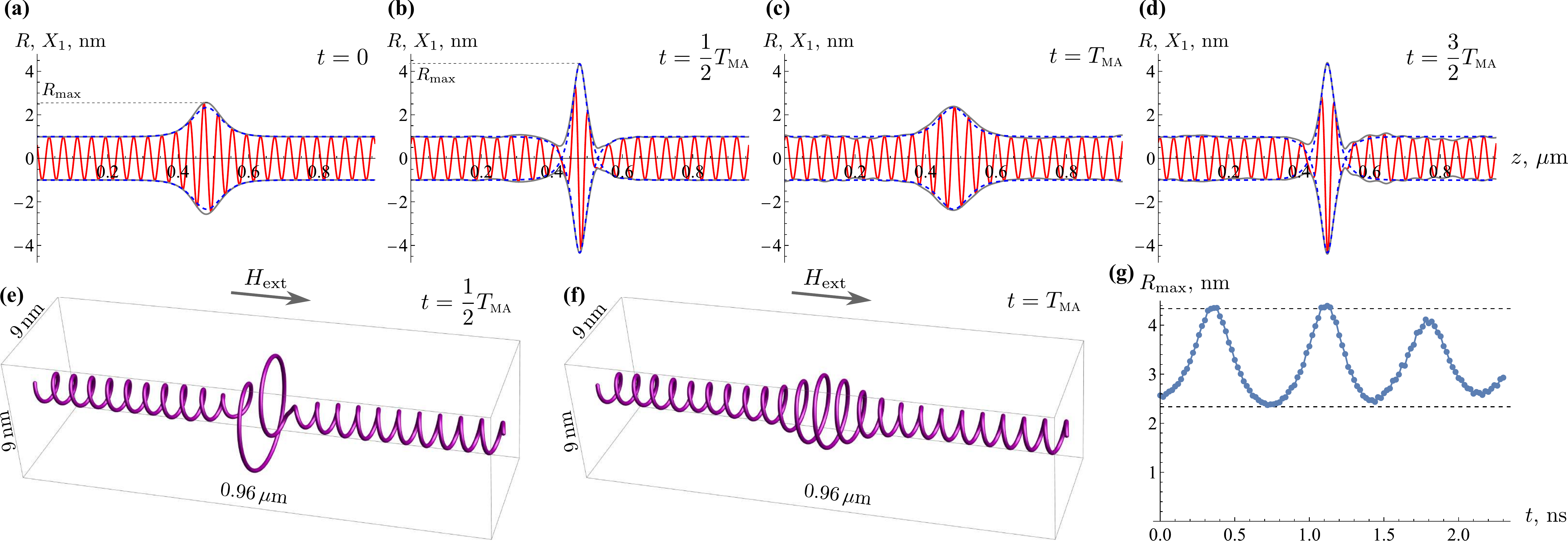}
	\caption{(a)-(d) -- micromagnetic simulations of time evolution of Ma breather along the skyrmion string in FeGe for $\mu_0H_{\text{ext}}=0.8$~T. The carrying helical wave has amplitude $R_0=1$~nm ($\mathcal{R}_0=0.01$) and pitch 40~nm ($k_0=1.75$). We consider the case $\varphi=1.1$. Dashed blue line shows the theoretically predicted profile $R=|\mathcal{A}_{\textsc{ma}}|/Q$ determined by \eqref{eq:Ma} for given parameters, while solid red and gray lines represent the simulation data, they have the same meaning as in Figs.~\ref{fig:instab},~\ref{fig:waves-simuls}. Insets (e) and (f) show 3D structure of the breather for the time moments corresponding to the maximal and minimal amplitude. Note the 30x magnification in the directions perpendicular to magnetic field.  Breathers moves with the group velocity $v_g(k_0)$, however, in each snapshot, the breather is shown in the center of the sample for better visual perception. Inset (g) shows time evolution of the breather amplitude $R_{\text{max}}=\mathcal{A}_{\textsc{ma}}^0/Q$. Dashed horizontal lines indicates the theoretically predicted boundaries for the breather amplitude, see discussion in the text. Period of the simulated breather is $T_{\textsc{ma}}=0.76$~ns. The breather dynamics is shown in supplemental Movie~6~\cite{Note2}. }\label{fig:breather}
\end{figure*}

In order to verify the considered above predictions of NLSE~\eqref{eq:NLS}, we performed a number of micromagnetic simulations~\cite{Vansteenkiste14}. The initial state was created programmatically in the form of a skyrmion string, whose shape approximately corresponds to the theoretically predicted solution for dn- or cn-wave with the profiles determined by \eqref{eq:dn} or \eqref{eq:cn}, respectively. Then for a short time ($\approx10$~ps) the micromagnetic dynamics was relaxed, i.e. run with high Gilbert damping ($\alpha=0.5$). On the next step, the damping was switched-off and the micromagnetic dynamics was simulated for a long time (10~ns). We observe the propagation of the created nonlinear wave with almost unchanged envelope profile, see Fig.~\ref{fig:waves-simuls}. The latter says that the initially created wave is a solution of the equations of string dynamics. The slight profiles deformation during dynamics takes place because of deviation of the initial states from the exact solutions. This deviation is approximately 10\%, it unavoidably appears due to the relaxation procedure applied on the second step. Dynamics of the cnoidal waves extracted from the simulations are demonstated in the supplemental Movies~3 and 4 \cite{Note2}.

Propagation of solitons along the skyrmion string was in detail considered in our previous work~\cite{Kravchuk20}, see also the supplemental movies in Ref.~\onlinecite{Kravchuk20}. However, for the sake of completeness, we present here an example of the soliton dynamics obtained by means of the micromagnetic simulations, see Fig.~\ref{fig:soliton}. The initial state was close to solution \eqref{eq:sol} with $\mathcal{R}_{\text{max}}=0.27$ (3~nm) and $\Delta_{\textsc{dn}}=5.17$ (57.6~nm). These parameters are consistent with the wave vector $k_0=1.17$ ($\frac{2\pi}{60\,\text{nm}}$) of the carrying wave. Note that soliton keeps its shape close to the initial profile (blue dashed line). Due to the inperfectness of the initial state and the discretness effects, the low-amplitude magnons are generated on background. The corresponding energy loss leads to the insignificant reduce of the soliton amplitude, see panels (a) and (b) in Fig.~\ref{fig:soliton}. The complete time evolution of the soliton propagation is shown in supplemental Movie~5 \cite{Note2}.

\subsection{A breather solutions}
The considered above solutions of NLSE \eqref{eq:NLS}  have form of the traveling waves, i.e. there is a frame of reference in which the envelope wave $|\mathcal{A}|$ is static. Here we consider breathers -- the family of solutions beyond the class of the traveling waves. There are several kinds of breathers of NLSE: localized in space and periodic in time Ma breathers~\cite{Ma79}, localized in time and periodic in space Akhmediev breather~\cite{Akhmediev87} and Peregrine solution which is localized both in space and in time \cite{Peregrine83}. In the following, we focus on Ma breather which is spatially localized periodically pulsating perturbation of the helical wave in form \cite{Ma79} $\mathcal{A}_{\textsc{ma}}=\mathcal{R}_0\varrho(\tilde{z}',\tilde{t})e^{i\Omega\tilde{t}}$, where $\Omega=\nu(k_0)\mathcal{R}_0^2$ is the same as for the helical wave and
\begin{equation}\label{eq:Ma}
	\varrho(\tilde{z}',\tilde{t})=\frac{\cos(\Omega_{\textsc{ma}}\tilde{t}-2i\varphi)+\cosh(\tilde{z}'/\Delta_{\textsc{ma}})\cosh\varphi}{\cos(\Omega_{\textsc{ma}}\tilde{t})+\cosh(\tilde{z}'/\Delta_{\textsc{ma}})\cosh\varphi}.
\end{equation}
Additionally to the characteristics of the carrying wave $\mathcal{R}_0$ and $k_0$, the breather solution is controlled also by the real-valued parameter $\varphi$. These parameters determine the pulsation frequency $\Omega_{\textsc{ma}}=\Omega\sinh(2\varphi)$ and the breather width $\Delta_{\textsc{ma}}=\Delta_{\textsc{dn}}/(2\sinh\varphi)$. The breather amplitude $\mathcal{A}_{\textsc{ma}}^0=|\mathcal{A}_{\textsc{ma}}(\tilde{z}'=0)|$ varies in range $\mathcal{R}_0(2\cosh\varphi-1)\le\mathcal{A}_{\textsc{ma}}^0\le\mathcal{R}_0(2\cosh\varphi+1)$ during the pulsations. The considered breather moves with the group velocity. The generalization for the case of arbitrary velocity $V$ is the same as for the traveling waves solutions.

Using micromagnetic simulations we found Ma breather for a skyrmion string in FeGe, see Fig.~\ref{fig:breather}.
A very good agreement between the simulated and theoretical breather profiles (Fig.~\ref{fig:breather}a-d) as well as the observed periodical breathing behavior prove that Ma breather is indeed a solution of the skyrmion string dynamics. Nevertheless one has to note that in simulations the breather develops instability after the first three breathing periods. This instability has several sources. In contrast to solitons, the breather is the excitation of the helical wave of a \emph{finite} amplitude. As it was shown above, such waves demonstrate modulational instability. Also, for technical reasons related to the limited computational resources, we were able to simulate breather for relatively large wave-vector of the carrying wave, $k_0=1.75$. Since $k_0\mathcal{A}_{\textsc{ma}}^0\approx0.71$, we are at the edge of applicability of the developed theory which implies $|k_0\mathcal{A}|\ll1$. Due to the large value of $k_0$ one can only approximately estimate linear $\omega_0(k_0)$ and nonlinear $\omega_1(k_0)$ parts of the dispersion if only a few first terms in \eqref{eq:helix-disp} are taken into account. This is the reason why the theoretically expected period of the breathing $2\pi/\Omega_{\textsc{ma}}\approx0.49$~ns differs from the period $T_{\textsc{ma}}\approx0.76$~ns obtained in the simulations. 

\section{Collective dynamics of generalized strings}\label{sec:general}
Previously we considered the case when the collective variables $X_i(z,t)$ have sense of the string displacement in $xy$-plane.
Let us now consider a generalized Ansatz
\begin{equation}\label{eq:ansatz-general}
	\vec{m}(\vec{r},t)=\vec{m}_0(x,y,X_i(z,t),X_i'(z,t)),
\end{equation}
where $\vec{m}_0$ is a known function and the string collective variables $X_i$ can have an arbitrary sense. In this section we use prime and the overdot for the derivatives with respect to $z$ and $t$, respectively. The equation of motion for $X_i$ can be formulated in general form $\delta\mathfrak{S}/\delta{X_i}=\delta\mathfrak{F}/\delta\dot{X}_i$, where $\mathfrak{S}=\int\dd t\int\dd z\mathcal{L}$ is the action with the Lagrangian $\mathcal{L}=\frac{M_s}{\gamma_0}\iint\left(\vec{\mathfrak{A}}(\vec{m})\cdot\dot{\vec{m}}\right)\dd x\dd y-\mathcal{H}$ and $\mathfrak{F}=\int\dd z\mathcal{F}$ is the dissipation function with $\mathcal{F}=\frac{\alpha}{2}\frac{M_s}{\gamma_0}\iint\dot{\vec{m}}^2\dd x\dd y$. Here the vector potential $\vec{\mathfrak{A}}$ is such that $\vec{m}\cdot\left(\vec{\nabla}_{\vec{m}}\times\vec{\mathfrak{A}}\right)=1$ and the system Hamiltonian is $H=\int\dd z \mathcal{H}$. For the model \eqref{eq:ansatz-general}, the equations of motion obtain form
\begin{equation}\label{eq:main-equation}
	\begin{split}
		&\left[G_{ij}^{(0)}+\!\left(G_{ji}^{(1)}\right)'\right]\dot{X}_j+\!
		\left[G_{ij}^{(1)}+G_{ji}^{(1)}-\!\left(G_{ij}^{(2)}\right)'\right]\dot{X}_j'\\
		&-G_{ij}^{(2)}\dot{X}_j''=\frac{\delta H}{\delta X_i}+\alpha\biggl\{\left[D_{ij}^{(0)}-\left(D_{ji}^{(1)}\right)'\right]\dot{X}_j\\
		&+\left[D_{ij}^{(1)}-D_{ji}^{(1)}-\left(D_{ij}^{(2)}\right)'\right]\dot{X}_j'-D_{ij}^{(2)}\dot{X}_j''\biggr\},
	\end{split}
\end{equation}
for details see Appendix~\ref{app:gen}. The gyroscopic $G_{ij}^{(n)}$ and dissipation $D_{ij}^{(n)}$ tensors are functionals of $X_i$ and their derivatives, they are listed in \eqref{eq:G}. Tensors $G_{ij}^{(0)}=-G_{ji}^{(0)}$ and $G_{ij}^{(2)}=-G_{ji}^{(2)}$ are asymmetrical by definition. While the tensors $D_{ij}^{(0)}=D_{ji}^{(0)}$ and $D_{ij}^{(2)}=D_{ji}^{(2)}$ are symmetrical. Note that $(G_{ij}^{(n)})'\propto X_i',X_i''$ and $(D_{ij}^{(n)})'\propto X_i',X_i''$, so these terms in \eqref{eq:main-equation} result in the nonlinear corrections.

\subsection{Radially symmetrical excitation of the skyrmion string}
As an example of application of the generalized equations \eqref{eq:main-equation} we consider dynamics of the radially symmetrical deformation of the string. It can be described by the following Ansatz
\begin{equation}\label{eq:ansatz-rad-sym}
	\theta=\theta_0\left(\rho/s\right),\qquad \phi=\chi+\frac{\pi}{2}+\varphi,
\end{equation}
where $\theta$ and $\phi$ are the spherical angles of the parameterization $\vec{m}=\sin\theta(\cos\phi\hat{\vec{x}}+\sin\phi\hat{\vec{y}})+\cos\theta\hat{\vec{z}}$, and $(\rho,\chi)$ are polar coordinates within $xy$-plane.
Here, $\theta_0(\rho)$ is profile of the unperturbed skyrmion, and $s(z,t)=X_1$ and $\varphi(z,t)=X_2$ are the collective string variables. According to \eqref{eq:G}, we have $G_{ij}^{(0)}=2\pi\epsilon_{ij}|\mathfrak{c}|sM_s/(\gamma_0Q^2)$, and $G_{ij}^{(1)}=G_{ij}^{(2)}=0$. Here $\mathfrak{c}=\int_0^\infty\tilde{\rho}^2\sin\theta_0\theta_0'(\tilde{\rho})\dd\tilde{\rho}<0$, where $\tilde{\rho}=Q\rho$. 

In what follows, we neglect damping for simplicity.
In terms of the dimensionless time $\tilde{t}=t\omega_{c2}$ and coordinate $\tilde{z}=zQ$ we write the equations of motion \eqref{eq:main-equation} in form
\begin{equation}\label{eq:s-phi}
	2|\mathfrak{c}|s\partial_{\tilde{t}}\varphi=\frac{\delta\tilde{H}}{\delta{s}},\qquad	-2|\mathfrak{c}|s\partial_{\tilde{t}}s=\frac{\delta\tilde{H}}{\delta{\varphi}},
\end{equation}
where the dimensionless Hamiltonian is
\begin{equation}\label{eq:H-s-phi}
	\begin{split}
	\tilde{H}=\!\!\int_{-\infty}^{\infty}\!\!\dd\tilde{z}&\bigl\{e_{\text{ex}}+\mathfrak{c}_1(\partial_{\tilde{z}}s)^2+\mathfrak{c}_2s^2\left[(\partial_{\tilde{z}}\varphi)^2-2\partial_{\tilde{z}}\varphi\right]\\
	&+2e_{\text{dmi}}s\cos\varphi+2e_{\text{z}}hs^2\bigr\}.
	\end{split}
\end{equation}
In order to obtain the effective Hamiltonian \eqref{eq:H-s-phi} for the collective string variables $s$ and $\varphi$, we substitute Ansatz \eqref{eq:ansatz-rad-sym} into the main Hamiltonian \eqref{eq:H} and perform the integration over $xy$-plane. Finally, we obtain $H=2\pi AQ^{-1}\tilde{H}$. The constants $e_{\text{ex}}=\mathscr{H}_{ex}^{(0)}/4$,  $e_{\text{dmi}}=\mathscr{H}_{\textsc{dmi}}^{(0)}/4$, and $e_{\text{z}}=\mathscr{H}_{z}^{(0)}/4$ represent (up to a constant multiplier) the exchange, DMI and Zeeman energies of the unperturbed skyrmion string, respectively, see Appendix~\ref{app:energies} for the definitions of $\mathscr{H}_{\bullet}^{(0)}$. The other constants are $\mathfrak{c}_1=\int_0^\infty\tilde{\rho}^3\theta_0'(\tilde{\rho})^2\dd\tilde{\rho}$ and $\mathfrak{c}_2=\int_0^\infty\tilde{\rho}\sin^2\theta_0\dd\tilde{\rho}$.

The explicit form of the equations of motion \eqref{eq:s-phi} is
\begin{equation}\label{eq:s-phi-expl}
	\begin{split}
		|\mathfrak{c}|s\partial_{\tilde{t}}s=&\mathfrak{c}_2s^2\partial_{\tilde{z}}^2\varphi-2\mathfrak{c}_2s\partial_{\tilde{z}}s(1-\partial_{\tilde{z}}\varphi)-|e_{\text{dmi}}|s\sin\varphi,\\
		|\mathfrak{c}|s\partial_{\tilde{t}}\varphi=&-\mathfrak{c}_1\partial_{\tilde{z}}^2s+\mathfrak{c}_2s\left[(\partial_{\tilde{z}}\varphi)^2-2\partial_{\tilde{z}}\varphi\right]\\
		&+|e_{\text{dmi}}|\left(s-\cos\varphi\right).
	\end{split}
\end{equation}
Writing \eqref{eq:s-phi-expl} we exclude $e_{\text{z}}$ by means of the virial relation $2he_{\text{z}}+e_{\text{dmi}}=0$ and we use that $e_{\text{dmi}}<0$. The solution $s=1$ and $\varphi=0$ corresponds to the ground state if $h>1$. 

\begin{figure}
	\includegraphics[width=\columnwidth]{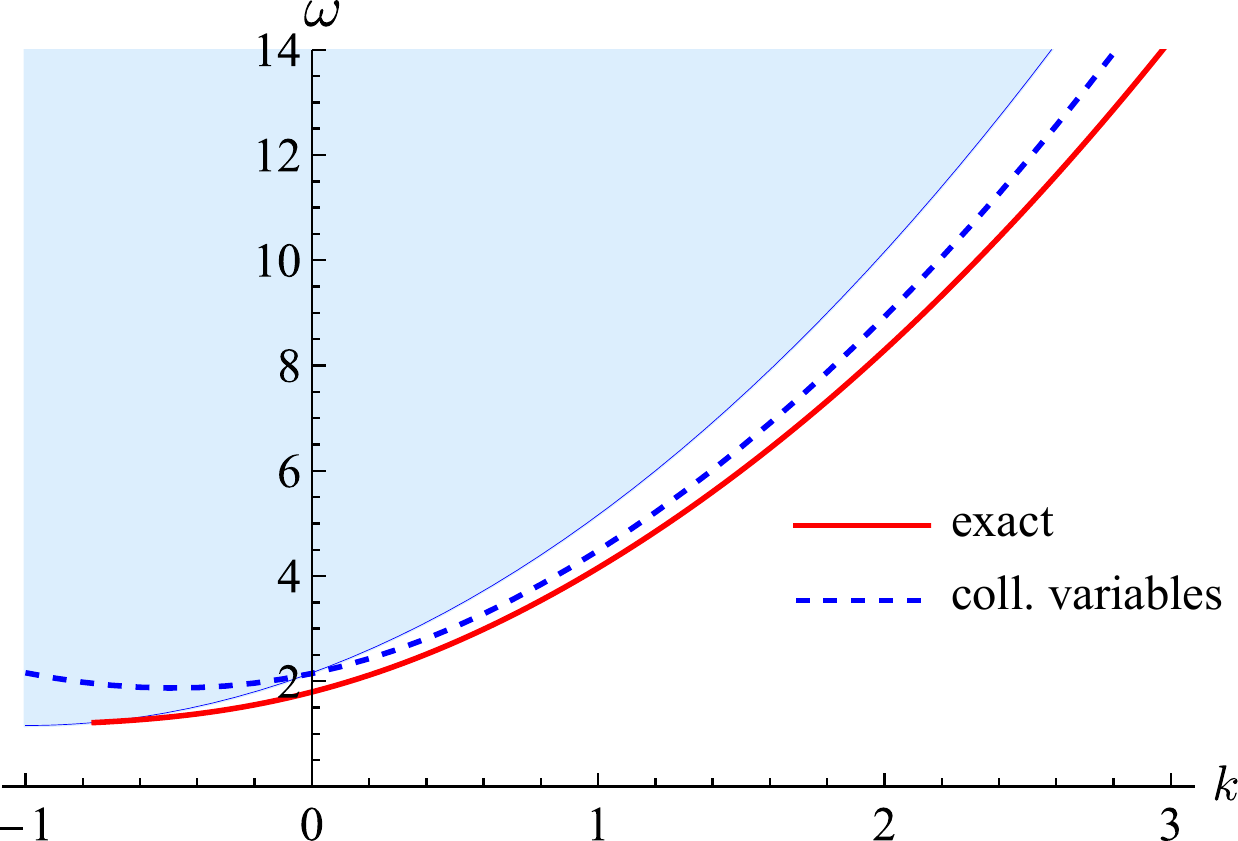}
	\caption{Dispersion relation of the breathing mode for $h=2.16$ ($\mu_0H_{\text{ext}}=0.8$~T) (solid line) is compared to the collective variables approximation \eqref{eq:disp-brth} (dashed line) with the following values of the constants $\mathfrak{c}\approx-0.684$, $\mathfrak{c}_1\approx1.299$, $\mathfrak{c}_2\approx0.397$, $e_{\text{dmi}}\approx-1.477$. The ``exact'' solution is the same as in Fig.~\ref{fig:kc}(a).}\label{fig:disp-brth}
\end{figure}

Next, we introduce small deviations $s=1+\tilde{s}$ and $\varphi=\tilde{\varphi}$ on the top of the ground state. The linearization of \eqref{eq:s-phi-expl} with respect to the deviations results in the planar wave solutions $\tilde{s},\tilde{\varphi}\propto e^{i(k\tilde{z}-\omega\tilde{t})}$ with the dispersion relation
\begin{equation}\label{eq:disp-brth}
	\begin{split}
	|\mathfrak{c}|\omega=&2\mathfrak{c}_2k+\sqrt{(\mathfrak{c}_1k^2+|e_{\text{dmi}}|)(\mathfrak{c}_2k^2+|e_{\text{dmi}}|)}\\
	&\approx|e_{\text{dmi}}|+2\mathfrak{c}_2k+\frac{\mathfrak{c}_1+\mathfrak{c}_2}{2}k^2.
	\end{split}
\end{equation}
Note that $\mathfrak{c}_1+\mathfrak{c}_2$ is the second moment of the exchange energy density of the unperturbed string. The comparison of the approximated dispersion \eqref{eq:disp-brth} with the exact one is shown in Fig.~\ref{fig:disp-brth}.

Equations of motion \eqref{eq:s-phi} are Euler-Lagrange equations of the Lagrange function 
\begin{equation}\label{eq:L}
	\mathfrak{L}=|\mathfrak{c}|\int_{-\infty}^{\infty}s^2\partial_{\tilde{t}}\varphi\,\dd\tilde{z}-\tilde{H}.
\end{equation}
The invariance of the Lagrange function with respect to translations along $\tilde{t}$ and $\tilde{z}$ results in two integrals of motion, namely energy $E=\tilde{H}$ and momentum $P=-|\mathfrak{c}|\int_{-\infty}^{\infty}s^2\partial_{\tilde{z}}\varphi\dd\tilde{z}$. From the latter expression, we derive $\delta P=2|\mathfrak{c}|\int_{-\infty}^{\infty}s\left[\partial_{\tilde{z}}s\delta\varphi-\partial_{\tilde{z}}\varphi\delta s\right]\dd\tilde{z}$. On the other hand, we can write $\delta E=\int_{-\infty}^{\infty}\left[\frac{\delta\tilde{H}}{\delta\varphi}\delta\varphi+\frac{\delta\tilde{H}}{\delta s}\delta s\right]\dd\tilde{z}$, and with the help of \eqref{eq:s-phi} we obtain $\delta E=V\delta P$ for the traveling-wave solutions $s=s(\tilde{z}-V\tilde{t})$ and $\varphi=\varphi(\tilde{z}-V\tilde{t})$. Thus, we obtain a Hamiltonian equation $V=\partial E/\partial P$ for a Newtonian particle. Note that the second Hamiltonian equation is $\partial_{\tilde{t}}P=0$ due to the momentum conservation.

For the traveling-wave solutions, functions $s(\tilde{z}')$ and $\varphi(\tilde{z}')$ are determined by equations
\begin{equation}\label{eq:s-phi-travel}
	\begin{split}
		&|\mathfrak{c}|Vss'=-\mathfrak{c}_2s^2\varphi''+2\mathfrak{c}_2ss'(1-\varphi')+|e_{\text{dmi}}|s\sin\varphi,\\
		&|\mathfrak{c}|Vs\varphi'=\mathfrak{c}_1s''-\mathfrak{c}_2s\left[\varphi'^2-2\varphi'\right]-|e_{\text{dmi}}|\left(s-\cos\varphi\right),
	\end{split}
\end{equation}
where prime denotes the derivative with respect to $\tilde{z}'=\tilde{z}-V\tilde{t}$. Note that it is technically easier to find the numerical solutions $s(\tilde{z}')$ and $\varphi(\tilde{z}')$ as minimizers of the effective energy $E_{eff}=E-VP$.

\begin{figure}
	\includegraphics[width=\columnwidth]{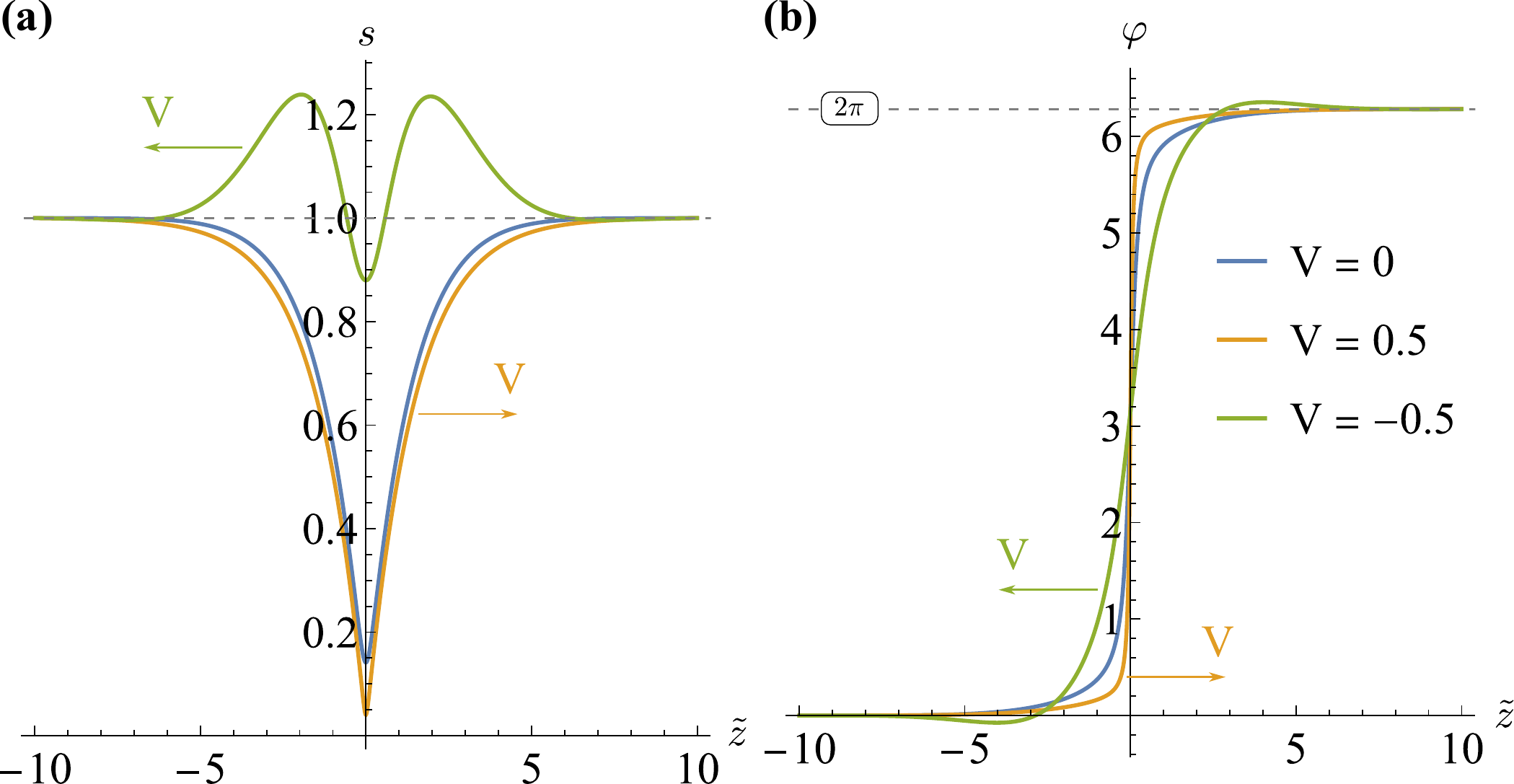}
	\caption{Topological solutions of Eqs.~\eqref{eq:s-phi-travel} in form of $2\pi$-domain wall are obtained for different $V$. We consider the case with  $h=1.05$ corresponding to $\mathfrak{c}\approx-3.34$, $\mathfrak{c}_1\approx7.4$, $\mathfrak{c}_2\approx1.67$, and $e_{\text{dmi}}\approx-3.51$. }\label{fig:dw}
\end{figure}

Eqs.~\eqref{eq:s-phi-travel} have solutions in form of $2\pi$-domain wall (DW), which encircles the skyrmion tube, see Fig.~\ref{fig:dw}. In the center of the DW, $\phi=\pi$, see Fig.~\ref{fig:dw}(b), meaning that skyrmion helicity there is opposite to the equilibrium one. The latter results in the significant increase of the DMI energy. As a result, the skyrmion tube shrinks at the DW position, in order to reduce the DMI energy, see Fig.~\ref{fig:dw}(a). On the other hand, the skyrmion tube shrinking leads to increase of the magnetization gradients and therefore -- to the increase of the exchange energy. The competition between exchange and DMI results in small but nonvanishing radius of the skyrmion tube in the domain wall center. Remarkably, that the DW motion against the applied magnetic field ($V<0$) can significantly increase the skyrmion tube radius in the DW position, see the case $V=-0.5$ in Fig.~\ref{fig:dw}(a). The latter effect is promising for avoiding the skyrmion string breaking in possible experimental realization of the considered DW.


\section{Conclusions}
We have shown that the low-energy dynamics of the translational mode propagating along the skyrmion string is captured by the Nonlinear Schr{\"o}dinger Equation (NLSE) of focusing type. As a result, a number of solutions of NLSE are found and confirmed by means of micromagnetic simulations, namely cnoidal waves, solitons, breathers. Finally, we proposed a generalized approach, which enables one to describe nonlinear dynamics of the modes of different symmetries, radially symmetrical, elliptical, etc.

The proposed here approach has a wide spectrum of the future perspectives: (i) it can be applied for antyskirmion string, as well as for meron strings, (ii) it can be adapted for antiferromagnetic skyrmion or meron string, (iii) it can be used for analysis of weakly nonlinear dynamics of higher modes of the strings.

\section{Acknowledgments}
I appreciate fruitful discussions with Markus Garst; and also discussions with Denis Sheka on the early stage of this work.

In part, this work was supported by the Program of Fundamental Research of the Department of Physics and Astronomy of the National Academy of Sciences of Ukraine (Project No. 0117U000240).

\appendix

\section{Poisson brackets for the string collective variables.}\label{app:PB}

As a direct consequence of \eqref{eq:m-poiss}, one can show that the Poisson brackets of any two functionals $F[\vec{m}]$, $G[\vec{m}]$ are~\cite{McKeever19}
\begin{equation}\label{eq:poisson-FG}
	\{F,G\}=-\frac{\gamma_0}{M_s}\int\vec{m}\cdot\left[\frac{\delta F}{\delta\vec{m}}\times \frac{\delta G}{\delta\vec{m}}\right]\dd\vec{r}.
\end{equation}
The string coordinates \eqref{eq:Xi} can be considered as functionals of magnetization:
\begin{equation}\label{eq:X-func}
	X_i(z)=\frac{1}{N_{\text{top}}}\int x'_i g_{z}(\vec{r}')\delta(z-z')\dd\vec{r}'.
\end{equation}
The straightforward calculation of the functional derivative results in 
\begin{equation}\label{eq:dXdm}
	\begin{split}
	\frac{\delta X_i(z)}{\delta\vec{m}(z')}=\frac{1}{4\pi N_{\text{top}}}\bigl\{&3x_i[\partial_x\vec{m}\times\partial_y\vec{m}]\\
	&+\epsilon_{ij}[\vec{m}\times\partial_{x_j}\vec{m}]\bigr\}\delta(z-z').
	\end{split}
\end{equation}
The usage of \eqref{eq:dXdm} in \eqref{eq:poisson-FG} results in \eqref{eq:X-poiss}.

\section{Definition and properties of the curvilinear coordinates.}\label{app:coords}

The aim of this section is to introduce a curvilinear frame of reference which follows the form of the string. Such a reference frame is convenient for formulation of the skyrmion string Ansatz and for calculation of the string energy. 

Let us consider a 3D curve $\vec{\gamma}: \; \mathbb{R}\to\mathbb{R}^3$ parametrized by means of the scalar parameter $-\infty<\zeta<\infty$ in the following manner: $\vec{\gamma}(\zeta)=\vec{X}(\zeta)+\zeta\hat{\vec{z}}$ where $\vec{X}(\zeta)=X_1(\zeta)\hat{\vec{x}}+X_2(\zeta)\hat{\vec{y}}$. The unit vector tangential to the string is
\begin{equation}\label{eq:et}
	\vec{e}_{\textsc{t}}(\zeta)=\frac{\vec{X}'+\hat{\vec{z}}}{\sqrt{1+|\vec{X}'|^2}}.
\end{equation}
Here and below, prime denotes derivative with respect to $\zeta$. Within the plane perpendicular to $\vec{e}_{\textsc{t}}$, we introduce two orthogonal unit vectors $\vec{e}_1$ and $\vec{e}_2$ as follows
\begin{equation}\label{eq:e1-e2}
	\vec{e}_1(\zeta)=\frac{\hat{\vec{x}}-\vec{e}_{\textsc{t}}(\vec{e}_{\textsc{t}}\cdot\hat{\vec{x}})}{\sqrt{1-(\vec{e}_{\textsc{t}}\cdot\hat{\vec{x}})^2}},\qquad \vec{e}_2(\zeta)=\vec{e}_{\textsc{t}}\times\vec{e}_1.
\end{equation}
I.e. vector $\vec{e}_1$ is the normalized projection of the Cartesian ort $\hat{\vec{x}}$ on the plane perpendicular to $\vec{e}_{\textsc{t}}$. The space domain which includes curve $\vec{\gamma}$ and its vicinity we parameterize as $\vec{r}(\xi_1,\xi_2,\zeta)=\vec{\gamma}(\zeta)+\xi_1\vec{e}_1(\zeta)+\xi_2\vec{e}_2(\zeta)$, where $\xi_1$, $\xi_2$, $\zeta$ are local coordinates of the frame of reference defined on the string. In the other words, $(\xi_1,\xi_2)$ are coordinates within the plane $\Pi_\zeta$ perpendicular to the string which crosses the string in point $\vec{r}=\vec{\gamma}(\zeta)$.

The explicit form of the relation between coordinates $\{x,y,z,t\}$ of the laboratory frame of reference and coordinates $\{\xi_1,\xi_2,\zeta,\tau\}$ of the string frame of reference is
\begin{equation}\label{eq:params}
	\begin{split}
		&x=X_1(\zeta,\tau)+(\vec{e}_1\cdot\hat{\vec{x}})\xi_1+(\vec{e}_2\cdot\hat{\vec{x}})\xi_2,\\
		&y=X_2(\zeta,\tau)+(\vec{e}_1\cdot\hat{\vec{y}})\xi_1+(\vec{e}_2\cdot\hat{\vec{y}})\xi_2,\\
		&z=\zeta+(\vec{e}_1\cdot\hat{\vec{z}})\xi_1+(\vec{e}_2\cdot\hat{\vec{z}})\xi_2,\\
		&t=\tau,
	\end{split}
\end{equation}
where $\vec{e}_i=\vec{e}_i(\zeta,\tau)$. So, on the curve $\vec{\gamma}$ one has $(\xi_1,\xi_2)=(0,0)$ and $\zeta=z$. 

\begin{widetext}
	 The  differentials of the coordinates are related via Jacobian $J$:
	\begin{equation}\label{eq:J}
		\begin{bmatrix}
			dx\\dy\\dz\\dt
		\end{bmatrix}=J\begin{bmatrix}
			d\xi_1\\d\xi_2\\d\zeta\\d\tau
		\end{bmatrix},\qquad
		J=\begin{bmatrix}
			(\vec{e}_1\cdot\hat{\vec{x}}) & (\vec{e}_2\cdot\hat{\vec{x}}) & X_1'+(\vec{e}_1'\cdot\hat{\vec{x}})\xi_1 + (\vec{e}_2'\cdot\hat{\vec{x}})\xi_2 & \partial_\tau{X}_1+(\partial_\tau{\vec{e}}_1\cdot\hat{\vec{x}})\xi_1 + (\partial_\tau{\vec{e}}_2\cdot\hat{\vec{x}})\xi_2 \\
			(\vec{e}_1\cdot\hat{\vec{y}}) & (\vec{e}_2\cdot\hat{\vec{y}}) & X_2'+(\vec{e}_1'\cdot\hat{\vec{y}})\xi_1 + (\vec{e}_2'\cdot\hat{\vec{y}})\xi_2 & \partial_\tau{X}_2+(\partial_\tau{\vec{e}}_1\cdot\hat{\vec{y}})\xi_1 + (\partial_\tau{\vec{e}}_2\cdot\hat{\vec{y}})\xi_2 \\
			(\vec{e}_1\cdot\hat{\vec{z}}) & (\vec{e}_2\cdot\hat{\vec{z}}) & 1+(\vec{e}_1'\cdot\hat{\vec{z}})\xi_1 + (\vec{e}_2'\cdot\hat{\vec{z}})\xi_2 & (\partial_\tau{\vec{e}}_1\cdot\hat{\vec{z}})\xi_1 + (\partial_\tau{\vec{e}}_2\cdot\hat{\vec{y}})\xi_2\\
			0 & 0 & 0 & 1
		\end{bmatrix}.
	\end{equation}
		\end{widetext}
	Determinant $|J|$ determines the volume element in the curvilinear frame of reference, namely $\dd x\,\dd y\,\dd z\,\dd t=|J|\dd\xi_1\,\dd\xi_2\,\dd\zeta\,\dd\tau$. For small derivatives of $X_i$ we can write $|J|=1-(\xi_1X_1''+\xi_2X_2'')+\frac12|\vec{X}'|^2+\dots$. Note that due to the structure of Jacobian \eqref{eq:J}, one has $\dd t=\dd\tau$ and therefore $\dd x\,\dd y\,\dd z=|J|\dd\xi_1\,\dd\xi_2\,\dd\zeta$. 
	
	Elements of the inverted Jacobian determine derivatives with respect to the Cartesian coordinates:
	\begin{equation}\label{eq:J-inv}
		\begin{split}
		&J^{-1}=\begin{bmatrix}
			\partial_x\xi_1 & \partial_y\xi_1 & \partial_z\xi_1 & \partial_t\xi_1 \\
			\partial_x\xi_2 & \partial_y\xi_2 & \partial_z\xi_2 & \partial_t\xi_2 \\
			\partial_x\zeta & \partial_y\zeta & \partial_z\zeta & \partial_t\zeta \\
			\partial_x\tau & \partial_y\tau & \partial_z\tau & \partial_t\tau
		\end{bmatrix}\\
	&=\begin{bmatrix}
			1 & 0 & -X_1' & -\partial_\tau{X}_1 \\
			0 & 1 & -X_2' & -\partial_\tau{X}_2 \\
			X_1' & X_2' & 1+\xi_1X_1''+\xi_2 X_2'' & \xi_1\partial_\tau{X}_1'+ \xi_2\partial_\tau{X}_2' \\
			0 & 0 & 0 & 1
		\end{bmatrix}+o(X).
	\end{split}
	\end{equation}
	
Thus, we find that the spatial derivatives are related as
	\begin{equation}\label{eq:derivs}
	\begin{split}
		\partial_x=&\partial_{\xi_1}+X_1'\partial_\zeta+o(X),\\ \partial_y=&\partial_{\xi_2}+X_2'\partial_\zeta+o(X),\\
		\partial_z=&-X_1'\partial_{\xi_1}-X_2'\partial_{\xi_2}\\
		&+(1+\xi_1X_1''+\xi_2 X_2'')\partial_\zeta+o(X).
	\end{split}
\end{equation} 
And the time derivative is 	
	\begin{equation}\label{eq:derivs-t}
			\begin{split}			
		\partial_t=&\partial_{\tau}-\partial_\tau{X}_1\partial_{\xi_1}-\partial_\tau{X}_2\partial_{\xi_2}\\
		&+(\xi_1\partial_\tau{X}_1'+ \xi_2\partial_\tau{X}_2')\partial_\zeta+o(X).
		\end{split}
	\end{equation} 
Quadratic and higher order terms in derivatives of $X_i$ are denoted as $o(X)$. 

\section{Ansatz for the skyrmion string}\label{app:Ansatz}

Let us first consider the magnetization distribution in form
\begin{equation}\label{eq:m-tilde}
	\begin{split}
	&\tilde{\vec{m}}_{\textsc{a}}(\xi_1,\xi_2,\zeta)=\sin\theta_0(\rho)\cos\phi_0(\xi_1,\xi_2)\,\vec{e}_1(\zeta)\\
	&+\sin\theta_0(\rho)\sin\phi_0(\xi_1,\xi_2)\,\vec{e}_2(\zeta)+\cos\theta_0(\rho)\vec{e}_{\textsc{t}}(\zeta),
	\end{split}
\end{equation}
where $\xi_1,\,\xi_2,\,\zeta$ are curvilinear coordinates introduced in Appendix~\ref{app:coords}. Namely, the coordinates $(\xi_1,\xi_2)$ sweep the plane of perpendicular cross-section $\Pi_\zeta$ of the string made in point $\vec{r}=\vec{\gamma}(\zeta)$.
Here $\theta_0(\rho)$ is profile of the vertical equilibrium string oriented along the applied magnetic field, and $\rho=\sqrt{\xi_1^2+\xi_2^2}$.
Function $\theta_0(\rho)$ is assumed to be known, it coincides with skyrmion profile in a 2D magnet. Angular variable $\phi_0(\xi_1,\xi_2)=\chi+\varphi_0$ determines the magnetization orientation within the plane $\Pi_\zeta$. Here $\cos\chi=\xi_1/\sqrt{\xi_1^2+\xi_2^2}$ and $\sin\chi=\xi_2/\sqrt{\xi_1^2+\xi_2^2}$, in the other words, $(\rho,\chi)$ are polar coordinates within $\Pi_\zeta$ in the same maner as fo a planar skyrmion. Constant $\varphi_0=\pm\pi/2$ and $\varphi_0=0,\pi$ for Bloch and Ne{\'e}el skyrmion strings, respectively. An example of the magnetization distribution \eqref{eq:m-tilde} for a Bloch string is shown in Fig.~\ref{fig:Ansatz}(a). 

So, for a given $\zeta$, model \eqref{eq:m-tilde} determines the magnetization within $\Pi_\zeta$. However, \eqref{eq:m-tilde} can not be used as a string Ansatz because it is ambiguous on large distances from the string where different planes $\Pi_\zeta$ for different $\zeta$ can intersect. In the other words, model \eqref{eq:m-tilde} makes sense only for $\rho\kappa\ll1$, where $\kappa$ the string curvature. Moreover, \eqref{eq:m-tilde} does not satisfy the required boundary condition $\lim_{\varrho\to\infty}{\vec{m}}=\hat{\vec{z}}$. For this reason we consider the modified Ansatz 
\begin{equation}\label{eq:Ansatz}
	\vec{m}_{\textsc{a}}(\xi_1,\xi_2,\zeta)=\vec{\mathfrak{R}}_{\vec{u}}(\alpha_f)\tilde{\vec{m}}_{\textsc{a}}(\xi_1,\xi_2,\zeta).
\end{equation} 
Here $\vec{\mathfrak{R}}_{\vec{u}}(\alpha_f)$ is Rodrigues rotation matrix
\begin{equation}\label{eq:R}
	\vec{\mathfrak{R}}_{\vec{u}}(\alpha_f)=\cos\alpha_f\, \vec{\mathrm{I}}+\sin\alpha_f\,[\vec{u}]_\times+(1-\cos\alpha_f)(\vec{u}\otimes\vec{u}),
\end{equation}
which rotates vector $\tilde{\vec{m}}$ around the unit vector $\vec{u}$ by angle $\alpha_f$. Here $\bf{I}$ is the identity matrix and $[\vec{u}]_\times$ is the cross-product matrix of $\vec{u}$. We choose $\vec{u}=\vec{e}_{\textsc{t}}\times\hat{\vec{z}}/|\vec{e}_{\textsc{t}}\times\hat{\vec{z}}|$ and 
\begin{equation}\label{eq:phi}
	\cos\alpha_f=\frac{\sqrt{1+[1-f^2(\rho)]|\vec{X}'|^2}}{\sqrt{1+|\vec{X}'|^2}},\quad \sin\alpha_f=\frac{f(\rho)|\vec{X}'|}{\sqrt{1+|\vec{X}'|^2}}.
\end{equation}
Here function $0\le f(\rho)\le1$ controls the magnitude of rotation: the case $f=0$ corresponds to the absence of rotation, while the case $f=1$ corresponds the complete rotation by angle $\vartheta_\tau=\angle(\vec{e}_{\textsc{t}},\hat{\vec{z}})$. The unknown function $f(\rho)$ must have the property $\lim_{\rho\to\infty}f(\rho)=1$. 
In the following we assume that $f^*=1-f$ is a localized function whose localization radius $\rho^*$ is much smaller than the curvature radius of the string, i.e. $\rho^*\kappa\ll1$.  A rough estimation of the function $f(\rho)$ is discussed in Appendix~\ref{app:energies}, see Fig.~\ref{fig:f}.

For $|\vec{X}'|\ll1$ the rotation matrix is approximated as
\begin{equation}\label{eq:R-Appr}
	\vec{\mathfrak{R}}_{\vec{u}}(\alpha_f)\approx\begin{bmatrix}
		1-\frac{f^2X_1'^2}{2} & -\frac{f^2X_1'X_2'}{2} & -fX_1' \\
		-\frac{f^2X_1'X_2'}{2} & 1-\frac{f^2X_2'^2}{2} & -fX_2' \\
		fX_1' & fX_2' & 1-\frac{f^2|\vec{X}'|^2}{2}
	\end{bmatrix}.
\end{equation}

An example of model \eqref{eq:Ansatz} is shown in Fig.~\ref{fig:Ansatz}(b).

\section{Magnetic energies of the string}\label{app:energies}

In this section we compute the contribution from each energy term separately.

The exchange interaction is 
\begin{equation}\label{eq:Hex-in}
	\begin{split}
	H_{ex}&=A\iiint\dd x\dd y\dd z(\partial_i\vec{m}\cdot\partial_i\vec{m})\\
	&=A\int\dd\zeta\iint\dd\xi_1\dd\xi_2|J|(\partial_i\vec{m}_{\textsc{a}}\cdot\partial_i\vec{m}_{\textsc{a}}),
	\end{split}
\end{equation}
where $i\in\{x,y,z\}$ are Cartesian coordinates. For $\vec{m}_{\textsc{a}}$ we use Ansatz \eqref{eq:Ansatz} for the case $\theta_0(0)=\pi$ and $\theta_0(\infty)=0$. The Jacobian $|J|$ is determined from \eqref{eq:J}. For the derivatives $\partial_i$ we use \eqref{eq:derivs} with the sufficient number of terms. Finally, the straightforward calculation enables us to write \eqref{eq:Hex-in} in form
\begin{subequations}\label{eq:Hex-in-tot}
	\begin{align}\nonumber
		&H_{ex}=8\pi AQ^{-1}\!\!\int\!\!\dd\tilde{\zeta}\left(\mathscr{H}_{ex}^{(0)}+\mathscr{H}_{ex}^{(2)}+\mathscr{H}_{ex}^{(4)}+\dots\right),\\\nonumber
		&\mathscr{H}_{ex}^{(0)}=c_0,\\ \nonumber
		&\mathscr{H}_{ex}^{(2)}=\frac{c_0+\tilde{c}_0}{2}|\vec{\mathcal{X}}'|^2+\frac{c_1}{2}\sin\varphi_0[\vec{\mathcal{X}}'\times\vec{\mathcal{X}}'']_z+\frac{c_2}{2}|\vec{\mathcal{X}}''|^2,\\ \nonumber
		&\mathscr{H}_{ex}^{(4)}=-\frac{\frac12c_0+\bar{c}_0}{4}|\vec{\mathcal{X}}'|^4-\frac{c_3}{4}\sin\varphi_0|\vec{\mathcal{X}}'|^2[\vec{\mathcal{X}}'\times\vec{\mathcal{X}}'']_z. \nonumber
	\end{align}
	Here $\vec{\mathcal{X}}=Q\vec{X}$, $\tilde{\zeta}=Q\zeta$ with $Q=D/(2A)$, and $\vec{\mathcal{X}}'=\partial_{\tilde{\zeta}}\vec{\mathcal{X}}$. The coefficients are functionals of $\theta_0(\tilde{\rho})$ and $f(\tilde{\rho})$:
	\begin{align}
		\nonumber&c_0=\frac14\int_0^\infty\!\left[{\theta'_0}^2+\frac{\sin^2\theta_0}{\tilde{\rho}^{2}}\right]\!\tilde{\rho}\,\dd\tilde{\rho},\\  \nonumber&\tilde{c}_0=\frac14\int_{0}^{\infty}f'^2\left(1+\cos^2\theta_0\right)\tilde{\rho}\,\dd\tilde{\rho},\\ \nonumber&c_1=-\frac{1}{2}\int_0^\infty\tilde{\rho}^2\theta_0'f'\dd\tilde{\rho}, \\
		\nonumber&c_2=\frac14\int\limits_0^\infty\tilde{\rho}  (1+\cos^2\theta_0)(1-f)^2\dd \tilde{\rho}, \\ \nonumber&\bar{c}_0=\frac14\int_0^\infty\tilde{\rho}(1+\cos^2\theta_0)f'^2(1-2f^2)\dd\tilde{\rho}, \\ \nonumber&c_3=-\frac12\int_0^\infty\tilde{\rho}^2\theta_0'f'\left(2-f^2\right)\dd\tilde{\rho}.
	\end{align}
\end{subequations}
where $\tilde{\rho}=Q\rho$,  $\theta'_0=\partial_{\tilde{\rho}}\theta_0$ and $f'=\partial_{\tilde{\rho}} f$. Note the absence of the cubic nonlinear term.

In the analogous manner, we determine the energy of DMI:
\begin{subequations}\label{eq:HdmiB-in-tot}
	\begin{align}
		\nonumber&H_{\textsc{dmi}}=D\int\dd\zeta\iint\dd\xi_1\dd\xi_2|J|\,\vec{m}_{\textsc{a}}\cdot[\vec{\nabla}\times\vec{m}_{\textsc{a}}]\\
		\nonumber&=8\pi DQ^{-2}\int\dd\tilde{\zeta}\left(\mathscr{H}_{\textsc{dmi}}^{(0)}+\mathscr{H}_{\textsc{dmi}}^{(2)}+\mathscr{H}_{\textsc{dmi}}^{(4)}+\dots\right).
\end{align}
Here $\mathscr{H}_{\textsc{dmi}}^{(0)}=\frac14\sin\varphi_0\int_{0}^{\infty}\mathscr{E}_{\textsc{dmi}}(\tilde{\rho})\tilde{\rho}\,\dd \tilde{\rho}=\text{const}$ with $\mathscr{E}_{\textsc{dmi}}=\theta_0'+\frac{1}{\tilde{\rho}}\sin\theta_0\cos\theta_0$ represents DMI energy density of the vertical equilibrium string.  The higher terms are as follows
\begin{align}	
		\nonumber&\mathscr{H}_{\textsc{dmi}}^{(2)}=\frac{c_6}{2}\sin\varphi_0|\vec{\mathcal{X}}'|^2+\frac{c_7}{2}[\vec{\mathcal{X}}'\times\vec{\mathcal{X}}'']_z,\\
		\nonumber&\mathscr{H}_{\textsc{dmi}}^{(4)}=-\frac{c_8}{4}\sin\varphi_0|\vec{\mathcal{X}}'|^4-\frac{c_9}{4}|\vec{\mathcal{X}}'|^2[\vec{\mathcal{X}}'\times\vec{\mathcal{X}}'']_z.
		\end{align}
	Note that the cubic terms are absent, similarly to the case of the exchange energy. The coefficients have the following form
	\begin{align}
		\nonumber&c_6=\frac{1}{8}\left[(2-f^2)\mathscr{E}_{\textsc{dmi}}+ff'\sin2\theta_0\right]\tilde{\rho}\dd\tilde{\rho},\\
		\nonumber&c_7=-\frac{1}{2}\int_0^\infty\dd\tilde{\rho}\tilde{\rho} (1-f)\left[1-f\cos^2\theta_0\right],\\ \nonumber&c_8=\frac{1}{16}\!\!\int_0^\infty\!\!\dd\tilde{\rho}\tilde{\rho}\left[\mathscr{E}_{\textsc{dmi}}\left(1+(1-f^2)^2\right)+2\sin2\theta_0f'f(1-f^2)\right],\\ \nonumber&c_9=-\frac{1}{4}\!\!\int_0^\infty\!\!\dd\tilde{\rho}\tilde{\rho}\biggl[\sin^2\theta_0\left(\!1-f^3+\frac12f^4\!\right)-2f(1-f)(2-f)\\
		\nonumber&-\tilde{\rho} f'\left(2-f^2+\frac14f^2\sin^2\theta_0\right)\biggr].
	\end{align}
\end{subequations}

For Zeeman energy we obtain the analogous series
\begin{subequations}\label{eq:Hz-in-tot}
	\begin{align}
		\nonumber&H_{z}=M_s\mu_0H_{\text{ext}}\int\dd\zeta\iint\dd\xi_1\dd\xi_2|J|(1-m_{\textsc{a}z})\\
		\nonumber&=8\pi M_s\mu_0H_{\text{ext}}Q^{-3}\int\dd\tilde{\zeta}\left(\mathscr{H}_{z}^{(0)}+\mathscr{H}_{z}^{(2)}+\mathscr{H}_{z}^{(4)}+\dots\right),\\
		\nonumber&\mathscr{H}_{z}^{(0)}=\frac14\int_{0}^{\infty}(1-\cos\theta_0)\tilde{\rho}\,\dd \tilde{\rho},\\
		\nonumber&\mathscr{H}_{z}^{(2)}=\frac{c_{10}}{2}|\vec{\mathcal{X}}'|^2+\frac{c_{11}}{2}\sin\varphi_0[\vec{\mathcal{X}}'\times\vec{\mathcal{X}}'']_z,\\
		\nonumber&\mathscr{H}_{z}^{(4)}=-\frac{c_{12}}{4}|\vec{\mathcal{X}}'|^4-\frac{c_{13}}{4}\sin\varphi_0|\vec{\mathcal{X}}'|^2[\vec{\mathcal{X}}'\times\vec{\mathcal{X}}'']_z,
		\end{align}
	with the coefficients
	\begin{align}
		\nonumber& c_{10}=\frac{1}{4}\int_0^\infty\dd\tilde{\rho}\tilde{\rho} \left[1-\cos\theta_0f(2-f)\right],\\
		\nonumber& c_{11}=\frac{1}{4}\int_0^\infty\dd\tilde{\rho}\tilde{\rho}^2 (1-f)\sin\theta_0,\\
		\nonumber& c_{12}=\frac{1}{8}\int_0^\infty\dd\tilde{\rho}\tilde{\rho} \left[1-\cos\theta_0f(4-4f+f^3)\right],\\
		\nonumber& c_{13}=-\frac{1}{4}\int_0^\infty\dd\tilde{\rho}\tilde{\rho}^2 f(1-f)\sin\theta_0
	\end{align}
\end{subequations}

Collecting all energy terms together we write the harmonic and nonlinear parts of the total longitudinal energy densities in form of \eqref{eq:H-ser} with the following harmonic and the leading nonlinear terms
\begin{equation}\label{eq:Bloch-H-X}
	\begin{split}
		&\mathscr{H}^{(2)}=\frac{a_1}{2}|\vec{\mathcal{X}}'|^2+\frac{a_2}{2}\sigma[\vec{\mathcal{X}}'\times\vec{\mathcal{X}}'']_z+\frac{a_3}{2}|\vec{\mathcal{X}}''|^2+\dots,\\
		&\mathscr{H}^{(4)}=-\frac{b_1}{4}|\vec{\mathcal{X}}'|^4-\frac{b_2}{4}\sigma|\vec{\mathcal{X}}'|^2[\vec{\mathcal{X}}'\times\vec{\mathcal{\mathcal{X}}}'']_z+\dots,
	\end{split}
\end{equation}
where $\sigma=\sin\varphi_0=\pm1$. For the chosen boundary conditions for $\theta_0$, we have $\sigma=\text{sign}(D)$. Since $\tilde{\zeta}=\tilde{z}$ on the string, we replace $\tilde{\zeta}\to\tilde{z}$ in the integral \eqref{eq:H-ser}. In terms of $\psi$, Eqs.~\eqref{eq:Bloch-H-X} obtain form \eqref{eq:Bloch-H}. The coefficients are as follows
\begin{subequations}
	\begin{align}
		\label{eq:a-Bloch}	
		&a_1=c_0+\tilde{c}_0+2c_6+2hc_{10},\\ \nonumber &a_2=c_1+2c_7+2hc_{11},\qquad a_3=c_2,\\
		\label{eq:b-Bloch}	&b_1=\frac{c_0}{2}+\bar{c}_0+2c_8+2hc_{12},\\ \nonumber &b_2=c_3+2c_9+2hc_{13}.
	\end{align}
\end{subequations}

\begin{figure}
	\includegraphics[width=\columnwidth]{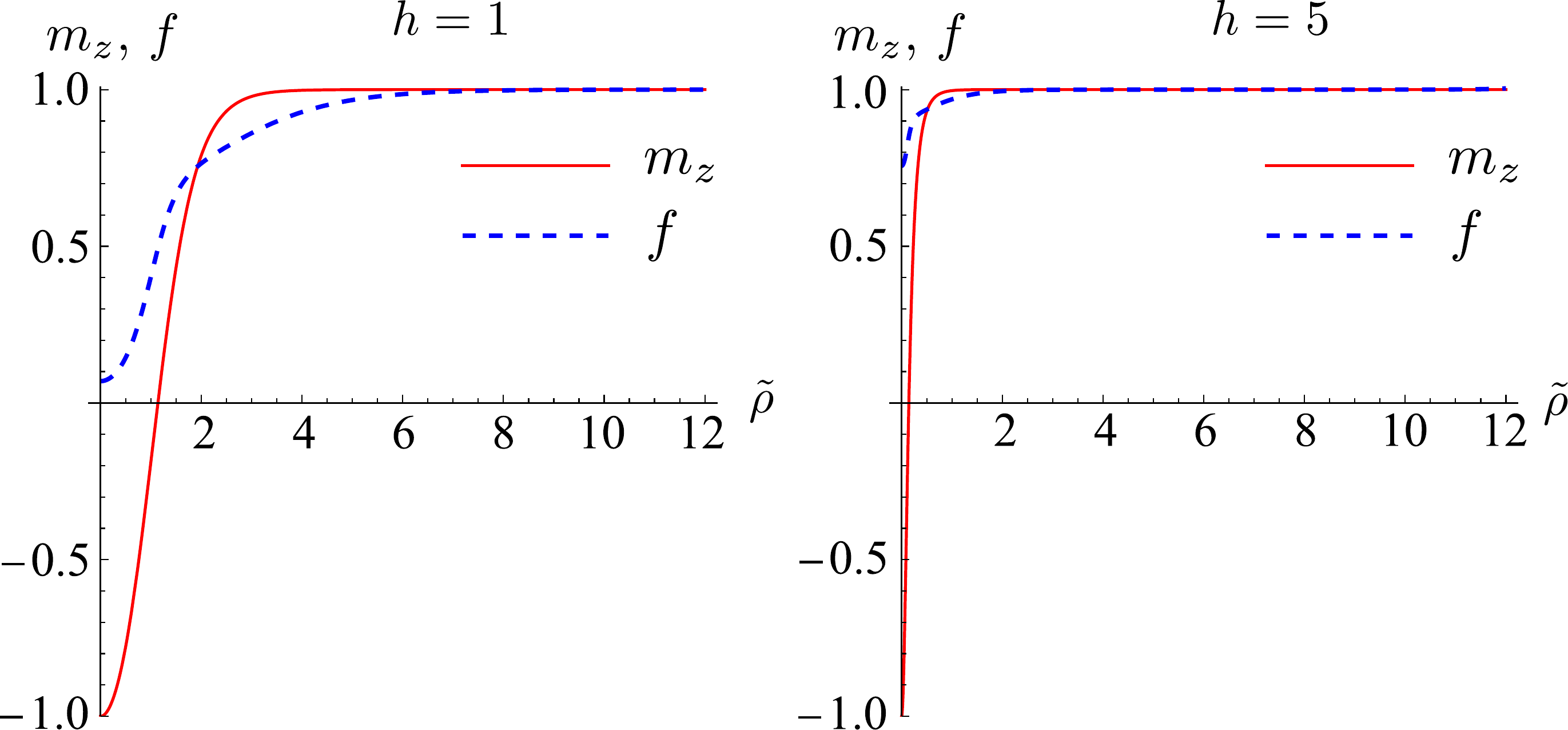}
	\caption{Functions $f$ obtained as solutions of Eq.~\eqref{eq:f} for different values of magnetic field $h$ are compared to the corresponding skyrmion profiles $m_z$.}\label{fig:f}
\end{figure}

Function $f(\tilde{\rho})$ which determines the coefficients $a_n$ and $b_n$ is unknown. However, it can be roughly estimated in the limit of the long-wave approximation, where the term $\frac12a_1|\vec{X}'|^2$ gives the main contribution to the energy. This enables us to estimate function $f(\tilde{\rho})$ as a minimizer of the coefficient $a_1$ which is a functional of $f(\tilde{\rho})$. The equation $\delta a_1/\delta f=0$ has the following explicit form
\begin{equation}\label{eq:f}
	\begin{split}
	&f''+\left[\frac{1}{\tilde\rho}-\theta_0'\frac{\sin2\theta_0}{1+\cos^2\theta_0}\right]f'+\frac{2b\cos\theta_0}{1+\cos^2\theta_0}(1-f)\\
	&+2f\frac{\theta_0'\cos^2\theta_0+\frac{1}{\tilde\rho}\sin\theta_0\cos\theta_0}{1+\cos^2\theta_0}=0
	\end{split}
\end{equation}
Equation \eqref{eq:f} is supplemented with the boundary conditions $f'(0)=0$ and $f(\infty)=1$. Some solutions for different skyrmion profiles are shown in Fig.~\ref{fig:f}.
An important consequence of the obtained solution is $f(0)\ne0$. This means that the magnetization in the string center is not tangential to the string. In the limit $h\to\infty$, we obtain $f(0)\to1$. This means that the magnetization of the string center is anti-parallel to the applied field for the infinitely thin string.

In the large field limit, we have $f\approx1$ and from the form of the coefficients $c$ we conclude that the contributions of all energies vanish except the exchange energy. In this case $c_0\to1$ and therefore $a_1\to1$ and $b_1\to1/2$.

\begin{figure}
	\includegraphics[width=\columnwidth]{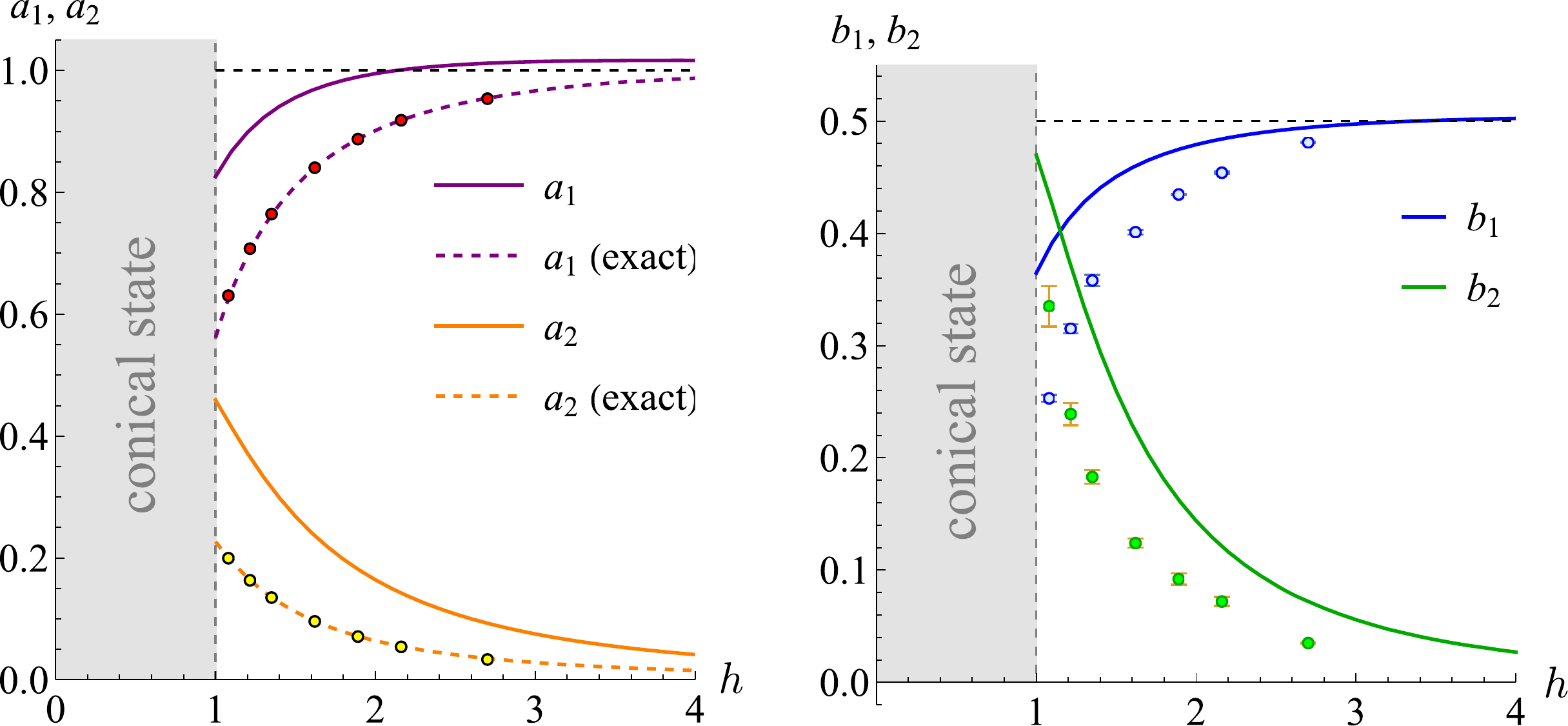}
	\caption{Field dependence of the first two coefficients of the linear ($a_1$, $a_2$) and nonlinear ($b_1$, $b_2$) parts of the Hamiltonian of the Bloch skyrmion string. Solid lines corresponds to expressions \eqref{eq:a-Bloch} and \eqref{eq:b-Bloch} with the use of function $f$ estimated as a solution of Eq.~\eqref{eq:f}. Dashed lines correspond to the exact values of $a_1$ and $a_2$ previously obtained from the linear dispersion of the translational magnon mode \cite{Kravchuk20}. Dots show the corresponding coefficients determined numerically from micromagnetic simulations, for detail see Appendix~\ref{app:simuls}. }\label{fig:ab_vs_b}
\end{figure}

Having function $f$ we estimate the coefficients $a_i$ and $b_i$, see Fig.~\ref{fig:ab_vs_b}.
The comparison with the corresponding coefficients obtained by means of micromagnetic simulations shows that the considered model and the estimation of function $f$ by means of Eq.~\eqref{eq:f} result in a qualitative agreement with the real dependencies $a_n(b)$ and $b_n(b)$. The best agreement is achieved for the case of large fields which correspond to thin strings. The quantitative analysis should be based on the values of $a_n$ and $b_n$ determined by means of micromagnetic simulation, see Appendix~\ref{app:simuls}.

\section{Determination of the coefficients $a_n$ and $b_n$ by means of micromagnetic simulations.}\label{app:simuls}
In order to determine coefficients $a_n$ and $b_n$, we simulated a damping-free dynamics of helical waves for different values of the helix radius $\mathcal{R}$ and wave vector $k_0$. For each simulation we determine frequency of the helix rotation $\omega=\omega(k_0,\mathcal{R})$. Based on the knowledge of the form of nonlinear dispersion \eqref{eq:helix-disp}, we extract coefficients $a_n$ and $b_n$ as explained below. 

Simulation of dynamics of magnetization media is based on numerical solution of Landau-Lifshits equation, it is performed with the help of mumax$^3$ code \cite{Vansteenkiste14}. We consider Hamiltonian \eqref{eq:H} with the material parameters of FeGe, namely $A=8.78$~pJ/m, $D=1.58$~mJ/m$^2$, $M_s=0.384$~MA/m. The scales of the length and time are determined by the wave-vector $Q=D/(2A)\approx2\pi/(70\,\text{nm})$ and frequency $\omega_{c2}=\gamma_0D^2/(2AM_s)\approx10.4~\text{GHz}$, respectively. 
It is important to note that the coefficients $a_n$ and $b_n$ are defined for the dimensionless dispersion \eqref{eq:helix-disp}, thus they depend on a single parameter $h$ only. This means that the extracted from simulations dependencies $a_n(h)$ and $b_n(h)$ are universal, they are valid for any cubic chiral magnet, not only FeGe.

\begin{figure*}
	\includegraphics[width=\textwidth]{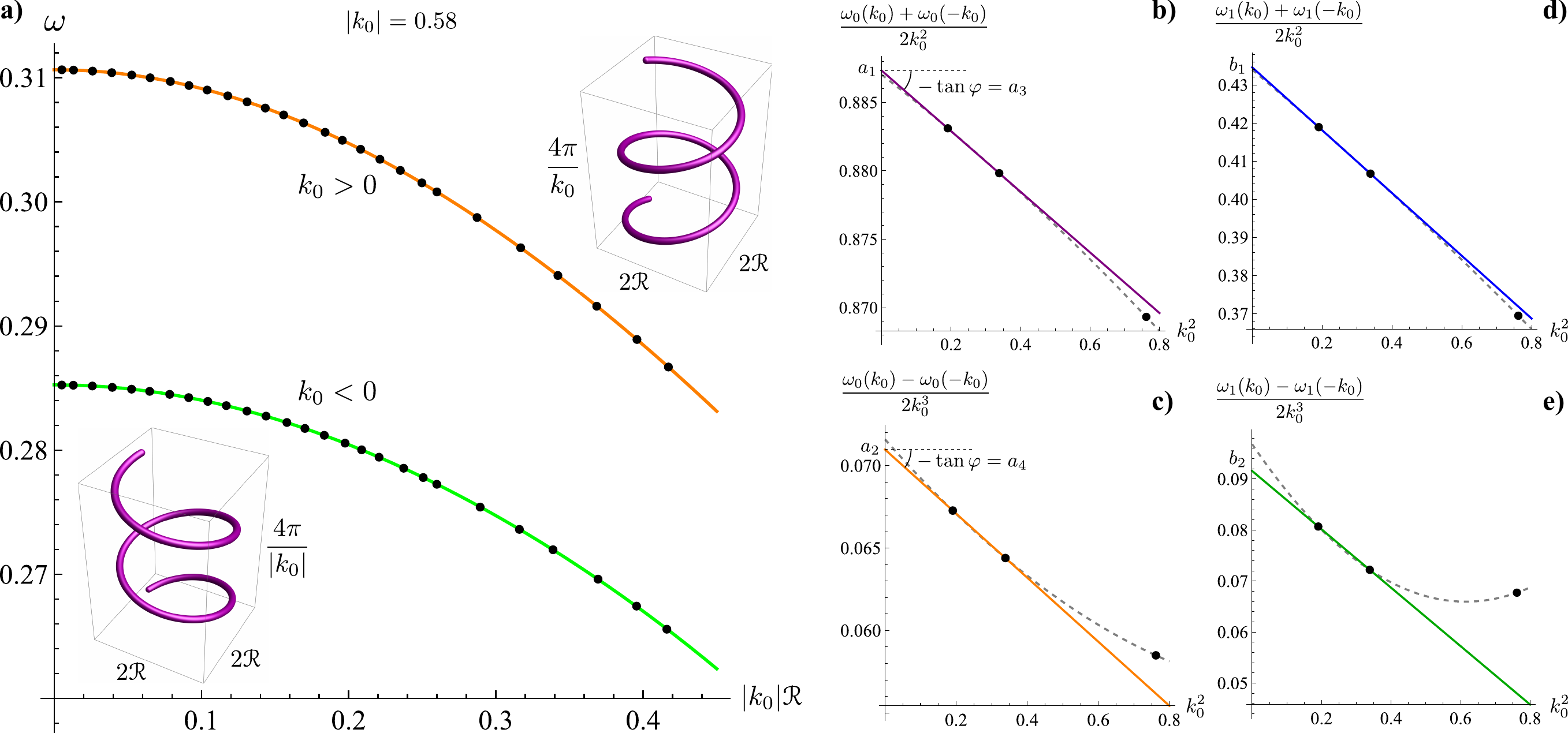}
	\caption{Determination of the coefficients $a_n$ and $b_n$ for the case $h=1.89$ ($0.7$~T). Panel a) shows the amplitude dependence of the helical wave frequency obtained by means of micromagnetic simulations for the wave-vector $k_0=0.58$ ($\frac{2\pi}{120\,\text{nm}}$). Solid lines correspond to the fitting $\omega=\omega_0-\omega_1\mathcal{R}^2k_0^2+\omega_2\mathcal{R}^4k_0^4+\omega_3\mathcal{R}^6k_0^6$. Next, the fitting coefficients $\omega_n$ are determined for several additional values of $k_0$, namely $k_0=0.44$ ($\frac{2\pi}{160\,\text{nm}}$), and $k_0=0.87$ ($\frac{2\pi}{80\,\text{nm}}$). The linear extrapolation of the dependence $[\omega_0(k_0)+\omega_0(-k_0)]/(2k_0^2)$ on $k_0^2$ to the region of vanishing $k_0$ enables one to determine coefficients $a_1$ and $a_3$, see panel b) and the corresponding explanations in the text. Similarly, we determine coefficients $a_2$ and $a_4$ using the dependence $[\omega_0(k_0)-\omega_0(-k_0)]/(2k_0^3)$ on $k_0^2$, see panel c). The parabolic extrapolation based on three points (dashed line) is used for the error bars estimation. Coefficients $b_n$ are determined analogously after the replacement $\omega_0\to\omega_1$, see the insets d),e). }\label{fig:w-vs-kR}
\end{figure*}

For a given value of $k_0$ we programmatically prepare an initial state close to a helix solution \eqref{eq:h-wave} with radius approximately 10~nm. The sample sizes are $L_x\times L_y \times L_z$ with $L_x=L_y=100-140$~ nm \footnote{For smaller fields (thicker strings), we use samples of larger lateral size.} and $L_z=2\pi/(|k_0|Q)$. The periodic boundary conditions are applied in all three directions. At the first step, we simulate the overdamped dynamics ($\alpha=0.5$) until the string reaches its equilibrium state in form of a vertical rectilinear line. During this overdamped dynamics we save several dozen of the magnetization snapshots. In this manner we obtain the helical waves of different radii in range 1 -- 8~nm. At the next step, we use these configurations as the initial states for the damping-free simulation of the helical wave dynamics. Using definition \eqref{eq:Xi} we extract the string central line $\vec{X}(\tilde{z},\tilde{t})$. For a fixed horizontal cross-section $\tilde{z}=\tilde{z}_0$, the linear time dependence of the phase $\arg(X_1(\tilde{z}_0,\tilde{t})+iX_2(\tilde{z}_0,\tilde{t}))=-\omega \tilde{t}$ enable us to extract the frequency $\omega$. As an example, in Fig.~\ref{fig:w-vs-kR}  we demonstrate the values of $\omega$, obtained for different helix radii and two different signs of $k_0$.

Performing the interpolation $\omega=\omega_0-\omega_1\mathcal{R}^2k_0^2+\omega_2\mathcal{R}^4k_0^4+\omega_3\mathcal{R}^6k_0^6$ we determine coefficients $\omega_n$. Repeating the described procedure for various $k_0$, we determine numerically the dependence $\omega_n(k_0)$. Since $\omega_0(k_0)=a_1k_0^2+a_2k_0^3+a_3k_0^4+\dots$, one can write $[\omega_0(k_0)+\omega_0(-k_0)]/(2k_0^2)=a_1+a_3k_0^2+\dots$. This enables us to determine coefficients $a_1$ and $a_3$ as it is explained in Fig.~\ref{fig:w-vs-kR}b). Similarly, we determine coefficients $a_2$ and $a_3$ using that $[\omega_0(k_0)-\omega_0(-k_0)]/(2k_0^3)=a_2+a_4k_0^2+\dots$, see Fig.~\ref{fig:w-vs-kR}c). Coefficients $b_n$ are determined analogously but the replacement $\omega_0(k_0)\to\omega_1(k_0)$, see  Fig.~\ref{fig:w-vs-kR}d,e). Coefficients $a_n$ and $b_n$ determined by this method for various fields are shown in Figs.~\ref{fig:an-bn-vs-b},~\ref{fig:ab_vs_b}. Note that the practically achievable accuracy of simulations does not allow us to determine the higher nonlinear coefficients $b_{n>2}$ with the acceptable precision.

\section{Method of multiple scales}\label{app:mult-scales}
Her we adapt to our case the derivation presented in Ref.~\onlinecite{Ryskin00}. We will look for solution of \eqref{eq:eq-motion} in form 
\begin{equation}\label{eq:mult-scales-ansatz}
	\psi=\sum_{i=1}^\infty u_i\varepsilon^i	
\end{equation}
where $\varepsilon$ is small parameter and $u_i=u_i(Z_0,Z_1,\dots,T_0,T_1,\dots)$ are functions of many space and time variables of different sales, namely $Z_i=\tilde{z}\varepsilon^i$ and $T_i=\tilde{t}\varepsilon^i$ with $i=0,1,\dots$. Thus $Z_0=\tilde{z}$ and $T_0=\tilde{t}$. 

Now we substitute \eqref{eq:mult-scales-ansatz} into \eqref{eq:eq-motion} and take into account the form of the derivatives $\partial_{\tilde{z}}=\varepsilon^i\partial_{Z_i}$ and $\partial_{\tilde{t}}=\varepsilon^i\partial_{T_i}$ where the summation over $i$ is assumed. Collecting only terms linear in $\varepsilon$ we obtain equation
\begin{equation}\label{eq:u1}
	\begin{split}
	&\hat{\mathbb{L}}u_1=0,\\ &\hat{\mathbb{L}}=i\partial_{T_0}+a_1\partial_{Z_0}^2-ia_2\partial_{Z_0}^3-a_3\partial_{Z_0}^4+\dots,
	\end{split}
\end{equation}
whose solution is $u_1=\tilde{\mathcal{A}}(Z_1,Z_2,\dots,T_1,T_2)e^{i\theta}$, where $\theta=k_0Z_0-\omega_0(k_0)T_0$. Here the complex-valued function $\tilde{\mathcal{A}}$ describes the slow-varying amplitude of the envelope wave, and $\omega_0$ is the linear part of the dispersion introduced in \eqref{eq:w0}. 

Collecting now terms proportional to $\varepsilon^2$ we obtain equation
\begin{equation}\label{eq:u2}
	\hat{\mathbb{L}}u_2=-i\left[\partial_{T_1}\mathcal{A}+v_g(k_0)\partial_{Z_1}\mathcal{A}\right]e^{i\theta},
\end{equation}
where $v_g=\partial_k\omega_0$ is group velocity of the linear wave. Inhomogeneous equation \eqref{eq:u2} has bounded solutions (without secular terms) for $u_2$ if the right-hand-side part of Eq.~\eqref{eq:u2} is orthogonal to solutions of the corresponding homogeneous equation. This means that $\partial_{T_1}\tilde{\mathcal{A}}+v_g(k_0)\partial_{Z_1}\tilde{\mathcal{A}}=0$. Thus, in the first approximation, the envelope $\tilde{\mathcal{A}}$ moves with the group velocity. In the other words $\tilde{\mathcal{A}}=\tilde{\mathcal{A}}(Z_1-v_g(k_0)T_1;Z_2,\dots,T_2,\dots)$. In this case, Eq.~\eqref{eq:u2} obtains the homogeneous form $\hat{\mathbb{L}}u_2=0$ which allows the trivial solution $u_2=0$.

Collecting now terms proportional to $\varepsilon^3$ we obtain equation 
\begin{equation}\label{eq:u3}
	\begin{split}
		\hat{\mathbb{L}}u_3=-\biggl\{&i\left[\partial_{T_2}\tilde{\mathcal{A}}+v_g(k_0)\partial_{Z_2}\tilde{\mathcal{A}}\right]+\frac{\mu(k_0)}{2}\partial^2_{Z_1}\tilde{\mathcal{A}}\\
		&+\nu(k_0)\tilde{\mathcal{A}}|\tilde{\mathcal{A}}|^2\biggr\}e^{i\theta},
	\end{split}
\end{equation}
where $\mu(k_0)$ and $\nu(k_0)$ are the same as in \eqref{eq:NLS-set}. The condition of absence of the secular terms in the solution for $u_3$ requires the vanishing curly brackets in the right-hand-side of Eq.~\eqref{eq:u3}. The latter results in NLSE \eqref{eq:NLS-set}, where we made the substitution $\tilde{\mathcal{A}}=\varepsilon\mathcal{A}$.

\section{General equation of the collective string dynamics}\label{app:gen}
It is convenient to proceed to the angular parameterization of the magnetization $\vec{m}=\sin\theta(\cos\phi\hat{\vec{x}}+\sin\phi\hat{\vec{y}})+\cos\theta\hat{\vec{z}}$. In this case the Lagrangian and the density of the dissipative function are 
\begin{subequations}\label{eq:L-R}
	\begin{align}
	&\mathcal{L}=\frac{M_s}{\gamma_0}\iint(1-\cos\theta)\dot\phi\,\dd x\dd y-\mathcal{H},\\ &\mathcal{F}=\frac{\alpha}{2}\frac{M_s}{\gamma_0}\iint\left(\dot{\theta}^2+\sin^2\theta\dot{\phi}^2\right)\dd x\dd y.
	\end{align}
\end{subequations}
Now we assume that $\theta=\theta_0(x,y,X_i(z,t),X_i'(z,t))$ and $\phi=\phi_0(x,y,X_i(z,t),X_i'(z,t))$ with $\theta_0$ and $\phi_0$ being the known functions. Taking into account that
\begin{equation}\label{eq:phi-derivs}
	\begin{split}
	&\dot{\phi}=\frac{\partial\phi_0}{\partial X_i}\dot{X}_i+\frac{\partial\phi_0}{\partial X_i'}\dot{X}_i',\quad \phi'=\frac{\partial\phi_0}{\partial X_i}X'_i+\frac{\partial\phi_0}{\partial X_i'}X_i'',\\ 
	&\dot{\theta}=\frac{\partial\theta_0}{\partial X_i}\dot{X}_i+\frac{\partial\theta_0}{\partial X_i'}\dot{X}_i',\quad \theta'=\frac{\partial\theta_0}{\partial X_i}X'_i+\frac{\partial\theta_0}{\partial X_i'}X_i'',
	\end{split}
\end{equation}
we write the general equations of motion $\delta \mathfrak{S}/\delta X_i=\delta \mathfrak{F}/\delta\dot{X}_i$ in form \eqref{eq:main-equation}, where
\begin{equation}\label{eq:G}
	\begin{split}
		&G_{ij}^{(0)}=\frac{M_s}{\gamma_0}\!\iint\!\sin\theta_0\!\left[\frac{\partial\theta_0}{\partial X_i}\frac{\partial\phi_0}{\partial X_j}-\frac{\partial\theta_0}{\partial X_j}\frac{\partial\phi_0}{\partial X_i}\right]\!\dd x \dd y,\\		&G_{ij}^{(1)}=\frac{M_s}{\gamma_0}\!\iint\!\sin\theta_0\!\left[\frac{\partial\theta_0}{\partial X_i}\frac{\partial\phi_0}{\partial X_j'}-\frac{\partial\theta_0}{\partial X_j'}\frac{\partial\phi_0}{\partial X_i}\right]\!\dd x \dd y,\\
		&G_{ij}^{(2)}=\frac{M_s}{\gamma_0}\!\iint\!\sin\theta_0\!\left[\frac{\partial\theta_0}{\partial X_i'}\frac{\partial\phi_0}{\partial X_j'}-\frac{\partial\theta_0}{\partial X_j'}\frac{\partial\phi_0}{\partial X_i'}\right]\!\dd x \dd y
		\end{split}
	\end{equation}
	are the gyrotensors, and 
	\begin{equation}\label{eq:D}
	\begin{split}
		 &D_{ij}^{(0)}\!=\!\frac{M_s}{\gamma_0}\!\iint\!\left[\frac{\partial\theta_0}{\partial X_i}\frac{\partial\theta_0}{\partial X_j}+\sin^2\theta_0\frac{\partial\phi_0}{\partial X_i}\frac{\partial\phi_0}{\partial X_j}\right]\!\!\dd x \dd y,\\
		&D_{ij}^{(1)}\!=\!\frac{M_s}{\gamma_0}\!\iint\!\left[\frac{\partial\theta_0}{\partial X_i}\frac{\partial\theta_0}{\partial X_j'}+\sin^2\theta_0\frac{\partial\phi_0}{\partial X_i}\frac{\partial\phi_0}{\partial X_j'}\right]\!\!\dd x \dd y,\\ &D_{ij}^{(2)}\!=\!\frac{M_s}{\gamma_0}\!\iint\!\left[\frac{\partial\theta_0}{\partial X_i'}\frac{\partial\theta_0}{\partial X_j'}+\sin^2\theta_0\frac{\partial\phi_0}{\partial X_i'}\frac{\partial\phi_0}{\partial X_j'}\right]\!\!\dd x \dd y
	\end{split}
\end{equation}
are the damping tensors. Note that due to the presence of terms with the mixed derivatives $\dot{X}_i'$ in \eqref{eq:phi-derivs}, it is required to work with the action, not with the Lagrange function. 


\section{Description of supplemental movies}
Movies~1 and 2 are the dynamical realization of Fig.~\ref{fig:helix} for the cases $k>0$ and $k<0$, respectively. All parameters and notations are the same as indicated in the caption of Fig.~\ref{fig:helix}. 

Movie 3 demonstrates dynamics of dn-cnoidal wave shown in Fig.~\ref{fig:waves-simuls}(a) in the time interval $t\in[10T-15T]=[5.06-7.6]$~ns. 

Movie 4 demonstrates dynamics of cn-cnoidal wave shown in Fig.~\ref{fig:waves-simuls}(b) in the time interval $t\in[10T-15T]=[3.78-5.7]$~ns. The parameters and notations are the same as in Fig.~\ref{fig:waves-simuls}.

Movie~5 shows the propagation of soliton Fig.~\ref{fig:soliton} along the string. The parameters and notations are the same as in Fig.~\ref{fig:soliton}.

Movie~6 shows the propagation of Ma-breather shown in Fig.~\ref{fig:breather} along the string. The parameters and notations are the same as in Fig.~\ref{fig:breather}.


\begin{thebibliography}{75}%
	\makeatletter
	\providecommand \@ifxundefined [1]{%
		\@ifx{#1\undefined}
	}%
	\providecommand \@ifnum [1]{%
		\ifnum #1\expandafter \@firstoftwo
		\else \expandafter \@secondoftwo
		\fi
	}%
	\providecommand \@ifx [1]{%
		\ifx #1\expandafter \@firstoftwo
		\else \expandafter \@secondoftwo
		\fi
	}%
	\providecommand \natexlab [1]{#1}%
	\providecommand \enquote  [1]{``#1''}%
	\providecommand \bibnamefont  [1]{#1}%
	\providecommand \bibfnamefont [1]{#1}%
	\providecommand \citenamefont [1]{#1}%
	\providecommand \href@noop [0]{\@secondoftwo}%
	\providecommand \href [0]{\begingroup \@sanitize@url \@href}%
	\providecommand \@href[1]{\@@startlink{#1}\@@href}%
	\providecommand \@@href[1]{\endgroup#1\@@endlink}%
	\providecommand \@sanitize@url [0]{\catcode `\\12\catcode `\$12\catcode
		`\&12\catcode `\#12\catcode `\^12\catcode `\_12\catcode `\%12\relax}%
	\providecommand \@@startlink[1]{}%
	\providecommand \@@endlink[0]{}%
	\providecommand \url  [0]{\begingroup\@sanitize@url \@url }%
	\providecommand \@url [1]{\endgroup\@href {#1}{\urlprefix }}%
	\providecommand \urlprefix  [0]{URL }%
	\providecommand \Eprint [0]{\href }%
	\providecommand \doibase [0]{https://doi.org/}%
	\providecommand \selectlanguage [0]{\@gobble}%
	\providecommand \bibinfo  [0]{\@secondoftwo}%
	\providecommand \bibfield  [0]{\@secondoftwo}%
	\providecommand \translation [1]{[#1]}%
	\providecommand \BibitemOpen [0]{}%
	\providecommand \bibitemStop [0]{}%
	\providecommand \bibitemNoStop [0]{.\EOS\space}%
	\providecommand \EOS [0]{\spacefactor3000\relax}%
	\providecommand \BibitemShut  [1]{\csname bibitem#1\endcsname}%
	\let\auto@bib@innerbib\@empty
	\bibitem [{\citenamefont {Seki}\ and\ \citenamefont
		{Mochizuki}(2016)}]{Seki16}%
	\BibitemOpen
	\bibfield  {author} {\bibinfo {author} {\bibfnamefont {S.}~\bibnamefont
			{Seki}}\ and\ \bibinfo {author} {\bibfnamefont {M.}~\bibnamefont
			{Mochizuki}},\ }\bibfield  {title} {\bibinfo {title} {Skyrmions in magnetic
			materials},\ }\bibfield  {journal} {\bibinfo  {journal} {SpringerBriefs in
			Physics}\ }\href {https://doi.org/10.1007/978-3-319-24651-2}
	{10.1007/978-3-319-24651-2} (\bibinfo {year} {2016})\BibitemShut {NoStop}%
	\bibitem [{\citenamefont {Liu}\ \emph {et~al.}(2020)\citenamefont {Liu},
		\citenamefont {Zhang},\ and\ \citenamefont {Zhao}}]{Liu20c}%
	\BibitemOpen
	\bibfield  {author} {\bibinfo {author} {\bibfnamefont {J.~P.}\ \bibnamefont
			{Liu}}, \bibinfo {author} {\bibfnamefont {Z.}~\bibnamefont {Zhang}},\ and\
		\bibinfo {author} {\bibfnamefont {G.}~\bibnamefont {Zhao}},\ }\href@noop {}
	{\emph {\bibinfo {title} {Skyrmions. Topological Structures, Properties, and
				Applications.}}}\ (\bibinfo  {publisher} {CRC Press},\ \bibinfo {year}
	{2020})\BibitemShut {NoStop}%
	\bibitem [{\citenamefont {Back}\ \emph {et~al.}(2020)\citenamefont {Back},
		\citenamefont {Cros}, \citenamefont {Ebert}, \citenamefont {Everschor-Sitte},
		\citenamefont {Fert}, \citenamefont {Garst}, \citenamefont {Ma},
		\citenamefont {Mankovsky}, \citenamefont {Monchesky}, \citenamefont
		{Mostovoy}, \citenamefont {Nagaosa}, \citenamefont {Parkin}, \citenamefont
		{Pfleiderer}, \citenamefont {Reyren}, \citenamefont {Rosch}, \citenamefont
		{Taguchi}, \citenamefont {Tokura}, \citenamefont {von Bergmann},\ and\
		\citenamefont {Zang}}]{Back20}%
	\BibitemOpen
	\bibfield  {author} {\bibinfo {author} {\bibfnamefont {C.}~\bibnamefont
			{Back}}, \bibinfo {author} {\bibfnamefont {V.}~\bibnamefont {Cros}}, \bibinfo
		{author} {\bibfnamefont {H.}~\bibnamefont {Ebert}}, \bibinfo {author}
		{\bibfnamefont {K.}~\bibnamefont {Everschor-Sitte}}, \bibinfo {author}
		{\bibfnamefont {A.}~\bibnamefont {Fert}}, \bibinfo {author} {\bibfnamefont
			{M.}~\bibnamefont {Garst}}, \bibinfo {author} {\bibfnamefont
			{T.}~\bibnamefont {Ma}}, \bibinfo {author} {\bibfnamefont {S.}~\bibnamefont
			{Mankovsky}}, \bibinfo {author} {\bibfnamefont {T.~L.}\ \bibnamefont
			{Monchesky}}, \bibinfo {author} {\bibfnamefont {M.}~\bibnamefont {Mostovoy}},
		\bibinfo {author} {\bibfnamefont {N.}~\bibnamefont {Nagaosa}}, \bibinfo
		{author} {\bibfnamefont {S.~S.~P.}\ \bibnamefont {Parkin}}, \bibinfo {author}
		{\bibfnamefont {C.}~\bibnamefont {Pfleiderer}}, \bibinfo {author}
		{\bibfnamefont {N.}~\bibnamefont {Reyren}}, \bibinfo {author} {\bibfnamefont
			{A.}~\bibnamefont {Rosch}}, \bibinfo {author} {\bibfnamefont
			{Y.}~\bibnamefont {Taguchi}}, \bibinfo {author} {\bibfnamefont
			{Y.}~\bibnamefont {Tokura}}, \bibinfo {author} {\bibfnamefont
			{K.}~\bibnamefont {von Bergmann}},\ and\ \bibinfo {author} {\bibfnamefont
			{J.}~\bibnamefont {Zang}},\ }\bibfield  {title} {\bibinfo {title} {The 2020
			skyrmionics roadmap},\ }\href {https://doi.org/10.1088/1361-6463/ab8418}
	{\bibfield  {journal} {\bibinfo  {journal} {Journal of Physics D: Applied
				Physics}\ }\textbf {\bibinfo {volume} {53}},\ \bibinfo {pages} {363001}
		(\bibinfo {year} {2020})}\BibitemShut {NoStop}%
	\bibitem [{\citenamefont {Nagaosa}\ and\ \citenamefont
		{Tokura}(2013)}]{Nagaosa13}%
	\BibitemOpen
	\bibfield  {author} {\bibinfo {author} {\bibfnamefont {N.}~\bibnamefont
			{Nagaosa}}\ and\ \bibinfo {author} {\bibfnamefont {Y.}~\bibnamefont
			{Tokura}},\ }\bibfield  {title} {\bibinfo {title} {Topological properties and
			dynamics of magnetic skyrmions},\ }\href
	{https://doi.org/10.1038/nnano.2013.243} {\bibfield  {journal} {\bibinfo
			{journal} {Nature Nanotechnology}\ }\textbf {\bibinfo {volume} {8}},\
		\bibinfo {pages} {899} (\bibinfo {year} {2013})}\BibitemShut {NoStop}%
	\bibitem [{\citenamefont {Fert}\ \emph {et~al.}(2017)\citenamefont {Fert},
		\citenamefont {Reyren},\ and\ \citenamefont {Cros}}]{Fert17}%
	\BibitemOpen
	\bibfield  {author} {\bibinfo {author} {\bibfnamefont {A.}~\bibnamefont
			{Fert}}, \bibinfo {author} {\bibfnamefont {N.}~\bibnamefont {Reyren}},\ and\
		\bibinfo {author} {\bibfnamefont {V.}~\bibnamefont {Cros}},\ }\bibfield
	{title} {\bibinfo {title} {Magnetic skyrmions: advances in physics and
			potential applications},\ }\href {https://doi.org/10.1038/natrevmats.2017.31}
	{\bibfield  {journal} {\bibinfo  {journal} {Nature Reviews Materials}\
		}\textbf {\bibinfo {volume} {2}},\ \bibinfo {pages} {17031} (\bibinfo {year}
		{2017})}\BibitemShut {NoStop}%
	\bibitem [{\citenamefont {Wiesendanger}(2016)}]{Wiesendanger16}%
	\BibitemOpen
	\bibfield  {author} {\bibinfo {author} {\bibfnamefont {R.}~\bibnamefont
			{Wiesendanger}},\ }\bibfield  {title} {\bibinfo {title} {Nanoscale magnetic
			skyrmions in metallic films and multilayers: a new twist for spintronics},\
	}\href {https://doi.org/10.1038/natrevmats.2016.44} {\bibfield  {journal}
		{\bibinfo  {journal} {Nature Reviews Materials}\ }\textbf {\bibinfo {volume}
			{1}},\ \bibinfo {pages} {16044} (\bibinfo {year} {2016})}\BibitemShut
	{NoStop}%
	\bibitem [{\citenamefont {Bogdanov}\ and\ \citenamefont
		{Panagopoulos}(2020)}]{Bogdanov20a}%
	\BibitemOpen
	\bibfield  {author} {\bibinfo {author} {\bibfnamefont {A.~N.}\ \bibnamefont
			{Bogdanov}}\ and\ \bibinfo {author} {\bibfnamefont {C.}~\bibnamefont
			{Panagopoulos}},\ }\bibfield  {title} {\bibinfo {title} {Physical foundations
			and basic properties of magnetic skyrmions},\ }\href
	{https://doi.org/10.1038/s42254-020-0203-7} {\bibfield  {journal} {\bibinfo
			{journal} {Nature Reviews Physics}\ }\textbf {\bibinfo {volume} {2}},\
		\bibinfo {pages} {492} (\bibinfo {year} {2020})}\BibitemShut {NoStop}%
	\bibitem [{\citenamefont {Sampaio}\ \emph {et~al.}(2013)\citenamefont
		{Sampaio}, \citenamefont {Cros}, \citenamefont {Rohart}, \citenamefont
		{Thiaville},\ and\ \citenamefont {Fert}}]{Sampaio13}%
	\BibitemOpen
	\bibfield  {author} {\bibinfo {author} {\bibfnamefont {J.}~\bibnamefont
			{Sampaio}}, \bibinfo {author} {\bibfnamefont {V.}~\bibnamefont {Cros}},
		\bibinfo {author} {\bibfnamefont {S.}~\bibnamefont {Rohart}}, \bibinfo
		{author} {\bibfnamefont {A.}~\bibnamefont {Thiaville}},\ and\ \bibinfo
		{author} {\bibfnamefont {A.}~\bibnamefont {Fert}},\ }\bibfield  {title}
	{\bibinfo {title} {Nucleation, stability and current-induced motion of
			isolated magnetic skyrmions in nanostructures},\ }\href
	{https://doi.org/10.1038/nnano.2013.210} {\bibfield  {journal} {\bibinfo
			{journal} {Nature Nanotechnology}\ }\textbf {\bibinfo {volume} {8}},\
		\bibinfo {pages} {839} (\bibinfo {year} {2013})}\BibitemShut {NoStop}%
	\bibitem [{\citenamefont {Zhang}\ \emph {et~al.}(2015)\citenamefont {Zhang},
		\citenamefont {Ezawa},\ and\ \citenamefont {Zhou}}]{Zhang15c}%
	\BibitemOpen
	\bibfield  {author} {\bibinfo {author} {\bibfnamefont {X.}~\bibnamefont
			{Zhang}}, \bibinfo {author} {\bibfnamefont {M.}~\bibnamefont {Ezawa}},\ and\
		\bibinfo {author} {\bibfnamefont {Y.}~\bibnamefont {Zhou}},\ }\bibfield
	{title} {\bibinfo {title} {Magnetic skyrmion logic gates: conversion,
			duplication and merging of skyrmions},\ }\href
	{https://doi.org/10.1038/srep09400} {\bibfield  {journal} {\bibinfo
			{journal} {Scientific Reports}\ }\textbf {\bibinfo {volume} {5}},\ \bibinfo
		{pages} {9400} (\bibinfo {year} {2015})}\BibitemShut {NoStop}%
	\bibitem [{\citenamefont {Fert}\ \emph {et~al.}(2013)\citenamefont {Fert},
		\citenamefont {Cros},\ and\ \citenamefont {Sampaio}}]{Fert13}%
	\BibitemOpen
	\bibfield  {author} {\bibinfo {author} {\bibfnamefont {A.}~\bibnamefont
			{Fert}}, \bibinfo {author} {\bibfnamefont {V.}~\bibnamefont {Cros}},\ and\
		\bibinfo {author} {\bibfnamefont {J.}~\bibnamefont {Sampaio}},\ }\bibfield
	{title} {\bibinfo {title} {Skyrmions on the track},\ }\href
	{https://doi.org/10.1038/nnano.2013.29} {\bibfield  {journal} {\bibinfo
			{journal} {Nature Nanotechnology}\ }\textbf {\bibinfo {volume} {8}},\
		\bibinfo {pages} {152} (\bibinfo {year} {2013})}\BibitemShut {NoStop}%
	\bibitem [{\citenamefont {M{\"u}hlbauer}\ \emph {et~al.}(2009)\citenamefont
		{M{\"u}hlbauer}, \citenamefont {Binz}, \citenamefont {Jonietz}, \citenamefont
		{Pfleiderer}, \citenamefont {Rosch}, \citenamefont {Neubauer}, \citenamefont
		{Georgii},\ and\ \citenamefont {B{\"o}ni}}]{Muehlbauer09}%
	\BibitemOpen
	\bibfield  {author} {\bibinfo {author} {\bibfnamefont {S.}~\bibnamefont
			{M{\"u}hlbauer}}, \bibinfo {author} {\bibfnamefont {B.}~\bibnamefont {Binz}},
		\bibinfo {author} {\bibfnamefont {F.}~\bibnamefont {Jonietz}}, \bibinfo
		{author} {\bibfnamefont {C.}~\bibnamefont {Pfleiderer}}, \bibinfo {author}
		{\bibfnamefont {A.}~\bibnamefont {Rosch}}, \bibinfo {author} {\bibfnamefont
			{A.}~\bibnamefont {Neubauer}}, \bibinfo {author} {\bibfnamefont
			{R.}~\bibnamefont {Georgii}},\ and\ \bibinfo {author} {\bibfnamefont
			{P.}~\bibnamefont {B{\"o}ni}},\ }\bibfield  {title} {\bibinfo {title}
		{Skyrmion lattice in a chiral magnet},\ }\href
	{https://doi.org/10.1126/science.1166767} {\bibfield  {journal} {\bibinfo
			{journal} {Science}\ }\textbf {\bibinfo {volume} {323}},\ \bibinfo {pages}
		{915} (\bibinfo {year} {2009})}\BibitemShut {NoStop}%
	\bibitem [{\citenamefont {Seki}\ \emph {et~al.}(2021)\citenamefont {Seki},
		\citenamefont {Suzuki}, \citenamefont {Ishibashi}, \citenamefont {Takagi},
		\citenamefont {Khanh}, \citenamefont {Shiota}, \citenamefont {Shibata},
		\citenamefont {Koshibae}, \citenamefont {Tokura},\ and\ \citenamefont
		{Ono}}]{Seki21}%
	\BibitemOpen
	\bibfield  {author} {\bibinfo {author} {\bibfnamefont {S.}~\bibnamefont
			{Seki}}, \bibinfo {author} {\bibfnamefont {M.}~\bibnamefont {Suzuki}},
		\bibinfo {author} {\bibfnamefont {M.}~\bibnamefont {Ishibashi}}, \bibinfo
		{author} {\bibfnamefont {R.}~\bibnamefont {Takagi}}, \bibinfo {author}
		{\bibfnamefont {N.~D.}\ \bibnamefont {Khanh}}, \bibinfo {author}
		{\bibfnamefont {Y.}~\bibnamefont {Shiota}}, \bibinfo {author} {\bibfnamefont
			{K.}~\bibnamefont {Shibata}}, \bibinfo {author} {\bibfnamefont
			{W.}~\bibnamefont {Koshibae}}, \bibinfo {author} {\bibfnamefont
			{Y.}~\bibnamefont {Tokura}},\ and\ \bibinfo {author} {\bibfnamefont
			{T.}~\bibnamefont {Ono}},\ }\bibfield  {title} {\bibinfo {title} {Direct
			visualization of the three-dimensional shape of skyrmion strings in a
			noncentrosymmetric magnet},\ }\bibfield  {journal} {\bibinfo  {journal}
		{Nature Materials}\ }\href {https://doi.org/10.1038/s41563-021-01141-w}
	{10.1038/s41563-021-01141-w} (\bibinfo {year} {2021})\BibitemShut {NoStop}%
	\bibitem [{\citenamefont {Birch}\ \emph {et~al.}(2020)\citenamefont {Birch},
		\citenamefont {Cort{\'{e}}s-Ortu{\~{n}}o}, \citenamefont {Turnbull},
		\citenamefont {Wilson}, \citenamefont {Gro{\ss}}, \citenamefont {Träger},
		\citenamefont {Laurenson}, \citenamefont {Bukin}, \citenamefont {Moody},
		\citenamefont {Weigand}, \citenamefont {Schütz}, \citenamefont {Popescu},
		\citenamefont {Fan}, \citenamefont {Steadman}, \citenamefont {Verezhak},
		\citenamefont {Balakrishnan}, \citenamefont {Loudon}, \citenamefont
		{Twitchett-Harrison}, \citenamefont {Hovorka}, \citenamefont {Fangohr},
		\citenamefont {Ogrin}, \citenamefont {Gräfe},\ and\ \citenamefont
		{Hatton}}]{Birch20}%
	\BibitemOpen
	\bibfield  {author} {\bibinfo {author} {\bibfnamefont {M.~T.}\ \bibnamefont
			{Birch}}, \bibinfo {author} {\bibfnamefont {D.}~\bibnamefont
			{Cort{\'{e}}s-Ortu{\~{n}}o}}, \bibinfo {author} {\bibfnamefont {L.~A.}\
			\bibnamefont {Turnbull}}, \bibinfo {author} {\bibfnamefont {M.~N.}\
			\bibnamefont {Wilson}}, \bibinfo {author} {\bibfnamefont {F.}~\bibnamefont
			{Gro{\ss}}}, \bibinfo {author} {\bibfnamefont {N.}~\bibnamefont {Träger}},
		\bibinfo {author} {\bibfnamefont {A.}~\bibnamefont {Laurenson}}, \bibinfo
		{author} {\bibfnamefont {N.}~\bibnamefont {Bukin}}, \bibinfo {author}
		{\bibfnamefont {S.~H.}\ \bibnamefont {Moody}}, \bibinfo {author}
		{\bibfnamefont {M.}~\bibnamefont {Weigand}}, \bibinfo {author} {\bibfnamefont
			{G.}~\bibnamefont {Schütz}}, \bibinfo {author} {\bibfnamefont
			{H.}~\bibnamefont {Popescu}}, \bibinfo {author} {\bibfnamefont
			{R.}~\bibnamefont {Fan}}, \bibinfo {author} {\bibfnamefont {P.}~\bibnamefont
			{Steadman}}, \bibinfo {author} {\bibfnamefont {J.~A.~T.}\ \bibnamefont
			{Verezhak}}, \bibinfo {author} {\bibfnamefont {G.}~\bibnamefont
			{Balakrishnan}}, \bibinfo {author} {\bibfnamefont {J.~C.}\ \bibnamefont
			{Loudon}}, \bibinfo {author} {\bibfnamefont {A.~C.}\ \bibnamefont
			{Twitchett-Harrison}}, \bibinfo {author} {\bibfnamefont {O.}~\bibnamefont
			{Hovorka}}, \bibinfo {author} {\bibfnamefont {H.}~\bibnamefont {Fangohr}},
		\bibinfo {author} {\bibfnamefont {F.~Y.}\ \bibnamefont {Ogrin}}, \bibinfo
		{author} {\bibfnamefont {J.}~\bibnamefont {Gräfe}},\ and\ \bibinfo {author}
		{\bibfnamefont {P.~D.}\ \bibnamefont {Hatton}},\ }\bibfield  {title}
	{\bibinfo {title} {Real-space imaging of confined magnetic skyrmion tubes},\
	}\bibfield  {journal} {\bibinfo  {journal} {Nature Communications}\ }\textbf
	{\bibinfo {volume} {11}},\ \href {https://doi.org/10.1038/s41467-020-15474-8}
	{10.1038/s41467-020-15474-8} (\bibinfo {year} {2020})\BibitemShut {NoStop}%
	\bibitem [{\citenamefont {Wolf}\ \emph {et~al.}(2021)\citenamefont {Wolf},
		\citenamefont {Schneider}, \citenamefont {Rö{\ss}ler}, \citenamefont
		{Kov{\'{a}}cs}, \citenamefont {Schmidt}, \citenamefont {Dunin-Borkowski},
		\citenamefont {Büchner}, \citenamefont {Rellinghaus},\ and\ \citenamefont
		{Lubk}}]{Wolf21}%
	\BibitemOpen
	\bibfield  {author} {\bibinfo {author} {\bibfnamefont {D.}~\bibnamefont
			{Wolf}}, \bibinfo {author} {\bibfnamefont {S.}~\bibnamefont {Schneider}},
		\bibinfo {author} {\bibfnamefont {U.~K.}\ \bibnamefont {Rö{\ss}ler}},
		\bibinfo {author} {\bibfnamefont {A.}~\bibnamefont {Kov{\'{a}}cs}}, \bibinfo
		{author} {\bibfnamefont {M.}~\bibnamefont {Schmidt}}, \bibinfo {author}
		{\bibfnamefont {R.~E.}\ \bibnamefont {Dunin-Borkowski}}, \bibinfo {author}
		{\bibfnamefont {B.}~\bibnamefont {Büchner}}, \bibinfo {author}
		{\bibfnamefont {B.}~\bibnamefont {Rellinghaus}},\ and\ \bibinfo {author}
		{\bibfnamefont {A.}~\bibnamefont {Lubk}},\ }\bibfield  {title} {\bibinfo
		{title} {Unveiling the three-dimensional magnetic texture of skyrmion
			tubes},\ }\href {https://doi.org/10.1038/s41565-021-01031-x} {\bibfield
		{journal} {\bibinfo  {journal} {Nature Nanotechnology}\ }\textbf {\bibinfo
			{volume} {17}},\ \bibinfo {pages} {250} (\bibinfo {year} {2021})}\BibitemShut
	{NoStop}%
	\bibitem [{\citenamefont {Pismen}(1999)}]{Pismen99}%
	\BibitemOpen
	\bibfield  {author} {\bibinfo {author} {\bibfnamefont {L.~M.}\ \bibnamefont
			{Pismen}},\ }\href@noop {} {\emph {\bibinfo {title} {Vortices in Nonlinear
				Field}}},\ edited by\ \bibinfo {editor} {\bibfnamefont {J.}~\bibnamefont
		{Birman}}\ (\bibinfo  {publisher} {Oxford University Press},\ \bibinfo {year}
	{1999})\BibitemShut {NoStop}%
	\bibitem [{\citenamefont {Sonin}(1987)}]{Sonin87}%
	\BibitemOpen
	\bibfield  {author} {\bibinfo {author} {\bibfnamefont {E.~B.}\ \bibnamefont
			{Sonin}},\ }\bibfield  {title} {\bibinfo {title} {Vortex oscillations and
			hydrodynamics of rotating superfluids},\ }\href
	{https://doi.org/10.1103/revmodphys.59.87} {\bibfield  {journal} {\bibinfo
			{journal} {Reviews of Modern Physics}\ }\textbf {\bibinfo {volume} {59}},\
		\bibinfo {pages} {87} (\bibinfo {year} {1987})}\BibitemShut {NoStop}%
	\bibitem [{\citenamefont {Blatter}\ \emph {et~al.}(1994)\citenamefont
		{Blatter}, \citenamefont {Feigel'man}, \citenamefont {Geshkenbein},
		\citenamefont {Larkin},\ and\ \citenamefont {Vinokur}}]{Blatter94}%
	\BibitemOpen
	\bibfield  {author} {\bibinfo {author} {\bibfnamefont {G.}~\bibnamefont
			{Blatter}}, \bibinfo {author} {\bibfnamefont {M.~V.}\ \bibnamefont
			{Feigel'man}}, \bibinfo {author} {\bibfnamefont {V.~B.}\ \bibnamefont
			{Geshkenbein}}, \bibinfo {author} {\bibfnamefont {A.~I.}\ \bibnamefont
			{Larkin}},\ and\ \bibinfo {author} {\bibfnamefont {V.~M.}\ \bibnamefont
			{Vinokur}},\ }\bibfield  {title} {\bibinfo {title} {Vortices in
			high-temperature superconductors},\ }\href
	{https://doi.org/10.1103/revmodphys.66.1125} {\bibfield  {journal} {\bibinfo
			{journal} {Reviews of Modern Physics}\ }\textbf {\bibinfo {volume} {66}},\
		\bibinfo {pages} {1125} (\bibinfo {year} {1994})}\BibitemShut {NoStop}%
	\bibitem [{\citenamefont {Madison}\ \emph {et~al.}(2000)\citenamefont
		{Madison}, \citenamefont {Chevy}, \citenamefont {Wohlleben},\ and\
		\citenamefont {Dalibard}}]{Madison00}%
	\BibitemOpen
	\bibfield  {author} {\bibinfo {author} {\bibfnamefont {K.~W.}\ \bibnamefont
			{Madison}}, \bibinfo {author} {\bibfnamefont {F.}~\bibnamefont {Chevy}},
		\bibinfo {author} {\bibfnamefont {W.}~\bibnamefont {Wohlleben}},\ and\
		\bibinfo {author} {\bibfnamefont {J.}~\bibnamefont {Dalibard}},\ }\bibfield
	{title} {\bibinfo {title} {Vortex formation in a stirred bose-einstein
			condensate},\ }\href {https://doi.org/10.1103/physrevlett.84.806} {\bibfield
		{journal} {\bibinfo  {journal} {Physical Review Letters}\ }\textbf {\bibinfo
			{volume} {84}},\ \bibinfo {pages} {806} (\bibinfo {year} {2000})}\BibitemShut
	{NoStop}%
	\bibitem [{\citenamefont {Schulz}\ \emph {et~al.}(2012)\citenamefont {Schulz},
		\citenamefont {Ritz}, \citenamefont {Bauer}, \citenamefont {Halder},
		\citenamefont {Wagner}, \citenamefont {Franz}, \citenamefont {Pfleiderer},
		\citenamefont {Everschor}, \citenamefont {Garst},\ and\ \citenamefont
		{Rosch}}]{Schulz12}%
	\BibitemOpen
	\bibfield  {author} {\bibinfo {author} {\bibfnamefont {T.}~\bibnamefont
			{Schulz}}, \bibinfo {author} {\bibfnamefont {R.}~\bibnamefont {Ritz}},
		\bibinfo {author} {\bibfnamefont {A.}~\bibnamefont {Bauer}}, \bibinfo
		{author} {\bibfnamefont {M.}~\bibnamefont {Halder}}, \bibinfo {author}
		{\bibfnamefont {M.}~\bibnamefont {Wagner}}, \bibinfo {author} {\bibfnamefont
			{C.}~\bibnamefont {Franz}}, \bibinfo {author} {\bibfnamefont
			{C.}~\bibnamefont {Pfleiderer}}, \bibinfo {author} {\bibfnamefont
			{K.}~\bibnamefont {Everschor}}, \bibinfo {author} {\bibfnamefont
			{M.}~\bibnamefont {Garst}},\ and\ \bibinfo {author} {\bibfnamefont
			{A.}~\bibnamefont {Rosch}},\ }\bibfield  {title} {\bibinfo {title} {Emergent
			electrodynamics of skyrmions in a chiral magnet},\ }\href
	{https://doi.org/10.1038/nphys2231} {\bibfield  {journal} {\bibinfo
			{journal} {Nature Physics}\ }\textbf {\bibinfo {volume} {8}},\ \bibinfo
		{pages} {301} (\bibinfo {year} {2012})}\BibitemShut {NoStop}%
	\bibitem [{\citenamefont {Bruno}\ \emph {et~al.}(2004)\citenamefont {Bruno},
		\citenamefont {Dugaev},\ and\ \citenamefont {Taillefumier}}]{Bruno04}%
	\BibitemOpen
	\bibfield  {author} {\bibinfo {author} {\bibfnamefont {P.}~\bibnamefont
			{Bruno}}, \bibinfo {author} {\bibfnamefont {V.~K.}\ \bibnamefont {Dugaev}},\
		and\ \bibinfo {author} {\bibfnamefont {M.}~\bibnamefont {Taillefumier}},\
	}\bibfield  {title} {\bibinfo {title} {Topological {H}all effect and {B}erry
			phase in magnetic nanostructures},\ }\href
	{https://doi.org/10.1103/physrevlett.93.096806} {\bibfield  {journal}
		{\bibinfo  {journal} {Physical Review Letters}\ }\textbf {\bibinfo {volume}
			{93}},\ \bibinfo {pages} {096806} (\bibinfo {year} {2004})}\BibitemShut
	{NoStop}%
	\bibitem [{\citenamefont {Lin}\ and\ \citenamefont {Saxena}(2016)}]{Lin16}%
	\BibitemOpen
	\bibfield  {author} {\bibinfo {author} {\bibfnamefont {S.-Z.}\ \bibnamefont
			{Lin}}\ and\ \bibinfo {author} {\bibfnamefont {A.}~\bibnamefont {Saxena}},\
	}\bibfield  {title} {\bibinfo {title} {Dynamics of {D}irac strings and
			monopolelike excitations in chiral magnets under a current drive},\ }\href
	{https://doi.org/10.1103/physrevb.93.060401} {\bibfield  {journal} {\bibinfo
			{journal} {Physical Review B}\ }\textbf {\bibinfo {volume} {93}},\ \bibinfo
		{pages} {060401} (\bibinfo {year} {2016})}\BibitemShut {NoStop}%
	\bibitem [{\citenamefont {Braun}(2012)}]{Braun12}%
	\BibitemOpen
	\bibfield  {author} {\bibinfo {author} {\bibfnamefont {H.-B.}\ \bibnamefont
			{Braun}},\ }\bibfield  {title} {\bibinfo {title} {Topological effects in
			nanomagnetism: from superparamagnetism to chiral quantum solitons},\ }\href
	{https://doi.org/10.1080/00018732.2012.663070} {\bibfield  {journal}
		{\bibinfo  {journal} {Advances in Physics}\ }\textbf {\bibinfo {volume}
			{61}},\ \bibinfo {pages} {1} (\bibinfo {year} {2012})}\BibitemShut {NoStop}%
	\bibitem [{\citenamefont {Milde}\ \emph {et~al.}(2013)\citenamefont {Milde},
		\citenamefont {Kohler}, \citenamefont {Seidel}, \citenamefont {Eng},
		\citenamefont {Bauer}, \citenamefont {Chacon}, \citenamefont {Kindervater},
		\citenamefont {Muhlbauer}, \citenamefont {Pfleiderer}, \citenamefont
		{Buhrandt},\ and\ \citenamefont {et~al.}}]{Milde13}%
	\BibitemOpen
	\bibfield  {author} {\bibinfo {author} {\bibfnamefont {P.}~\bibnamefont
			{Milde}}, \bibinfo {author} {\bibfnamefont {D.}~\bibnamefont {Kohler}},
		\bibinfo {author} {\bibfnamefont {J.}~\bibnamefont {Seidel}}, \bibinfo
		{author} {\bibfnamefont {L.~M.}\ \bibnamefont {Eng}}, \bibinfo {author}
		{\bibfnamefont {A.}~\bibnamefont {Bauer}}, \bibinfo {author} {\bibfnamefont
			{A.}~\bibnamefont {Chacon}}, \bibinfo {author} {\bibfnamefont
			{J.}~\bibnamefont {Kindervater}}, \bibinfo {author} {\bibfnamefont
			{S.}~\bibnamefont {Muhlbauer}}, \bibinfo {author} {\bibfnamefont
			{C.}~\bibnamefont {Pfleiderer}}, \bibinfo {author} {\bibfnamefont
			{S.}~\bibnamefont {Buhrandt}},\ and\ \bibinfo {author} {\bibnamefont
			{et~al.}},\ }\bibfield  {title} {\bibinfo {title} {Unwinding of a skyrmion
			lattice by magnetic monopoles},\ }\href
	{https://doi.org/10.1126/science.1234657} {\bibfield  {journal} {\bibinfo
			{journal} {Science}\ }\textbf {\bibinfo {volume} {340}},\ \bibinfo {pages}
		{1076} (\bibinfo {year} {2013})}\BibitemShut {NoStop}%
	\bibitem [{\citenamefont {Sch{\"u}tte}\ and\ \citenamefont
		{Rosch}(2014)}]{Schuette14b}%
	\BibitemOpen
	\bibfield  {author} {\bibinfo {author} {\bibfnamefont {C.}~\bibnamefont
			{Sch{\"u}tte}}\ and\ \bibinfo {author} {\bibfnamefont {A.}~\bibnamefont
			{Rosch}},\ }\bibfield  {title} {\bibinfo {title} {Dynamics and energetics of
			emergent magnetic monopoles in chiral magnets},\ }\href
	{https://doi.org/10.1103/physrevb.90.174432} {\bibfield  {journal} {\bibinfo
			{journal} {Physical Review B}\ }\textbf {\bibinfo {volume} {90}},\ \bibinfo
		{pages} {174432} (\bibinfo {year} {2014})}\BibitemShut {NoStop}%
	\bibitem [{\citenamefont {Seki}\ \emph {et~al.}(2020)\citenamefont {Seki},
		\citenamefont {Garst}, \citenamefont {Waizner}, \citenamefont {Takagi},
		\citenamefont {Khanh}, \citenamefont {Okamura}, \citenamefont {Kondou},
		\citenamefont {Kagawa}, \citenamefont {Otani},\ and\ \citenamefont
		{Tokura}}]{Seki20}%
	\BibitemOpen
	\bibfield  {author} {\bibinfo {author} {\bibfnamefont {S.}~\bibnamefont
			{Seki}}, \bibinfo {author} {\bibfnamefont {M.}~\bibnamefont {Garst}},
		\bibinfo {author} {\bibfnamefont {J.}~\bibnamefont {Waizner}}, \bibinfo
		{author} {\bibfnamefont {R.}~\bibnamefont {Takagi}}, \bibinfo {author}
		{\bibfnamefont {N.~D.}\ \bibnamefont {Khanh}}, \bibinfo {author}
		{\bibfnamefont {Y.}~\bibnamefont {Okamura}}, \bibinfo {author} {\bibfnamefont
			{K.}~\bibnamefont {Kondou}}, \bibinfo {author} {\bibfnamefont
			{F.}~\bibnamefont {Kagawa}}, \bibinfo {author} {\bibfnamefont
			{Y.}~\bibnamefont {Otani}},\ and\ \bibinfo {author} {\bibfnamefont
			{Y.}~\bibnamefont {Tokura}},\ }\bibfield  {title} {\bibinfo {title}
		{Propagation dynamics of spin excitations along skyrmion strings},\
	}\bibfield  {journal} {\bibinfo  {journal} {Nature Communications}\ }\textbf
	{\bibinfo {volume} {11}},\ \href {https://doi.org/10.1038/s41467-019-14095-0}
	{10.1038/s41467-019-14095-0} (\bibinfo {year} {2020})\BibitemShut {NoStop}%
	\bibitem [{\citenamefont {Mathur}\ \emph {et~al.}(2021)\citenamefont {Mathur},
		\citenamefont {Yasin}, \citenamefont {Stolt}, \citenamefont {Nagai},
		\citenamefont {Kimoto}, \citenamefont {Du}, \citenamefont {Tian},
		\citenamefont {Tokura}, \citenamefont {Yu},\ and\ \citenamefont
		{Jin}}]{Mathur21}%
	\BibitemOpen
	\bibfield  {author} {\bibinfo {author} {\bibfnamefont {N.}~\bibnamefont
			{Mathur}}, \bibinfo {author} {\bibfnamefont {F.~S.}\ \bibnamefont {Yasin}},
		\bibinfo {author} {\bibfnamefont {M.~J.}\ \bibnamefont {Stolt}}, \bibinfo
		{author} {\bibfnamefont {T.}~\bibnamefont {Nagai}}, \bibinfo {author}
		{\bibfnamefont {K.}~\bibnamefont {Kimoto}}, \bibinfo {author} {\bibfnamefont
			{H.}~\bibnamefont {Du}}, \bibinfo {author} {\bibfnamefont {M.}~\bibnamefont
			{Tian}}, \bibinfo {author} {\bibfnamefont {Y.}~\bibnamefont {Tokura}},
		\bibinfo {author} {\bibfnamefont {X.}~\bibnamefont {Yu}},\ and\ \bibinfo
		{author} {\bibfnamefont {S.}~\bibnamefont {Jin}},\ }\bibfield  {title}
	{\bibinfo {title} {In-plane magnetic field-driven creation and annihilation
			of magnetic skyrmion strings in nanostructures},\ }\href
	{https://doi.org/10.1002/adfm.202008521} {\bibfield  {journal} {\bibinfo
			{journal} {Advanced Functional Materials}\ }\textbf {\bibinfo {volume}
			{31}},\ \bibinfo {pages} {2008521} (\bibinfo {year} {2021})}\BibitemShut
	{NoStop}%
	\bibitem [{\citenamefont {Birch}\ \emph {et~al.}(2022)\citenamefont {Birch},
		\citenamefont {Cort{\'{e}}s-Ortu{\~{n}}o}, \citenamefont {Litzius},
		\citenamefont {Wintz}, \citenamefont {Schulz}, \citenamefont {Weigand},
		\citenamefont {{\v{S}}tefan{\v{c}}i{\v{c}}}, \citenamefont {Mayoh},
		\citenamefont {Balakrishnan}, \citenamefont {Hatton},\ and\ \citenamefont
		{Schütz}}]{Birch22}%
	\BibitemOpen
	\bibfield  {author} {\bibinfo {author} {\bibfnamefont {M.~T.}\ \bibnamefont
			{Birch}}, \bibinfo {author} {\bibfnamefont {D.}~\bibnamefont
			{Cort{\'{e}}s-Ortu{\~{n}}o}}, \bibinfo {author} {\bibfnamefont
			{K.}~\bibnamefont {Litzius}}, \bibinfo {author} {\bibfnamefont
			{S.}~\bibnamefont {Wintz}}, \bibinfo {author} {\bibfnamefont
			{F.}~\bibnamefont {Schulz}}, \bibinfo {author} {\bibfnamefont
			{M.}~\bibnamefont {Weigand}}, \bibinfo {author} {\bibfnamefont
			{A.}~\bibnamefont {{\v{S}}tefan{\v{c}}i{\v{c}}}}, \bibinfo {author}
		{\bibfnamefont {D.~A.}\ \bibnamefont {Mayoh}}, \bibinfo {author}
		{\bibfnamefont {G.}~\bibnamefont {Balakrishnan}}, \bibinfo {author}
		{\bibfnamefont {P.~D.}\ \bibnamefont {Hatton}},\ and\ \bibinfo {author}
		{\bibfnamefont {G.}~\bibnamefont {Schütz}},\ }\bibfield  {title} {\bibinfo
		{title} {Toggle-like current-induced bloch point dynamics of 3d skyrmion
			strings in a room temperature nanowire},\ }\bibfield  {journal} {\bibinfo
		{journal} {Nature Communications}\ }\textbf {\bibinfo {volume} {13}},\ \href
	{https://doi.org/10.1038/s41467-022-31335-y} {10.1038/s41467-022-31335-y}
	(\bibinfo {year} {2022})\BibitemShut {NoStop}%
	\bibitem [{\citenamefont {Yokouchi}\ \emph {et~al.}(2018)\citenamefont
		{Yokouchi}, \citenamefont {Hoshino}, \citenamefont {Kanazawa}, \citenamefont
		{Kikkawa}, \citenamefont {Morikawa}, \citenamefont {Shibata}, \citenamefont
		{hisa Arima}, \citenamefont {Taguchi}, \citenamefont {Kagawa}, \citenamefont
		{Nagaosa},\ and\ \citenamefont {Tokura}}]{Yokouchi18}%
	\BibitemOpen
	\bibfield  {author} {\bibinfo {author} {\bibfnamefont {T.}~\bibnamefont
			{Yokouchi}}, \bibinfo {author} {\bibfnamefont {S.}~\bibnamefont {Hoshino}},
		\bibinfo {author} {\bibfnamefont {N.}~\bibnamefont {Kanazawa}}, \bibinfo
		{author} {\bibfnamefont {A.}~\bibnamefont {Kikkawa}}, \bibinfo {author}
		{\bibfnamefont {D.}~\bibnamefont {Morikawa}}, \bibinfo {author}
		{\bibfnamefont {K.}~\bibnamefont {Shibata}}, \bibinfo {author} {\bibfnamefont
			{T.}~\bibnamefont {hisa Arima}}, \bibinfo {author} {\bibfnamefont
			{Y.}~\bibnamefont {Taguchi}}, \bibinfo {author} {\bibfnamefont
			{F.}~\bibnamefont {Kagawa}}, \bibinfo {author} {\bibfnamefont
			{N.}~\bibnamefont {Nagaosa}},\ and\ \bibinfo {author} {\bibfnamefont
			{Y.}~\bibnamefont {Tokura}},\ }\bibfield  {title} {\bibinfo {title}
		{Current-induced dynamics of skyrmion strings},\ }\href
	{https://doi.org/10.1126/sciadv.aat1115} {\bibfield  {journal} {\bibinfo
			{journal} {Science Advances}\ }\textbf {\bibinfo {volume} {4}},\ \bibinfo
		{pages} {eaat1115} (\bibinfo {year} {2018})}\BibitemShut {NoStop}%
	\bibitem [{\citenamefont {Jonietz}\ \emph {et~al.}(2010)\citenamefont
		{Jonietz}, \citenamefont {Muhlbauer}, \citenamefont {Pfleiderer},
		\citenamefont {Neubauer}, \citenamefont {Munzer}, \citenamefont {Bauer},
		\citenamefont {Adams}, \citenamefont {Georgii}, \citenamefont {Boni},
		\citenamefont {Duine}, \citenamefont {Everschor}, \citenamefont {Garst},\
		and\ \citenamefont {Rosch}}]{Jonietz10}%
	\BibitemOpen
	\bibfield  {author} {\bibinfo {author} {\bibfnamefont {F.}~\bibnamefont
			{Jonietz}}, \bibinfo {author} {\bibfnamefont {S.}~\bibnamefont {Muhlbauer}},
		\bibinfo {author} {\bibfnamefont {C.}~\bibnamefont {Pfleiderer}}, \bibinfo
		{author} {\bibfnamefont {A.}~\bibnamefont {Neubauer}}, \bibinfo {author}
		{\bibfnamefont {W.}~\bibnamefont {Munzer}}, \bibinfo {author} {\bibfnamefont
			{A.}~\bibnamefont {Bauer}}, \bibinfo {author} {\bibfnamefont
			{T.}~\bibnamefont {Adams}}, \bibinfo {author} {\bibfnamefont
			{R.}~\bibnamefont {Georgii}}, \bibinfo {author} {\bibfnamefont
			{P.}~\bibnamefont {Boni}}, \bibinfo {author} {\bibfnamefont {R.~A.}\
			\bibnamefont {Duine}}, \bibinfo {author} {\bibfnamefont {K.}~\bibnamefont
			{Everschor}}, \bibinfo {author} {\bibfnamefont {M.}~\bibnamefont {Garst}},\
		and\ \bibinfo {author} {\bibfnamefont {A.}~\bibnamefont {Rosch}},\ }\bibfield
	{title} {\bibinfo {title} {Spin transfer torques in {MnSi} at ultralow
			current densities},\ }\href {https://doi.org/10.1126/science.1195709}
	{\bibfield  {journal} {\bibinfo  {journal} {Science}\ }\textbf {\bibinfo
			{volume} {330}},\ \bibinfo {pages} {1648} (\bibinfo {year}
		{2010})}\BibitemShut {NoStop}%
	\bibitem [{\citenamefont {Yu}\ \emph {et~al.}(2012)\citenamefont {Yu},
		\citenamefont {Kanazawa}, \citenamefont {Zhang}, \citenamefont {Nagai},
		\citenamefont {Hara}, \citenamefont {Kimoto}, \citenamefont {Matsui},
		\citenamefont {Onose},\ and\ \citenamefont {Tokura}}]{Yu12}%
	\BibitemOpen
	\bibfield  {author} {\bibinfo {author} {\bibfnamefont {X.}~\bibnamefont
			{Yu}}, \bibinfo {author} {\bibfnamefont {N.}~\bibnamefont {Kanazawa}},
		\bibinfo {author} {\bibfnamefont {W.}~\bibnamefont {Zhang}}, \bibinfo
		{author} {\bibfnamefont {T.}~\bibnamefont {Nagai}}, \bibinfo {author}
		{\bibfnamefont {T.}~\bibnamefont {Hara}}, \bibinfo {author} {\bibfnamefont
			{K.}~\bibnamefont {Kimoto}}, \bibinfo {author} {\bibfnamefont
			{Y.}~\bibnamefont {Matsui}}, \bibinfo {author} {\bibfnamefont
			{Y.}~\bibnamefont {Onose}},\ and\ \bibinfo {author} {\bibfnamefont
			{Y.}~\bibnamefont {Tokura}},\ }\bibfield  {title} {\bibinfo {title} {Skyrmion
			flow near room temperature in an ultralow current density},\ }\href
	{https://doi.org/10.1038/ncomms1990} {\bibfield  {journal} {\bibinfo
			{journal} {Nature Communications}\ }\textbf {\bibinfo {volume} {3}},\
		\bibinfo {pages} {988} (\bibinfo {year} {2012})}\BibitemShut {NoStop}%
	\bibitem [{\citenamefont {Tang}\ \emph {et~al.}(2021)\citenamefont {Tang},
		\citenamefont {Wu}, \citenamefont {Wang}, \citenamefont {Kong}, \citenamefont
		{Lv}, \citenamefont {Wei}, \citenamefont {Zang}, \citenamefont {Tian},\ and\
		\citenamefont {Du}}]{Tang21}%
	\BibitemOpen
	\bibfield  {author} {\bibinfo {author} {\bibfnamefont {J.}~\bibnamefont
			{Tang}}, \bibinfo {author} {\bibfnamefont {Y.}~\bibnamefont {Wu}}, \bibinfo
		{author} {\bibfnamefont {W.}~\bibnamefont {Wang}}, \bibinfo {author}
		{\bibfnamefont {L.}~\bibnamefont {Kong}}, \bibinfo {author} {\bibfnamefont
			{B.}~\bibnamefont {Lv}}, \bibinfo {author} {\bibfnamefont {W.}~\bibnamefont
			{Wei}}, \bibinfo {author} {\bibfnamefont {J.}~\bibnamefont {Zang}}, \bibinfo
		{author} {\bibfnamefont {M.}~\bibnamefont {Tian}},\ and\ \bibinfo {author}
		{\bibfnamefont {H.}~\bibnamefont {Du}},\ }\bibfield  {title} {\bibinfo
		{title} {Magnetic skyrmion bundles and their current-driven dynamics},\
	}\href {https://doi.org/10.1038/s41565-021-00954-9} {\bibfield  {journal}
		{\bibinfo  {journal} {Nature Nanotechnology}\ }\textbf {\bibinfo {volume}
			{16}},\ \bibinfo {pages} {1086} (\bibinfo {year} {2021})}\BibitemShut
	{NoStop}%
	\bibitem [{\citenamefont {Okumura}\ \emph {et~al.}(2023)\citenamefont
		{Okumura}, \citenamefont {Kravchuk},\ and\ \citenamefont
		{Garst}}]{Okumura23}%
	\BibitemOpen
	\bibfield  {author} {\bibinfo {author} {\bibfnamefont {S.}~\bibnamefont
			{Okumura}}, \bibinfo {author} {\bibfnamefont {V.~P.}\ \bibnamefont
			{Kravchuk}},\ and\ \bibinfo {author} {\bibfnamefont {M.}~\bibnamefont
			{Garst}},\ }\bibfield  {title} {\bibinfo {title} {Instability of magnetic
			skyrmion strings induced by longitudinal spin currents},\ }\href@noop {}
	{\bibfield  {journal} {\bibinfo  {journal} {ArXiv e-prints}\ } (\bibinfo
		{year} {2023})},\ \Eprint
	{https://arxiv.org/abs/http://arxiv.org/abs/2303.08532v1}
	{http://arxiv.org/abs/2303.08532v1} \BibitemShut {NoStop}%
	\bibitem [{\citenamefont {Kagawa}\ \emph {et~al.}(2017)\citenamefont {Kagawa},
		\citenamefont {Oike}, \citenamefont {Koshibae}, \citenamefont {Kikkawa},
		\citenamefont {Okamura}, \citenamefont {Taguchi}, \citenamefont {Nagaosa},\
		and\ \citenamefont {Tokura}}]{Kagawa17}%
	\BibitemOpen
	\bibfield  {author} {\bibinfo {author} {\bibfnamefont {F.}~\bibnamefont
			{Kagawa}}, \bibinfo {author} {\bibfnamefont {H.}~\bibnamefont {Oike}},
		\bibinfo {author} {\bibfnamefont {W.}~\bibnamefont {Koshibae}}, \bibinfo
		{author} {\bibfnamefont {A.}~\bibnamefont {Kikkawa}}, \bibinfo {author}
		{\bibfnamefont {Y.}~\bibnamefont {Okamura}}, \bibinfo {author} {\bibfnamefont
			{Y.}~\bibnamefont {Taguchi}}, \bibinfo {author} {\bibfnamefont
			{N.}~\bibnamefont {Nagaosa}},\ and\ \bibinfo {author} {\bibfnamefont
			{Y.}~\bibnamefont {Tokura}},\ }\bibfield  {title} {\bibinfo {title}
		{Current-induced viscoelastic topological unwinding of metastable skyrmion
			strings},\ }\bibfield  {journal} {\bibinfo  {journal} {Nature
			Communications}\ }\textbf {\bibinfo {volume} {8}},\ \href
	{https://doi.org/10.1038/s41467-017-01353-2} {10.1038/s41467-017-01353-2}
	(\bibinfo {year} {2017})\BibitemShut {NoStop}%
	\bibitem [{\citenamefont {Yu}\ \emph {et~al.}(2020)\citenamefont {Yu},
		\citenamefont {Masell}, \citenamefont {Yasin}, \citenamefont {Karube},
		\citenamefont {Kanazawa}, \citenamefont {Nakajima}, \citenamefont {Nagai},
		\citenamefont {Kimoto}, \citenamefont {Koshibae}, \citenamefont {Taguchi},
		\citenamefont {Nagaosa},\ and\ \citenamefont {Tokura}}]{Yu20}%
	\BibitemOpen
	\bibfield  {author} {\bibinfo {author} {\bibfnamefont {X.}~\bibnamefont
			{Yu}}, \bibinfo {author} {\bibfnamefont {J.}~\bibnamefont {Masell}}, \bibinfo
		{author} {\bibfnamefont {F.~S.}\ \bibnamefont {Yasin}}, \bibinfo {author}
		{\bibfnamefont {K.}~\bibnamefont {Karube}}, \bibinfo {author} {\bibfnamefont
			{N.}~\bibnamefont {Kanazawa}}, \bibinfo {author} {\bibfnamefont
			{K.}~\bibnamefont {Nakajima}}, \bibinfo {author} {\bibfnamefont
			{T.}~\bibnamefont {Nagai}}, \bibinfo {author} {\bibfnamefont
			{K.}~\bibnamefont {Kimoto}}, \bibinfo {author} {\bibfnamefont
			{W.}~\bibnamefont {Koshibae}}, \bibinfo {author} {\bibfnamefont
			{Y.}~\bibnamefont {Taguchi}}, \bibinfo {author} {\bibfnamefont
			{N.}~\bibnamefont {Nagaosa}},\ and\ \bibinfo {author} {\bibfnamefont
			{Y.}~\bibnamefont {Tokura}},\ }\bibfield  {title} {\bibinfo {title}
		{Real-space observation of topological defects in extended
			skyrmion-strings},\ }\href {https://doi.org/10.1021/acs.nanolett.0c02708}
	{\bibfield  {journal} {\bibinfo  {journal} {Nano Letters}\ }\textbf {\bibinfo
			{volume} {20}},\ \bibinfo {pages} {7313} (\bibinfo {year}
		{2020})}\BibitemShut {NoStop}%
	\bibitem [{\citenamefont {Xia}\ \emph {et~al.}(2022)\citenamefont {Xia},
		\citenamefont {Zhang}, \citenamefont {Tretiakov}, \citenamefont {Diep},
		\citenamefont {Yang}, \citenamefont {Zhao}, \citenamefont {Ezawa},
		\citenamefont {Zhou},\ and\ \citenamefont {Liu}}]{Xia22}%
	\BibitemOpen
	\bibfield  {author} {\bibinfo {author} {\bibfnamefont {J.}~\bibnamefont
			{Xia}}, \bibinfo {author} {\bibfnamefont {X.}~\bibnamefont {Zhang}}, \bibinfo
		{author} {\bibfnamefont {O.~A.}\ \bibnamefont {Tretiakov}}, \bibinfo {author}
		{\bibfnamefont {H.~T.}\ \bibnamefont {Diep}}, \bibinfo {author}
		{\bibfnamefont {J.}~\bibnamefont {Yang}}, \bibinfo {author} {\bibfnamefont
			{G.}~\bibnamefont {Zhao}}, \bibinfo {author} {\bibfnamefont {M.}~\bibnamefont
			{Ezawa}}, \bibinfo {author} {\bibfnamefont {Y.}~\bibnamefont {Zhou}},\ and\
		\bibinfo {author} {\bibfnamefont {X.}~\bibnamefont {Liu}},\ }\bibfield
	{title} {\bibinfo {title} {Bifurcation of a topological skyrmion string},\
	}\href {https://doi.org/10.1103/physrevb.105.214402} {\bibfield  {journal}
		{\bibinfo  {journal} {Physical Review B}\ }\textbf {\bibinfo {volume}
			{105}},\ \bibinfo {pages} {214402} (\bibinfo {year} {2022})}\BibitemShut
	{NoStop}%
	\bibitem [{\citenamefont {Zheng}\ \emph {et~al.}(2021)\citenamefont {Zheng},
		\citenamefont {Rybakov}, \citenamefont {Kiselev}, \citenamefont {Song},
		\citenamefont {Kov{\'{a}}cs}, \citenamefont {Du}, \citenamefont {Blügel},\
		and\ \citenamefont {Dunin-Borkowski}}]{Zheng21}%
	\BibitemOpen
	\bibfield  {author} {\bibinfo {author} {\bibfnamefont {F.}~\bibnamefont
			{Zheng}}, \bibinfo {author} {\bibfnamefont {F.~N.}\ \bibnamefont {Rybakov}},
		\bibinfo {author} {\bibfnamefont {N.~S.}\ \bibnamefont {Kiselev}}, \bibinfo
		{author} {\bibfnamefont {D.}~\bibnamefont {Song}}, \bibinfo {author}
		{\bibfnamefont {A.}~\bibnamefont {Kov{\'{a}}cs}}, \bibinfo {author}
		{\bibfnamefont {H.}~\bibnamefont {Du}}, \bibinfo {author} {\bibfnamefont
			{S.}~\bibnamefont {Blügel}},\ and\ \bibinfo {author} {\bibfnamefont {R.~E.}\
			\bibnamefont {Dunin-Borkowski}},\ }\bibfield  {title} {\bibinfo {title}
		{Magnetic skyrmion braids},\ }\bibfield  {journal} {\bibinfo  {journal}
		{Nature Communications}\ }\textbf {\bibinfo {volume} {12}},\ \href
	{https://doi.org/10.1038/s41467-021-25389-7} {10.1038/s41467-021-25389-7}
	(\bibinfo {year} {2021})\BibitemShut {NoStop}%
	\bibitem [{\citenamefont {Yan}\ \emph {et~al.}(2007)\citenamefont {Yan},
		\citenamefont {Hertel},\ and\ \citenamefont {Schneider}}]{Yan07}%
	\BibitemOpen
	\bibfield  {author} {\bibinfo {author} {\bibfnamefont {M.}~\bibnamefont
			{Yan}}, \bibinfo {author} {\bibfnamefont {R.}~\bibnamefont {Hertel}},\ and\
		\bibinfo {author} {\bibfnamefont {C.~M.}\ \bibnamefont {Schneider}},\
	}\bibfield  {title} {\bibinfo {title} {Calculations of three-dimensional
			magnetic normal modes in mesoscopic permalloy prisms with vortex structure},\
	}\href {https://doi.org/10.1103/physrevb.76.094407} {\bibfield  {journal}
		{\bibinfo  {journal} {Physical Review B}\ }\textbf {\bibinfo {volume} {76}},\
		\bibinfo {pages} {094407} (\bibinfo {year} {2007})}\BibitemShut {NoStop}%
	\bibitem [{\citenamefont {Ding}\ \emph {et~al.}(2014)\citenamefont {Ding},
		\citenamefont {Kakazei}, \citenamefont {Liu}, \citenamefont {Guslienko},\
		and\ \citenamefont {Adeyeye}}]{Ding14}%
	\BibitemOpen
	\bibfield  {author} {\bibinfo {author} {\bibfnamefont {J.}~\bibnamefont
			{Ding}}, \bibinfo {author} {\bibfnamefont {G.~N.}\ \bibnamefont {Kakazei}},
		\bibinfo {author} {\bibfnamefont {X.}~\bibnamefont {Liu}}, \bibinfo {author}
		{\bibfnamefont {K.~Y.}\ \bibnamefont {Guslienko}},\ and\ \bibinfo {author}
		{\bibfnamefont {A.~O.}\ \bibnamefont {Adeyeye}},\ }\bibfield  {title}
	{\bibinfo {title} {Higher order vortex gyrotropic modes in circular
			ferromagnetic nanodots},\ }\href {https://doi.org/10.1038/srep04796}
	{\bibfield  {journal} {\bibinfo  {journal} {Scientific Reports}\ }\textbf
		{\bibinfo {volume} {4}},\ \bibinfo {pages} {4796} (\bibinfo {year}
		{2014})}\BibitemShut {NoStop}%
	\bibitem [{\citenamefont {Guslienko}(2022)}]{Guslienko22}%
	\BibitemOpen
	\bibfield  {author} {\bibinfo {author} {\bibfnamefont {K.}~\bibnamefont
			{Guslienko}},\ }\bibfield  {title} {\bibinfo {title} {Magnetic vortex core
			string gyrotropic oscillations in thick cylindrical dots},\ }\href
	{https://doi.org/10.3390/magnetism2030018} {\bibfield  {journal} {\bibinfo
			{journal} {Magnetism}\ }\textbf {\bibinfo {volume} {2}},\ \bibinfo {pages}
		{239} (\bibinfo {year} {2022})}\BibitemShut {NoStop}%
	\bibitem [{\citenamefont {Finizio}\ \emph {et~al.}(2022)\citenamefont
		{Finizio}, \citenamefont {Donnelly}, \citenamefont {Mayr}, \citenamefont
		{Hrabec},\ and\ \citenamefont {Raabe}}]{Finizio22}%
	\BibitemOpen
	\bibfield  {author} {\bibinfo {author} {\bibfnamefont {S.}~\bibnamefont
			{Finizio}}, \bibinfo {author} {\bibfnamefont {C.}~\bibnamefont {Donnelly}},
		\bibinfo {author} {\bibfnamefont {S.}~\bibnamefont {Mayr}}, \bibinfo {author}
		{\bibfnamefont {A.}~\bibnamefont {Hrabec}},\ and\ \bibinfo {author}
		{\bibfnamefont {J.}~\bibnamefont {Raabe}},\ }\bibfield  {title} {\bibinfo
		{title} {Three-dimensional vortex gyration dynamics unraveled by
			time-resolved soft x-ray laminography with freely selectable excitation
			frequencies},\ }\href {https://doi.org/10.1021/acs.nanolett.1c04662}
	{\bibfield  {journal} {\bibinfo  {journal} {Nano Letters}\ }\textbf {\bibinfo
			{volume} {22}},\ \bibinfo {pages} {1971} (\bibinfo {year}
		{2022})}\BibitemShut {NoStop}%
	\bibitem [{\citenamefont {Volkov}\ \emph {et~al.}(2023)\citenamefont {Volkov},
		\citenamefont {Wolf}, \citenamefont {Pylypovskyi}, \citenamefont
		{K{\'{a}}kay}, \citenamefont {Sheka}, \citenamefont {Büchner}, \citenamefont
		{Fassbender}, \citenamefont {Lubk},\ and\ \citenamefont
		{Makarov}}]{Volkov23}%
	\BibitemOpen
	\bibfield  {author} {\bibinfo {author} {\bibfnamefont {O.~M.}\ \bibnamefont
			{Volkov}}, \bibinfo {author} {\bibfnamefont {D.}~\bibnamefont {Wolf}},
		\bibinfo {author} {\bibfnamefont {O.~V.}\ \bibnamefont {Pylypovskyi}},
		\bibinfo {author} {\bibfnamefont {A.}~\bibnamefont {K{\'{a}}kay}}, \bibinfo
		{author} {\bibfnamefont {D.~D.}\ \bibnamefont {Sheka}}, \bibinfo {author}
		{\bibfnamefont {B.}~\bibnamefont {Büchner}}, \bibinfo {author}
		{\bibfnamefont {J.}~\bibnamefont {Fassbender}}, \bibinfo {author}
		{\bibfnamefont {A.}~\bibnamefont {Lubk}},\ and\ \bibinfo {author}
		{\bibfnamefont {D.}~\bibnamefont {Makarov}},\ }\bibfield  {title} {\bibinfo
		{title} {Chirality coupling in topological magnetic textures with multiple
			magnetochiral parameters},\ }\href
	{https://doi.org/10.1038/s41467-023-37081-z} {\bibfield  {journal} {\bibinfo
			{journal} {Nature Communications}\ }\textbf {\bibinfo {volume} {14}},\
		\bibinfo {pages} {1491} (\bibinfo {year} {2023})}\BibitemShut {NoStop}%
	\bibitem [{\citenamefont {Azhar}\ \emph {et~al.}(2022)\citenamefont {Azhar},
		\citenamefont {Kravchuk},\ and\ \citenamefont {Garst}}]{Azhar22}%
	\BibitemOpen
	\bibfield  {author} {\bibinfo {author} {\bibfnamefont {M.}~\bibnamefont
			{Azhar}}, \bibinfo {author} {\bibfnamefont {V.~P.}\ \bibnamefont
			{Kravchuk}},\ and\ \bibinfo {author} {\bibfnamefont {M.}~\bibnamefont
			{Garst}},\ }\bibfield  {title} {\bibinfo {title} {Screw dislocations in
			chiral magnets},\ }\bibfield  {journal} {\bibinfo  {journal} {Physical Review
			Letters}\ }\textbf {\bibinfo {volume} {128}},\ \href
	{https://doi.org/10.1103/physrevlett.128.157204}
	{10.1103/physrevlett.128.157204} (\bibinfo {year} {2022})\BibitemShut
	{NoStop}%
	\bibitem [{\citenamefont {Kravchuk}\ \emph {et~al.}(2020)\citenamefont
		{Kravchuk}, \citenamefont {Rö{\ss}ler}, \citenamefont {van~den Brink},\ and\
		\citenamefont {Garst}}]{Kravchuk20}%
	\BibitemOpen
	\bibfield  {author} {\bibinfo {author} {\bibfnamefont {V.~P.}\ \bibnamefont
			{Kravchuk}}, \bibinfo {author} {\bibfnamefont {U.~K.}\ \bibnamefont
			{Rö{\ss}ler}}, \bibinfo {author} {\bibfnamefont {J.}~\bibnamefont {van~den
				Brink}},\ and\ \bibinfo {author} {\bibfnamefont {M.}~\bibnamefont {Garst}},\
	}\bibfield  {title} {\bibinfo {title} {Solitary wave excitations of skyrmion
			strings in chiral magnets},\ }\bibfield  {journal} {\bibinfo  {journal}
		{Physical Review B}\ }\textbf {\bibinfo {volume} {102}},\ \href
	{https://doi.org/10.1103/physrevb.102.220408} {10.1103/physrevb.102.220408}
	(\bibinfo {year} {2020})\BibitemShut {NoStop}%
	\bibitem [{\citenamefont {Ryskin}\ and\ \citenamefont
		{Trubetskov}(2000)}]{Ryskin00}%
	\BibitemOpen
	\bibfield  {author} {\bibinfo {author} {\bibfnamefont {N.~M.}\ \bibnamefont
			{Ryskin}}\ and\ \bibinfo {author} {\bibfnamefont {D.~I.}\ \bibnamefont
			{Trubetskov}},\ }\href@noop {} {\emph {\bibinfo {title} {Nonlinear Waves}}}\
	(\bibinfo  {publisher} {Moscow, Nauka},\ \bibinfo {year} {2000})\BibitemShut
	{NoStop}%
	\bibitem [{\citenamefont {Nayfeh}(2008)}]{Nayfeh08}%
	\BibitemOpen
	\bibfield  {author} {\bibinfo {author} {\bibfnamefont {A.}~\bibnamefont
			{Nayfeh}},\ }\href {http://books.google.com.ua/books?id=eh6RmWZ51NIC} {\emph
		{\bibinfo {title} {Perturbation Methods}}},\ Physics textbook\ (\bibinfo
	{publisher} {John Wiley \& Sons},\ \bibinfo {year} {2008})\BibitemShut
	{NoStop}%
	\bibitem [{\citenamefont {Bogdanov}\ and\ \citenamefont
		{Yablonski\u{\i}}(1989)}]{Bogdanov89r}%
	\BibitemOpen
	\bibfield  {author} {\bibinfo {author} {\bibfnamefont {A.~N.}\ \bibnamefont
			{Bogdanov}}\ and\ \bibinfo {author} {\bibfnamefont {D.~A.}\ \bibnamefont
			{Yablonski\u{\i}}},\ }\bibfield  {title} {\bibinfo {title} {Thermodynamically
			stable ``vortices'' in magnetically ordered crystals. {T}he mixed state of
			magnets},\ }\href {http://jetp.ac.ru/cgi-bin/e/index/e/68/1/p101?a=list}
	{\bibfield  {journal} {\bibinfo  {journal} {Zh. Eksp. Teor. Fiz.}\ }\textbf
		{\bibinfo {volume} {95}},\ \bibinfo {pages} {178} (\bibinfo {year}
		{1989})}\BibitemShut {NoStop}%
	\bibitem [{\citenamefont {Bogdanov}\ and\ \citenamefont
		{Hubert}(1994)}]{Bogdanov94}%
	\BibitemOpen
	\bibfield  {author} {\bibinfo {author} {\bibfnamefont {A.}~\bibnamefont
			{Bogdanov}}\ and\ \bibinfo {author} {\bibfnamefont {A.}~\bibnamefont
			{Hubert}},\ }\bibfield  {title} {\bibinfo {title} {Thermodynamically stable
			magnetic vortex states in magnetic crystals},\ }\href
	{https://doi.org/10.1016/0304-8853(94)90046-9} {\bibfield  {journal}
		{\bibinfo  {journal} {Journal of Magnetism and Magnetic Materials}\ }\textbf
		{\bibinfo {volume} {138}},\ \bibinfo {pages} {255} (\bibinfo {year}
		{1994})}\BibitemShut {NoStop}%
	\bibitem [{\citenamefont {Papanicolaou}\ and\ \citenamefont
		{Tomaras}(1991)}]{Papanicolaou91}%
	\BibitemOpen
	\bibfield  {author} {\bibinfo {author} {\bibfnamefont {N.}~\bibnamefont
			{Papanicolaou}}\ and\ \bibinfo {author} {\bibfnamefont {T.~N.}\ \bibnamefont
			{Tomaras}},\ }\bibfield  {title} {\bibinfo {title} {Dynamics of magnetic
			vortices},\ }\href
	{http://www.sciencedirect.com/science/article/B6TVC-470F3HY-36/2/69a5e1a128fb5b7ef2ee0c512c3d78fc}
	{\bibfield  {journal} {\bibinfo  {journal} {Nuclear Physics B}\ }\textbf
		{\bibinfo {volume} {360}},\ \bibinfo {pages} {425} (\bibinfo {year}
		{1991})}\BibitemShut {NoStop}%
	\bibitem [{\citenamefont {Thiele}(1973)}]{Thiele73}%
	\BibitemOpen
	\bibfield  {author} {\bibinfo {author} {\bibfnamefont {A.~A.}\ \bibnamefont
			{Thiele}},\ }\bibfield  {title} {\bibinfo {title} {Steady--state motion of
			magnetic of magnetic domains},\ }\href
	{http://link.aps.org/abstract/PRL/v30/p230} {\bibfield  {journal} {\bibinfo
			{journal} {Physical Review Letters}\ }\textbf {\bibinfo {volume} {30}},\
		\bibinfo {pages} {230} (\bibinfo {year} {1973})}\BibitemShut {NoStop}%
	\bibitem [{\citenamefont {Lighthill}(1965)}]{LIGHTHILL65}%
	\BibitemOpen
	\bibfield  {author} {\bibinfo {author} {\bibfnamefont {M.~J.}\ \bibnamefont
			{Lighthill}},\ }\bibfield  {title} {\bibinfo {title} {Contributions to the
			theory of waves in non-linear dispersive systems},\ }\href
	{https://doi.org/10.1093/imamat/1.3.269} {\bibfield  {journal} {\bibinfo
			{journal} {{IMA} Journal of Applied Mathematics}\ }\textbf {\bibinfo {volume}
			{1}},\ \bibinfo {pages} {269} (\bibinfo {year} {1965})}\BibitemShut {NoStop}%
	\bibitem [{\citenamefont {Zakharov}\ and\ \citenamefont
		{Ostrovsky}(2009)}]{Zakharov09}%
	\BibitemOpen
	\bibfield  {author} {\bibinfo {author} {\bibfnamefont {V.}~\bibnamefont
			{Zakharov}}\ and\ \bibinfo {author} {\bibfnamefont {L.}~\bibnamefont
			{Ostrovsky}},\ }\bibfield  {title} {\bibinfo {title} {Modulation instability:
			The beginning},\ }\href {https://doi.org/10.1016/j.physd.2008.12.002}
	{\bibfield  {journal} {\bibinfo  {journal} {Physica D: Nonlinear Phenomena}\
		}\textbf {\bibinfo {volume} {238}},\ \bibinfo {pages} {540} (\bibinfo {year}
		{2009})}\BibitemShut {NoStop}%
	\bibitem [{Note1()}]{Note1}%
	\BibitemOpen
	\bibinfo {note} {Note that the nonlinear part of the dispersion relation
		\protect \textup {\hbox {\mathsurround \z@ \protect \normalfont
				(\ignorespaces \ref {eq:disp-main}\unskip \@@italiccorr )}} comes with the
		negative sign.}\BibitemShut {Stop}%
	\bibitem [{Note2()}]{Note2}%
	\BibitemOpen
	\bibinfo {note} {Reference to the supplemental materials is provided by the
		Publisher.}\BibitemShut {Stop}%
	\bibitem [{\citenamefont {Komineas}\ and\ \citenamefont
		{Papanicolaou}(2015)}]{Komineas15c}%
	\BibitemOpen
	\bibfield  {author} {\bibinfo {author} {\bibfnamefont {S.}~\bibnamefont
			{Komineas}}\ and\ \bibinfo {author} {\bibfnamefont {N.}~\bibnamefont
			{Papanicolaou}},\ }\bibfield  {title} {\bibinfo {title} {Skyrmion dynamics in
			chiral ferromagnets},\ }\href {https://doi.org/10.1103/physrevb.92.064412}
	{\bibfield  {journal} {\bibinfo  {journal} {Physical Review B}\ }\textbf
		{\bibinfo {volume} {92}},\ \bibinfo {pages} {064412} (\bibinfo {year}
		{2015})}\BibitemShut {NoStop}%
	\bibitem [{\citenamefont {Kravchuk}\ \emph {et~al.}(2018)\citenamefont
		{Kravchuk}, \citenamefont {Sheka}, \citenamefont {R\"o\ss{}ler},
		\citenamefont {van~den Brink},\ and\ \citenamefont {Gaididei}}]{Kravchuk18}%
	\BibitemOpen
	\bibfield  {author} {\bibinfo {author} {\bibfnamefont {V.~P.}\ \bibnamefont
			{Kravchuk}}, \bibinfo {author} {\bibfnamefont {D.~D.}\ \bibnamefont {Sheka}},
		\bibinfo {author} {\bibfnamefont {U.~K.}\ \bibnamefont {R\"o\ss{}ler}},
		\bibinfo {author} {\bibfnamefont {J.}~\bibnamefont {van~den Brink}},\ and\
		\bibinfo {author} {\bibfnamefont {Y.}~\bibnamefont {Gaididei}},\ }\bibfield
	{title} {\bibinfo {title} {Spin eigenmodes of magnetic skyrmions and the
			problem of the effective skyrmion mass},\ }\href
	{https://doi.org/10.1103/PhysRevB.97.064403} {\bibfield  {journal} {\bibinfo
			{journal} {Physical Review B}\ }\textbf {\bibinfo {volume} {97}},\ \bibinfo
		{pages} {064403} (\bibinfo {year} {2018})}\BibitemShut {NoStop}%
	\bibitem [{\citenamefont {Wu}\ and\ \citenamefont
		{Tchernyshyov}(2022)}]{Wu22a}%
	\BibitemOpen
	\bibfield  {author} {\bibinfo {author} {\bibfnamefont {X.}~\bibnamefont
			{Wu}}\ and\ \bibinfo {author} {\bibfnamefont {O.}~\bibnamefont
			{Tchernyshyov}},\ }\bibfield  {title} {\bibinfo {title} {How a skyrmion can
			appear both massive and massless},\ }\bibfield  {journal} {\bibinfo
		{journal} {{SciPost} Physics}\ }\textbf {\bibinfo {volume} {12}},\ \href
	{https://doi.org/10.21468/scipostphys.12.5.159}
	{10.21468/scipostphys.12.5.159} (\bibinfo {year} {2022})\BibitemShut
	{NoStop}%
	\bibitem [{\citenamefont {Mertens}\ \emph {et~al.}(1997)\citenamefont
		{Mertens}, \citenamefont {Schnitzer},\ and\ \citenamefont
		{Bishop}}]{Mertens97}%
	\BibitemOpen
	\bibfield  {author} {\bibinfo {author} {\bibfnamefont {F.~G.}\ \bibnamefont
			{Mertens}}, \bibinfo {author} {\bibfnamefont {H.~J.}\ \bibnamefont
			{Schnitzer}},\ and\ \bibinfo {author} {\bibfnamefont {A.~R.}\ \bibnamefont
			{Bishop}},\ }\bibfield  {title} {\bibinfo {title} {Hierarchy of equations of
			motion for nonlinear coherent excitations applied to magnetic vortices},\
	}\href {https://doi.org/10.1103/PhysRevB.56.2510} {\bibfield  {journal}
		{\bibinfo  {journal} {Physical Review B}\ }\textbf {\bibinfo {volume} {56}},\
		\bibinfo {pages} {2510} (\bibinfo {year} {1997})}\BibitemShut {NoStop}%
	\bibitem [{\citenamefont {Hertel}\ and\ \citenamefont
		{Schneider}(2006)}]{Hertel06}%
	\BibitemOpen
	\bibfield  {author} {\bibinfo {author} {\bibfnamefont {R.}~\bibnamefont
			{Hertel}}\ and\ \bibinfo {author} {\bibfnamefont {C.~M.}\ \bibnamefont
			{Schneider}},\ }\bibfield  {title} {\bibinfo {title} {Exchange explosions:
			Magnetization dynamics during vortex-antivortex annihilation},\ }\href
	{http://link.aps.org/abstract/PRL/v97/e177202} {\bibfield  {journal}
		{\bibinfo  {journal} {Physical Review Letters}\ }\textbf {\bibinfo {volume}
			{97}},\ \bibinfo {eid} {177202} (\bibinfo {year} {2006})}\BibitemShut
	{NoStop}%
	\bibitem [{\citenamefont {Hertel}\ \emph {et~al.}(2007)\citenamefont {Hertel},
		\citenamefont {Gliga}, \citenamefont {F\"ahnle},\ and\ \citenamefont
		{Schneider}}]{Hertel07}%
	\BibitemOpen
	\bibfield  {author} {\bibinfo {author} {\bibfnamefont {R.}~\bibnamefont
			{Hertel}}, \bibinfo {author} {\bibfnamefont {S.}~\bibnamefont {Gliga}},
		\bibinfo {author} {\bibfnamefont {M.}~\bibnamefont {F\"ahnle}},\ and\
		\bibinfo {author} {\bibfnamefont {C.~M.}\ \bibnamefont {Schneider}},\
	}\bibfield  {title} {\bibinfo {title} {Ultrafast nanomagnetic toggle
			switching of vortex cores},\ }\href
	{http://link.aps.org/abstract/PRL/v98/e117201} {\bibfield  {journal}
		{\bibinfo  {journal} {Physical Review Letters}\ }\textbf {\bibinfo {volume}
			{98}},\ \bibinfo {eid} {117201} (\bibinfo {year} {2007})}\BibitemShut
	{NoStop}%
	\bibitem [{\citenamefont {Kovalev}\ \emph {et~al.}(2003)\citenamefont
		{Kovalev}, \citenamefont {Mertens},\ and\ \citenamefont
		{Schnitzer}}]{Kovalev03a}%
	\BibitemOpen
	\bibfield  {author} {\bibinfo {author} {\bibfnamefont {A.~S.}\ \bibnamefont
			{Kovalev}}, \bibinfo {author} {\bibfnamefont {F.~G.}\ \bibnamefont
			{Mertens}},\ and\ \bibinfo {author} {\bibfnamefont {H.~J.}\ \bibnamefont
			{Schnitzer}},\ }\bibfield  {title} {\bibinfo {title} {Cycloidal vortex motion
			in easy--plane ferromagnets due to interaction with spin waves},\ }\href
	{http://dx.doi.org/10.1140/epjb/e2003-00150-3} {\bibfield  {journal}
		{\bibinfo  {journal} {Eur.~Phys.~J.}\ }\textbf {\bibinfo {volume} {B 33}},\
		\bibinfo {pages} {133} (\bibinfo {year} {2003})}\BibitemShut {NoStop}%
	\bibitem [{\citenamefont {Whitham}(1965)}]{Whitham65}%
	\BibitemOpen
	\bibfield  {author} {\bibinfo {author} {\bibfnamefont {G.~B.}\ \bibnamefont
			{Whitham}},\ }\bibfield  {title} {\bibinfo {title} {A general approach to
			linear and non-linear dispersive waves using a lagrangian},\ }\href
	{https://doi.org/10.1017/s0022112065000745} {\bibfield  {journal} {\bibinfo
			{journal} {Journal of Fluid Mechanics}\ }\textbf {\bibinfo {volume} {22}},\
		\bibinfo {pages} {273} (\bibinfo {year} {1965})}\BibitemShut {NoStop}%
	\bibitem [{\citenamefont {Whitham}(1999)}]{Whitham99}%
	\BibitemOpen
	\bibfield  {author} {\bibinfo {author} {\bibfnamefont {G.~B.}\ \bibnamefont
			{Whitham}},\ }\href
	{https://www.ebook.de/de/product/3602002/gerald_b_whitham_whitham_linear_and_nonlinear_waves.html}
	{\emph {\bibinfo {title} {Linear and Nonlinear Waves}}}\ (\bibinfo
	{publisher} {John Wiley \& Sons},\ \bibinfo {year} {1999})\BibitemShut
	{NoStop}%
	\bibitem [{\citenamefont {Karpman}\ and\ \citenamefont
		{Krushkal'}(1969)}]{Karpman69}%
	\BibitemOpen
	\bibfield  {author} {\bibinfo {author} {\bibfnamefont {V.~I.}\ \bibnamefont
			{Karpman}}\ and\ \bibinfo {author} {\bibfnamefont {E.~M.}\ \bibnamefont
			{Krushkal'}},\ }\bibfield  {title} {\bibinfo {title} {Modulated waves in
			nonlinear dispersive media},\ }\href@noop {} {\bibfield  {journal} {\bibinfo
			{journal} {Soviet Physics JETP}\ }\textbf {\bibinfo {volume} {28}},\ \bibinfo
		{pages} {277} (\bibinfo {year} {1969})}\BibitemShut {NoStop}%
	\bibitem [{\citenamefont {Karpman}(1975)}]{Karpman75}%
	\BibitemOpen
	\bibfield  {author} {\bibinfo {author} {\bibfnamefont {V.~I.}\ \bibnamefont
			{Karpman}},\ }\href@noop {} {\emph {\bibinfo {title} {Non-Linear Waves in
				Dispersive Media}}},\ \bibinfo {edition} {1st}\ ed.\ (\bibinfo  {publisher}
	{Pergamon Press},\ \bibinfo {year} {1975})\BibitemShut {NoStop}%
	\bibitem [{\citenamefont {Zakharov}\ and\ \citenamefont
		{B.}(1972)}]{Zakharov72}%
	\BibitemOpen
	\bibfield  {author} {\bibinfo {author} {\bibfnamefont {V.~E.}\ \bibnamefont
			{Zakharov}}\ and\ \bibinfo {author} {\bibfnamefont {S.~A.}\ \bibnamefont
			{B.}},\ }\bibfield  {title} {\bibinfo {title} {Exact theory of
			two-dimensional self-focusing and one-dimensional self-modulation of waves in
			nonlinear media},\ }\href@noop {} {\bibfield  {journal} {\bibinfo  {journal}
			{Sov. Phys. JETP}\ }\textbf {\bibinfo {volume} {34}},\ \bibinfo {pages} {62}
		(\bibinfo {year} {1972})}\BibitemShut {NoStop}%
	\bibitem [{\citenamefont {Akhmediev}\ \emph {et~al.}(1987)\citenamefont
		{Akhmediev}, \citenamefont {Eleonskii},\ and\ \citenamefont
		{Kulagin}}]{Akhmediev87}%
	\BibitemOpen
	\bibfield  {author} {\bibinfo {author} {\bibfnamefont {N.~N.}\ \bibnamefont
			{Akhmediev}}, \bibinfo {author} {\bibfnamefont {V.~M.}\ \bibnamefont
			{Eleonskii}},\ and\ \bibinfo {author} {\bibfnamefont {N.~E.}\ \bibnamefont
			{Kulagin}},\ }\bibfield  {title} {\bibinfo {title} {Exact first-order
			solutions of the nonlinear schr{\"o}dinger equation},\ }\href
	{https://doi.org/10.1007/bf01017105} {\bibfield  {journal} {\bibinfo
			{journal} {Theoretical and Mathematical Physics}\ }\textbf {\bibinfo {volume}
			{72}},\ \bibinfo {pages} {809} (\bibinfo {year} {1987})}\BibitemShut
	{NoStop}%
	\bibitem [{\citenamefont {Dysthe}\ and\ \citenamefont
		{Trulsen}(1999)}]{Dysthe99}%
	\BibitemOpen
	\bibfield  {author} {\bibinfo {author} {\bibfnamefont {K.~B.}\ \bibnamefont
			{Dysthe}}\ and\ \bibinfo {author} {\bibfnamefont {K.}~\bibnamefont
			{Trulsen}},\ }\bibfield  {title} {\bibinfo {title} {Note on breather type
			solutions of the {NLS} as models for freak-waves},\ }\href
	{https://doi.org/10.1238/physica.topical.082a00048} {\bibfield  {journal}
		{\bibinfo  {journal} {Physica Scripta}\ }\textbf {\bibinfo {volume} {T82}},\
		\bibinfo {pages} {48} (\bibinfo {year} {1999})}\BibitemShut {NoStop}%
	\bibitem [{mum()}]{mumax3}%
	\BibitemOpen
	\href {https://mumax.github.io/} {\bibinfo {title} {mumax3}},\ \bibinfo
	{note} {developed by DyNaMat group of Prof. Van Waeyenberge at Ghent
		University.}\BibitemShut {Stop}%
	\bibitem [{\citenamefont {Olver}\ \emph {et~al.}(2010)\citenamefont {Olver},
		\citenamefont {Lozier}, \citenamefont {Boisvert},\ and\ \citenamefont
		{Clark}}]{NIST10}%
	\BibitemOpen
	\bibinfo {editor} {\bibfnamefont {F.~W.~J.}\ \bibnamefont {Olver}}, \bibinfo
	{editor} {\bibfnamefont {D.~W.}\ \bibnamefont {Lozier}}, \bibinfo {editor}
	{\bibfnamefont {R.~F.}\ \bibnamefont {Boisvert}},\ and\ \bibinfo {editor}
	{\bibfnamefont {C.~W.}\ \bibnamefont {Clark}},\ eds.,\ \href
	{http://www.cambridge.org/us/academic/subjects/mathematics/abstract-analysis/nist-handbook-mathematical-functions}
	{\emph {\bibinfo {title} {NIST Handbook of Mathematical Functions}}}\
	(\bibinfo  {publisher} {Cambridge University Press},\ \bibinfo {address} {New
		York, NY},\ \bibinfo {year} {2010})\BibitemShut {NoStop}%
	\bibitem [{\citenamefont {Skyllingstad}(1991)}]{Skyllingstad91}%
	\BibitemOpen
	\bibfield  {author} {\bibinfo {author} {\bibfnamefont {E.~D.}\ \bibnamefont
			{Skyllingstad}},\ }\bibfield  {title} {\bibinfo {title} {Critical layer
			effects on atmospheric solitary and cnoidal waves},\ }\href
	{https://doi.org/10.1175/1520-0469(1991)048<1613:cleoas>2.0.co;2} {\bibfield
		{journal} {\bibinfo  {journal} {Journal of the Atmospheric Sciences}\
		}\textbf {\bibinfo {volume} {48}},\ \bibinfo {pages} {1613} (\bibinfo {year}
		{1991})}\BibitemShut {NoStop}%
	\bibitem [{\citenamefont {Vansteenkiste}\ \emph {et~al.}(2014)\citenamefont
		{Vansteenkiste}, \citenamefont {Leliaert}, \citenamefont {Dvornik},
		\citenamefont {Helsen}, \citenamefont {Garcia-Sanchez},\ and\ \citenamefont
		{Van~Waeyenberge}}]{Vansteenkiste14}%
	\BibitemOpen
	\bibfield  {author} {\bibinfo {author} {\bibfnamefont {A.}~\bibnamefont
			{Vansteenkiste}}, \bibinfo {author} {\bibfnamefont {J.}~\bibnamefont
			{Leliaert}}, \bibinfo {author} {\bibfnamefont {M.}~\bibnamefont {Dvornik}},
		\bibinfo {author} {\bibfnamefont {M.}~\bibnamefont {Helsen}}, \bibinfo
		{author} {\bibfnamefont {F.}~\bibnamefont {Garcia-Sanchez}},\ and\ \bibinfo
		{author} {\bibfnamefont {B.}~\bibnamefont {Van~Waeyenberge}},\ }\bibfield
	{title} {\bibinfo {title} {The design and verification of {MuMax3}},\ }\href
	{https://doi.org/10.1063/1.4899186} {\bibfield  {journal} {\bibinfo
			{journal} {AIP Advances}\ }\textbf {\bibinfo {volume} {4}},\ \bibinfo {pages}
		{107133} (\bibinfo {year} {2014})}\BibitemShut {NoStop}%
	\bibitem [{\citenamefont {Ma}(1979)}]{Ma79}%
	\BibitemOpen
	\bibfield  {author} {\bibinfo {author} {\bibfnamefont {Y.-C.}\ \bibnamefont
			{Ma}},\ }\bibfield  {title} {\bibinfo {title} {The perturbed plane-wave
			solutions of the cubic schrödinger equation},\ }\href
	{https://doi.org/10.1002/sapm197960143} {\bibfield  {journal} {\bibinfo
			{journal} {Stud. Appl. Math.}\ }\textbf {\bibinfo {volume} {60}},\ \bibinfo
		{pages} {43} (\bibinfo {year} {1979})}\BibitemShut {NoStop}%
	\bibitem [{\citenamefont {Peregrine}(1983)}]{Peregrine83}%
	\BibitemOpen
	\bibfield  {author} {\bibinfo {author} {\bibfnamefont {D.~H.}\ \bibnamefont
			{Peregrine}},\ }\bibfield  {title} {\bibinfo {title} {Water waves, nonlinear
			schrödinger equations and their solutions},\ }\href
	{https://doi.org/10.1017/s0334270000003891} {\bibfield  {journal} {\bibinfo
			{journal} {The Journal of the Australian Mathematical Society. Series B.
				Applied Mathematics}\ }\textbf {\bibinfo {volume} {25}},\ \bibinfo {pages}
		{16} (\bibinfo {year} {1983})}\BibitemShut {NoStop}%
	\bibitem [{\citenamefont {McKeever}\ \emph {et~al.}(2019)\citenamefont
		{McKeever}, \citenamefont {Rodrigues}, \citenamefont {Pinna}, \citenamefont
		{Abanov}, \citenamefont {Sinova},\ and\ \citenamefont
		{Everschor-Sitte}}]{McKeever19}%
	\BibitemOpen
	\bibfield  {author} {\bibinfo {author} {\bibfnamefont {B.~F.}\ \bibnamefont
			{McKeever}}, \bibinfo {author} {\bibfnamefont {D.~R.}\ \bibnamefont
			{Rodrigues}}, \bibinfo {author} {\bibfnamefont {D.}~\bibnamefont {Pinna}},
		\bibinfo {author} {\bibfnamefont {A.}~\bibnamefont {Abanov}}, \bibinfo
		{author} {\bibfnamefont {J.}~\bibnamefont {Sinova}},\ and\ \bibinfo {author}
		{\bibfnamefont {K.}~\bibnamefont {Everschor-Sitte}},\ }\bibfield  {title}
	{\bibinfo {title} {Characterizing breathing dynamics of magnetic skyrmions
			and antiskyrmions within the hamiltonian formalism},\ }\href
	{https://doi.org/10.1103/physrevb.99.054430} {\bibfield  {journal} {\bibinfo
			{journal} {Physical Review B}\ }\textbf {\bibinfo {volume} {99}},\ \bibinfo
		{pages} {054430} (\bibinfo {year} {2019})}\BibitemShut {NoStop}%
	\bibitem [{Note3()}]{Note3}%
	\BibitemOpen
	\bibinfo {note} {For smaller fields (thicker strings), we use samples of
		larger lateral size.}\BibitemShut {Stop}%
\end{thebibliography}

%

\end{document}